\documentclass[journal]{IEEEtran}
\usepackage{latexsym}

\usepackage{enumerate}
\usepackage{graphicx}
\usepackage{subfigure}
\usepackage{multirow}

\usepackage{framed,multirow}
\usepackage{times}
\usepackage{graphicx} 
\usepackage{subfigure}
\usepackage{amssymb}
\usepackage{latexsym}

\usepackage{algorithm}
\usepackage{algorithmic}

\usepackage{hyperref}

\usepackage{url}
\usepackage{xcolor}
\usepackage{amsmath}
\usepackage{amsxtra}
\usepackage{braket}
\usepackage{array}
\usepackage{color}
\usepackage{url}
\usepackage{placeins}
\usepackage{threeparttable}

\usepackage{times}
\usepackage{balance}
\usepackage{enumerate}
\usepackage{multirow}
\usepackage{pifont}
\usepackage{marvosym}
\usepackage{ifsym}

\begin{document}
\title{Entropic Dynamic Time Warping Kernels for Co-evolving Financial Time Series Analysis}
\author{Lu~Bai,~\IEEEmembership{}Lixin~Cui,~\IEEEmembership{}Lixiang~Xu,~\IEEEmembership{}Yue Wang,~\IEEEmembership{}Zhihong Zhang,~\IEEEmembership{}Edwin~R.~Hancock,~\IEEEmembership{IEEE~Fellow}

\thanks{Lu Bai, Lixin Cui, and Yue Wang are with Central University of Finance and Economics, Beijing, China. L. Bai (bailucs@cufe.edu.cn) and L. Cui (cuilixin@cufe.edu.cn) have equal contributions and are co-first authors.}
\thanks{Lixiang Xu is with Hefei University, Anhui, China.}
\thanks{Zhihong Zhang is with Xiamen University, Fujian, China. zhihong@xmu.edu.cn}
\thanks{Edwin R. Hancock is with University of York, York, UK. Email: edwin.hancock@york.ac.uk }}

\markboth{Journal of \LaTeX\ Class Files,~Vol.~6, No.~1, January~2007}%
{Shell \MakeLowercase{\textit{et al.}}: Bare Demo of IEEEtran.cls
for Journals}
\maketitle

\begin{abstract}
Network representations are powerful tools for the analysis of time-varying financial complex systems consisting of multiple co-evolving financial time series, e.g., stock prices, etc. In this work, we develop a novel framework to measure the similarity between dynamic financial networks, i.e., time-varying financial networks. Particularly, we explore whether the proposed similarity measure can be employed to understand the structural evolution of the financial networks with time. For a set of time-varying financial networks with each vertex representing the individual time series of a different stock and each edge between a pair of time series representing the absolute value of their Pearson correlation, our start point is to compute the commute time matrix associated with the weighted adjacency matrix of the network structures, where each element of the matrix can be seen as the enhanced correlation value between pairwise stocks. For each network, we show how the commute time matrix allows us to identify a reliable set of dominant correlated time series as well as an associated dominant probability distribution of the stock belonging to this set. Furthermore, we represent each original network as a discrete dominant Shannon entropy time series computed from the dominant probability distribution. With the dominant entropy time series for each pair of financial networks to hand, we develop a similarity measure based on the classical dynamic time warping framework, for analyzing the financial time-varying networks. We show that the proposed similarity measure is positive definite and thus corresponds to a kernel measure on graphs. The proposed kernel bridges the gap between graph kernels and the classical dynamic time warping framework for multiple financial time series analysis. Experiments on time-varying networks extracted through New York Stock Exchange (NYSE) database demonstrate that the proposed method can effectively detect abrupt changes in networks as time series structures and can be used to characterize different stages in time-varying financial network evolutions.
\end{abstract}


\IEEEpeerreviewmaketitle

\section{Introduction}\label{s1}

The financial market can be considered as a complex time-varying system consisting of multiple interacting financial components~\cite{DBLP:journals/amc/YinSX15}, e.g., the stock trade price and return rate. Due to the evolution of these financial variables with time, multiple co-evolving financial time series can be generated from the original data. For the objective of analyzing the time-varying financial market, a variety of time series analysis methods have been developed for anomaly detection applications. These include change point detection, sequence detection, and pattern detection in the time series evolution~\cite{DBLP:journals/kbs/RenLLP17,DBLP:journals/tsmc/ChaovalitwongseFS07,DBLP:journals/jcam/ChenZ09}. Among these applications, change point detection plays an important role for financial risk analysis and aims to identify abrupt changes in the time series properties~\cite{DBLP:journals/nn/LiuYCS13}. Unfortunately, detecting such crucial points remains challenging, since it is difficult to detect the changes that cannot be easily observed for a system consisting of complex inteactions between its constituent co-evolving time series~\cite{silva2015modular}. One way to overcome this problem is to represent multiple co-evolving financial time series as a family of time-varying financial networks, \textbf{with each vertex representing an individual time series of a stock (e.g., stock trading price) and each edge between a pair of co-evolving financial time series representing their degree of correlation (i.e., the absolute value of their Pearson correlation)}. As a result, network-based methods can be directly employed for analysis.

The aim of this paper is to define a new kernel-based approach for analyzing multiple co-evolving financial time series that are represented as network structures. Our work is based on representing each financial network as discrete entropy time series as well as the classical dynamic time warping measure between the series. The proposed approach bridges the gap between graph kernels and the classical dynamic time warping framework for time series analysis.

\begin{figure*}
\centering
\subfigure{\includegraphics[width=1\linewidth]{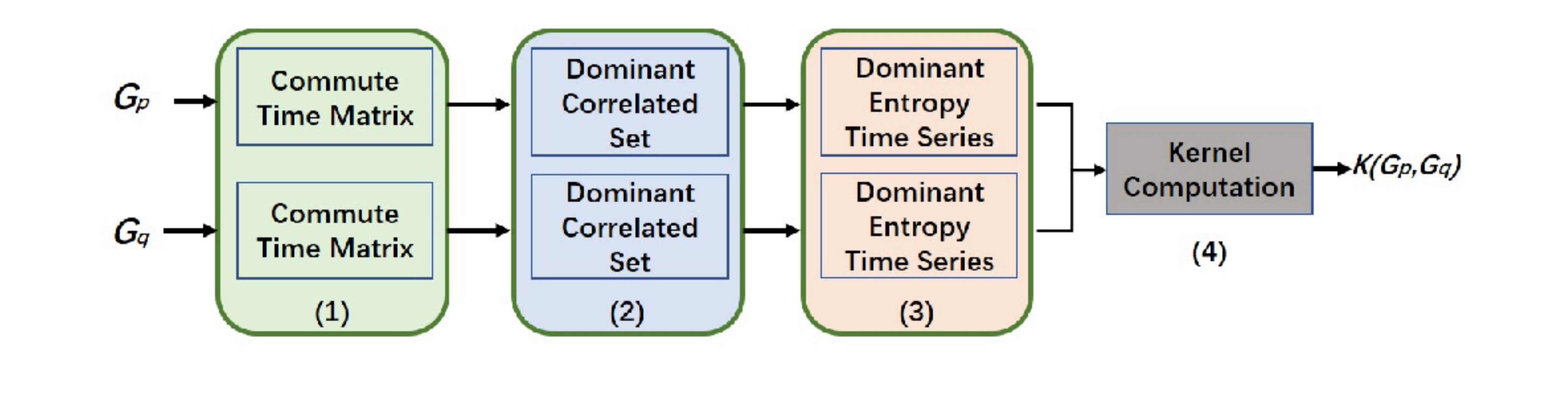}}
\vspace{-25pt}
\caption{\small{The architecture to compute the proposed kernel. Given a pair of time-varying financial networks, for each of them we (1) compute the commute time (CT) matrix, (2) identify the set of dominant correlated time series represented by vertices through CT, (3) compute the dominant probability distribution and represent each network as dominant entropy time series, and finally (4) compute the kernel between the dominant entropy time series of two networks.}} \label{flow}
\vspace{-10pt}
\end{figure*}

\subsection{Literature Review}
Network representations are powerful tools that can be employed for the analysis of time-varying complex systems consisting of multiple co-evolving time series~\cite{NetworkScience1,NetworkScience2,NetworkScience3,silva2015modular,ye2015thermodynamic}, e.g., the stock market with trade price, climate data, and functional magnetic resonance images. This approach is based on the idea that the structure of so-called time-varying complex networks~\cite{bullmore2009complex} inferred from the corresponding time series of the system can represent physical interactions between system entities that are richer than the original individual time series. According to this approach, one of the main objectives is to identify the extreme events which may considerably change the network structure. For example, in time-varying financial networks, extreme events corresponding to the financial instability of the stock are of particularly interest~\cite{silva2015modular} and can be inferred by detecting the anomalies in the corresponding networks~\cite{ye2015thermodynamic}. The network structure before and after an extreme event should be significantly different.

Broadly speaking, most existing approaches aim to characterize networks based on two principle approaches, namely a) derive network characteristics using connectivity structures, or statistics capturing connectivity structures and b) characterize the networks using statistical physics. Proponents of the former approach focus on capturing network substructures using communities, hubs and clusters~\cite{feldman1998measures,anand2011shannon,anand2014entropy}. On the other hand, proponents of the latter approach describe the network structures based on the partition function to characterize the network structures, and the corresponding temperature, energy, and entropy measures can be calculated in terms of this function~\cite{BOOK1,javarone2013quantum,delvenne2011centrality,fronczak2007thermodynamic,ye2015thermodynamic}. Unfortunately, both approaches tend to approximate structural relationships of networks in a low dimensional pattern space, hence leading to substantial loss of information. This shortcoming affects the effectiveness of existing network methods for time series analysis. One principle approach to address this drawback is to adopt graph kernels. In pattern recognition, graph kernels are powerful tools for analyzing graph-based structural data. The main advantage of adopting graph kernels is that they provide an effective way of mapping graph structures into a high dimensional Hilbert space and thus better encapsulate the structural information.

Most existing state-of-the-art graph kernels fall into the scenario of R-convolution kernels, that are originally proposed by Haussler in 1999~\cite{haussler99convolution}. The main idea underpinning R-convolution kernels is based on decomposing graphs into substructures and measuring the similarity between each pair of input graphs in terms of their isomorphic substructures, e.g., graph kernels based on comparing pairs of isomorphic a) walks, b) subgraphs, and c) subtrees. Representative R-convolution graph kernels based on substructures include the Weisfeiler-Lehman subtree kernel~\cite{shervashidze2010weisfeiler}, the tree-based continuous attributed kernel~\cite{DBLP:journals/tnn/MartinoNS18}, the aligned subtree kernel~\cite{DBLP:conf/icml/Bai0ZH15}, the Jensen-Tsallis q-difference graph kernel~\cite{DBLP:conf/pkdd/Bai0BH14}, the optima assignment Weisfeiler-Lehman kernel~\cite{DBLP:conf/nips/KriegeGW16}, the core variants-based shortest path kernel~\cite{DBLP:conf/ijcai/NikolentzosMLV18}, the random walk graph kernel~\cite{DBLP:conf/icml/KashimaTI03}, etc. Unfortunately, directly employing these graph kernels to analyze time-varying network structures inferred from the original time series tends to be elusive. Because such network structures in most real-world applications are by nature complete weighted graphs, i.e., each vertex is adjacent to all remainder vertices, whereas the edge weights between the vertices may be rather different. It is difficult to decompose such a graph into substructures. This in turn influences the effectiveness of most existing graph kernels.

One way to address the problem is to discard the less interacted information between a pair of vertices and adopt the sparser versions of original time-varying networks, i.e., the sparser networks only preserve the original edges indicating pairs of more interacted vertices. Under this scenario, Ye et al.~\cite{ye2015thermodynamic}, Silva et al.~\cite{silva2015modular} and Wang et al.~\cite{wang2018PRL} have taken the widely adopted threshold-based methods and preserved the edges whose weights fall into the larger $10\%$ of correlation-based weights. Although this strategy provides a way of directly employing existing graph kernels to accommodate time-varying networks for multiple co-evolving time series analysis, these sparse structures rely on the selection of the threshold. Thus, it is not clear how to preliminarily select a suitable threshold. Moreover, these sparser structures also lead to significant information loss, because many weighted edges are discarded. In summary, analyzing time-varying networks associated with state-of-the-art graph kernels remains challenges.


\subsection{Contributions}
The objective of this paper is to address the aforementioned problems and develop a new kernel-based approach for analyzing multiple co-evolving financial time series. Specifically, we propose an Entropic Dynamic Time Warping Kernel (EDTWK) for time-varying financial networks, \textbf{with each vertex representing the individual time series of a different stock (e.g., stock trading price) and each edge between a pair of co-evolving financial time series representing the absolute value of their Pearson correlation}. One key innovation of the proposed EDTWK kernel is the automatic identification of the dominant correlated vertex subset for each of the financial networks, \textbf{i.e., the proposed kernel incorporates the process of identifying the most mutually correlated stocks specified by the vertex subset}. In contrast, the aforementioned methods through the threshold-based strategy cannot guarantee that the preserved vertices correspond to a more mutually correlated vertex subset. This is because these methods tend to individually select each edge with a higher correlation weight and many edges between the preserved vertices may not exist. Based on financial risk theory~\cite{finance}, \textbf{the financial crises are usually caused by a set of the most mutually correlated stocks while having less uncertainty}. As a result, \textbf{the proposed EDTWK kernel cannot only overcome the shortcoming of heuristically selecting the threshold value that arises in the threshold-based approach for time-varying network analysis~\cite{ye2015thermodynamic,silva2015modular}, but also capture more reliable information concerning the evolution of the financial system to hand}. The computational framework of the proposed EDTWK kernel is shown in Fig.\ref{flow}. Specifically, the main contributions of this work are threefold.

\textbf{First}, for a family of time-varying financial networks, our start point is to compute the commute time matrices associated with their original weighted adjacency matrices, i.e., the absolute Pearson correlation based matrices. The reason of using the commute time matrix as the representation of each network structure is that each element of this matrix represents the average path length between a pair of vertices over all possible paths residing on the original weighted edges~\cite{DBLP:journals/pami/QiuH07a}. Thus, \textbf{the commute time can be seen as the enhanced absolute Pearson correlation value between the time series of pairwise stocks}, i.e., it integrates the effectiveness of all possible correlation-based paths between a pair of vertices of the original network. Moreover, the commute time is robust under the perturbation
of the network structure (e.g., the changes of edges or paths~\cite{DBLP:journals/pami/QiuH07a}). As a result, the commute time matrix can provide a more stable representation for the financial network structure that may accumulate a lot of noises over time. In summary, the commute time matrix offers an elegant way to probe the original structure of the time-varying financial networks (see details in Section~\ref{s2:1}). More specifically, \textbf{the proposed approach associated with the commute time matrix will be more effective than that associated with the original absolute Pearson correlation matrix} (see details in Section~\ref{exp:s2} and Section~\ref{exp:s3}).


\textbf{Second}, with the commute time matrix of each time-varying financial network to hand, we employ this matrix to automatically identify a set of dominant correlated vertices in the network structure (i.e., a set of the most mutually correlated time series represented by the set of vertices), by maximizing a quadratic programming problem associated with the commute time matrix. Specifically, we compute a dominant probability distribution of these time series belonging to the most mutually correlated set. We show that this strategy not only overcomes the shortcoming of existing threshold-based approaches~\cite{ye2015thermodynamic,silva2015modular} that roughly select pairs of relatively more correlated time series, but also encapsulates reliable information in terms of the evolution of the financial system to hand. Furthermore, we transform each original time-varying financial network into a discrete dominant entropy time series associated with the dominant probability distribution, i.e., we characterize the uncertainty of each network structure within the financial system to hand in terms of the classical Shannon entropy associated with the probability distribution. With each pair of entropy time series to hand, we compute the EDTWK kernel through the classical dynamic time warping framework. We show that the proposed kernel not only accommodates the complete weighted graphs through the commute time matrix, but also bridges the gap between graph kernels and the classical dynamic time warping framework for time series analysis (see details in Section~\ref{s3.2}).

\textbf{Third}, we perform the proposed kernel on time-varying financial networks extracted from New York Stock Exchange (NYSE) data. Experimental results demonstrate that \textbf{the proposed method can preserve the ordinal arrangement of the time-varying financial networks}, and thus well understand the structural evolution of the networks with time, i.e., the proposed kernel can effectively detect abrupt changes in networks as time series structures and can be used to characterize different stages in time-varying financial network evolutions.


\subsection{Paper Outline}

This paper is organized as follows. Section~\ref{s2} reviews the preliminary concepts. Section~\ref{s3} defines the EDTWK kernel for time series analysis. Section~\ref{s4} provides the empirical evaluation results. Section~\ref{s5} provides the conclusion and future work of this paper.

\section{Preliminary Concepts}\label{s2}
In this section, we briefly review preliminary concepts which will be utilized in this paper. We first review the concept of the commute time. Furthermore, we review the concept of a dynamic time warping framework inspired kernel.
\subsection{Commute Time on Graphs}\label{s2:1}
As we have stated, one main objective of this work is to automatically identify a set of most mutually correlated stocks in terms of their time series. To this end, we require a correlation matrix as the structural representation of the corresponding time-varying financial network (i.e., the weighted adjacency matrix of the network), with each vertex representing the individual time series of a different stock and each edge representing the correlation between a pair of co-evolving financial time series. Broadly speaking, most state-of-the-art approaches usually adopt the absolute Pearson correlation based matrix as the network representation~\cite{ye2015thermodynamic,wang2018PRL,silva2015modular}. In order to capture a reliable and robust mutually correlated stock set, in this work we propose to utilize the commute time matrix associated with the original correlation matrix as the network representation.

The main reasons of employing the commute time matrix are threefold. First, the commute time averages the time taken for a random walk to travel between a pair of vertices over all connecting paths residing on the original correlation based weighted adjacency matrix. Thus, the commute time can be considered as the enhanced correlation matrix. Second, since the commute time amplifies the correlation based affinity between a pair of vertices, it is robust under the perturbation of the graph structure, e.g., the changes of edges or paths. Thus, the commute time based enhanced correlation matrix is robust and provides a stable correlation representation for the time-varying financial network that may accumulate a lot of noises over time. Third, the commute time is calculated through the Laplacian matrix of the original correlation based weighted adjacency matrix. In Section~\ref{s3}, we will show how the commute time matrix can be employed to identify a set of most mutually correlation stocks specified by a set of dominant vertices, associated with a quadratic problem.

In this subsection, we briefly introduce the concept of the commute time. Assume $G(V,E,A)$ is a complete weighted graph, where $E$ is the edge set, $V$ is the vertex set $V$, and each vertex of $V$ is connected by all the remainder vertices,. Let $A$ be the associated weighted adjacency matrix of $G$. If $A(u,v)=A(v,u)>0$, we say that the vertices $v\in V$ and $u\in V$ are adjacent. Let $D$ denote the degree matrix of $G$. $D$ is a diagonal matrix and each of its diagonal element $D(u,u)$ corresponds to the sum of the corresponding row or column of $A$, i.e., $D(u,u) = \sum_v A(u,v)$. Then, the graph Laplacian matrix $L$ is computed by $L = D-A$. The spectral decomposition of $L$ is defined as $L=\Phi \Lambda \Phi^T$, where $\Lambda = \mbox{diag}(\lambda_1, \lambda_2, ..., \lambda_n)$ is a $|V| \times |V|$ diagonal matrix with ascending eigenvalues as elements, i.e., $0 = \lambda_1 \leq \lambda_2 \leq ... \leq \lambda_{|V|}$, and $\Phi$ is a $|V| \times |V|$ matrix $\Phi=(\phi_1 | \phi_2 | ... | \phi_{|V|})$ with the corresponding ordered eigenvectors as columns. For $G$, the hitting time $H(u,v)$ between each pair of vertices $v$ and $u$ is computed as the expected number of steps taken by a classical random walk commencing from $u$ and ending at $v$. Likewise, the commute time $C(u,v)$ is defined as the expected number of steps of the random walk commencing from $u$ and ending at $v$, and then coming back to $u$ again, i.e., $C(u,v)=H(u,v)+H(v,u)$. Thus, the commute time $C(u,v)$ can be calculated through the unnormalized Laplacian eigenvalues and eigenvectors~\cite{DBLP:journals/pami/QiuH07a} as
\begin{equation}
C(u,v) = \sum_{u\in V} D(u,u) \sum_{j=2}^{|V|} \frac{1}{\lambda_j}(\phi_j(u)-\phi_j(v))^2.
\end{equation}

\noindent\textbf{Remarks:} The commute time has been proven to be a powerful tool to extract rich characteristics from complete weighted graphs. In previous works, Bai et al.~\cite{DBLP:journals/ijon/CuiBZWH19} have employed the commute time matrix to develop a new quantum-inspired kernel for dynamic financial network analysis. Specifically, for the original complete weighted adjacency matrix of each financial network, they commence by abstracting the minimum or maximum spanning tree associated with the commute time matrix. For a pair of complete weighted graphs to be compared, the resulting quantum kernel is defined by measuring the similarity between their associated commute time spanning tree structures in terms of a new developed evolving model of discrete-time quantum walks. This approach significantly reduces the problem of information loss that arises in previously mentioned threshold-based methods for financial network analysis~\cite{ye2015thermodynamic,wang2018PRL,silva2015modular}. This is because the weights of the preserved edges on spanning tree structures correspond to the commute time values between corresponding pairs of vertices, and the commute time values integrates the effectiveness over all possible paths residing on the original weighted edges. However, similar to these threshold-based approaches~\cite{ye2015thermodynamic,wang2018PRL,silva2015modular}, the quantum kernel~\cite{DBLP:journals/ijon/CuiBZWH19} cannot guarantee that the preserved vertices correspond to a more mutually correlated vertex subset, since the spanning tree is a very sparse structure (only $n-1$ edges preserved for the network with $n$ vertices) and many edges between the vertices do not exist. In other words, this kernel approach cannot reflect the most mutually correlated time series specified by the vertices, and will influence the effectiveness. To overcome this problem, in Section~\ref{s3}, we will develop a new kernel-based approach for financial network analysis that can integrate the process of adaptively identifying the most mutually correlated financial time series of stocks associated with the commute time matrix.

\subsection{The Dynamic Time Warping Framework}\label{DTW}
We review the global alignment kernel that is defined through the classical dynamic time warping framework~\cite{DBLP:conf/icml/Cuturi11}. Assume $\mathbf{T}$ is a set of discrete time series that take values in a space $\mathcal{X}$. For each pair of discrete time series $\mathbf{P}=(p_1,\ldots,p_m)\in \mathbf{T}$ and $\mathbf{Q}=(q_1,\ldots,q_n)\in \mathbf{T}$ with lengths $m$ and $n$ respectively, the alignment $\pi$ between $\mathbf{P}$ and $\mathbf{Q}$ is computed as a pair of increasing integral vectors $(\pi_p,\pi_q)$ of length $l\leq m+n-1$, where $$1=\pi_p(1)\leq \cdots \leq \pi_p(l)=m$$ and $$1=\pi_q(1)\leq \cdots \leq \pi_q(l)=n$$ such that $(\pi_p,\pi_q)$ is assumed to possess unitary increments and no simultaneous repetitions. For $\mathbf{P}$ and $\mathbf{Q}$, each of their elements can be \textbf{an observation vector with fixed dimensions} at a corresponding time step. For any index $i$ that is between $1$ and $l-1$ (i.e., $1\leq i \leq l-1$), the following condition holds for the increment vector of $\pi=(\pi_p,\pi_q)$, i.e.,
\begin{equation}
\left(
\begin{array}{l}
\pi_p(i+1)-\pi_p(i)\\
\pi_q(i+1)-\pi_q(i)
\end{array}\right) \in
\left\{
\left(
\begin{array}{l}
0\\
1
\end{array}\right),
\left(
\begin{array}{l}
1\\
0
\end{array}\right),
\left(
\begin{array}{l}
1\\
1
\end{array}\right)
\right\}.
\end{equation}
Within the framework of the classical dynamic time warping~\cite{DBLP:conf/icml/Cuturi11}, the coordinates $\pi_p$ and $\pi_q$ of the alignment $\pi$ define the warping function. Assume $\mathcal{A}(m,n)$ corresponds to a set of all possible alignments between $\mathbf{P}$ and $\mathbf{Q}$, Cuturi~\cite{DBLP:conf/icml/Cuturi11} has proposed a dynamic time warping inspired kernel, namely the Global Alignment Kernel, by considering all the possible alignments in $\mathcal{A}(m,n)$. The kernel is defined as
\begin{equation}
k_{\mathrm{GA}}(\mathbf{P},\mathbf{Q})=\sum_{\pi\in \mathcal{A}(m,n)} e^{-{D}_{\mathbf{P},\mathbf{Q}}(\pi)},\label{GAK}
\end{equation}
where $D_{\mathbf{P},\mathbf{Q}}(\pi)$ is the alignment cost given by
\begin{equation}
D_{\mathbf{P},\mathbf{Q}}(\pi)=\sum_{i=1}^{|\pi|} \varphi (p_{\pi_p(i)},q_{\pi_q(i)}),\label{DTWD}
\end{equation}
and is defined through a local divergence $\varphi$ that quantifies the discrepancy between each pair of elements $p_i\in \mathbf{P}$ and $q_i\in \mathbf{Q}$. In general, $\varphi$ is defined as the squared Euclidean distance. Note that,
the kernel $k_{\mathrm{GA}}$ measures the quality of both the optimal alignment and all other alignments $\pi\in \mathcal{A}(m,n)$, thus it is positive definite. Moreover, $k_{\mathrm{GA}}$ provides richer statistical measures of similarity by encapsulating the overall spectrum of the alignment costs $\{{D}_{\mathbf{P},\mathbf{Q}}(\pi),\pi \in \mathcal{A}(m,n)\}$.\\

\noindent\textbf{Remarks:} The dynamic time warping based global alignment kernel $k_{\mathrm{GA}}$ has been proven to be a powerful tool of analyzing vectorial time series~\cite{DBLP:conf/icml/Cuturi11}. To extend $k_{\mathrm{GA}}$ into the graph kernel domain, Bai et al.~\cite{DBLP:journals/prl/Bai0CH18} have developed a family of nested graph kernels through $k_{\mathrm{GA}}$. Specifically, they commenced by decomposing each graph structure into a family of $K$-layer expansion subgraphs rooted at the centroid vertex. The nested depth-based complexity trace of each graph is computed by measuring the entropy on the family of $K$-layer expansion subgraphs. Since the parameter $K$ varies from $1$ to $K$, this complexity trace naturally forms a one-dimensional sequence-based characterization vector, that is similar to the one-dimensional time series vector. As a result, for a pair of graphs the resulting dynamic time warping based kernel can be directly computed by measuring the global alignment kernel $k_{\mathrm{GA}}$ between their complexity traces. Although, they demonstrated that the nested graph kernels outperform state-of-the-art graph kernels~\cite{DBLP:conf/icml/KashimaTI03,DBLP:journals/jmlr/ShervashidzeVPMB09,DBLP:conf/icml/JohanssonJDB14} on graph classification tasks. Unfortunately, as we have stated, the financial networks are by nature complete weighted graphs and it is difficult to decompose such network structures into the required expansion subgraphs rooted at the centroid vertex. As a result, directly preforming the dynamic time warping inspired kernel $k_{\mathrm{GA}}$ for time-varying financial networks tends to be elusive and remains challenges.

\section{The Kernel for Time-varying Networks}\label{s3}
In this section, we propose a kernel-based similarity measure for time-varying networks representing multiple co-evolving financial time series. Specifically, we commence by identifying a set of most mutually correlated time series through maximizing a quadratic programming method on the commute time matrix. We exhibit how this allows us to compute a probability distribution for the time series belonging to the dominant set. Finally, we characterize each time-varying network as a discrete dominant entropy time series through the Shannon entropy associated with the probability distribution, and in turn develop a new kernel-based approach in terms of the classical dynamic time warping framework~\cite{DBLP:conf/icml/Cuturi11}.  


\subsection{Identifying Dominant Correlated Time Series}\label{s3.1}

We identify a set of the most mutually correlated time series for each time-varying financial network. Let $\mathbf{G}=\{G_1,\ldots,G_t,\ldots,G_T\}$ be a family of time-varying financial networks extracted from a complex system $\mathbf{S}$ and $G_t(V_t,E_t, A_t)$ be the sample network extracted from the system at time $t$. For $G_t$, each vertex $v_t\in V_t$ represents the time series of a different stock (e.g., the stock price), each edge $e_t\in E_t$ represents the absolute Pearson correlation between a pair of time series, and $A_t$ is the absolute Pearson correlation based weighted adjacency matrix. In fact, this manner of constructing each network $G_t$ is a popular way to represent multiple co-evolving financial time series~\cite{NetworkScience1,NetworkScience2,NetworkScience3,silva2015modular,ye2015thermodynamic}. Note that, in this paper \textbf{we assume that the time-varying network structures have fixed numbers of vertices}, i.e., these networks have the same vertex set $V_t$, whereas the edge sets $E_t$ are quite different with time $t$. In real-world application, this a very common situation and usually appears where the time-varying networks are extracted from complex systems with a specified set of co-evolving time series, i.e., the system $\mathbf{S}$ has a fixing number of components co-evolving with time.


\begin{figure}
\centering
\subfigure{\includegraphics[width=0.7\linewidth]{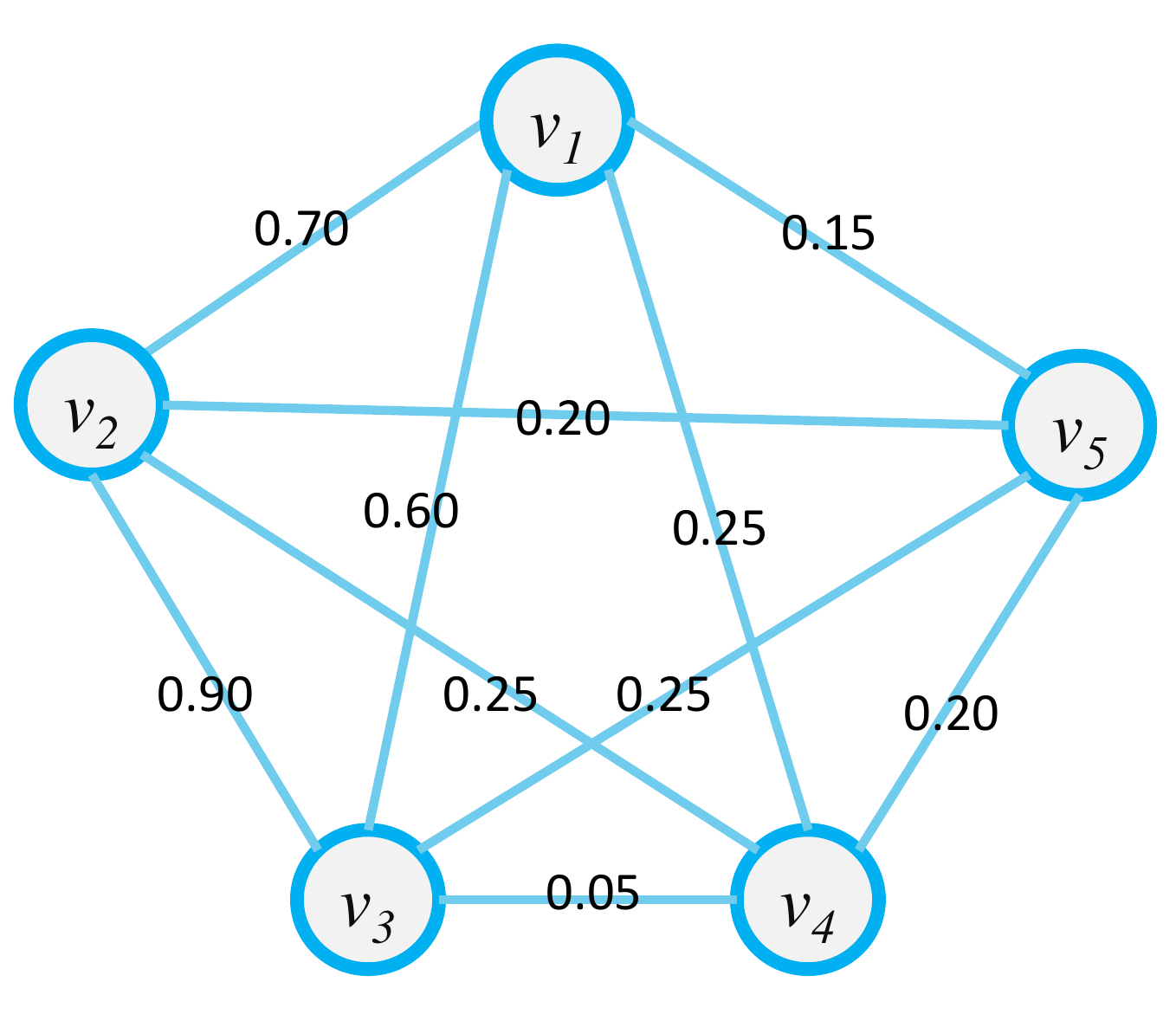}}
\vspace{-10pt}
\caption{\small{The subset of financial time series specified by vertices\{$v_1,v_2,v_3$\} is dominant.}} \label{ds}
\vspace{-10pt}
\end{figure}

For each network $G_t$, we first compute its commute time matrix as ${\mathrm{C}}_t$ associated with its original absolute Pearson correlation based adjacency matrix. As we have stated previously, the commute time not only reflects the integrated effectiveness of all possible weighted paths between a pair of vertices of the original network structure, but is also robust with the perturbation of the network structure (i.e., the changes of edge weight on the original weighted adjacency matrix). As a result, the commute time matrix ${\mathrm{C}}_t$ can be seen as a reliable enhanced absolute Pearson correlation matrix for $G_t$. In other words, the commute time matrix provides a stable representation to further characterize the dynamic network $G_t$ associated with time-varying correlations between vertices.

With the commute time matrix $\mathrm{C}_t$ of each network $G_t$ to hand, we automatically identify a set of dominant correlated time series through the dominant set problem proposed by Pavan et al.~\cite{DBLP:journals/pami/PavanP07}. The definition of the dominant set simultaneously emphasizes internal homogeneity and together with external inhomogeneity, and can be employed as a general definition of a cluster. An instance is exhibited in Fig.\ref{ds}. Here, assume a time-varying financial network consisting of $5$ vertices denoted as $v_1$, $v_2$, $v_3$, $v_4$ and $v_5$. Each weight of this network represents the correlation between pairwise vertices. For this instance, the subset ${\mathrm{DS}}=\{v_1,v_2,v_3\}$ forms the dominant set, i.e., the internal set. This is because the sum of the edge weights between the internal set $\{v_1,v_2,v_3\}$ is larger than the sum of those between the internal and external sets. As a result, the time series specified by ${\mathrm{DS}}$ can seen as the set of the most mutually correlated time series. To automatically identify the most mutually correlated time series from $G_t$, we can solve the corresponding dominant set problem by maximizing a quadratic program problem~\cite{DBLP:journals/pami/PavanP07}. More specifically, associated with $\mathrm{C}_t$, we compute the solution $\mathbf{a}$ of the following quadratic program problem~\cite{DBLP:journals/pami/PavanP07}
\begin{align}\label{solution vector}
&\mathrm{argmax}\  \frac{1}{2}\mathbf{a}^{T}\mathbf{C}_t\mathbf{a}
\end{align}
subject to $\mathbf{a} \in \mathbb{R}^{|V_t|}$, ${a_i} \geq 0$ and $\sum_{i=1}^{|V_t|}a_{i} = 1$. The solution vector $\mathbf{a}$ of Eq.(\ref{solution vector}) is an $|V_t|$-dimensional vector. When $a_{i}>0$, the $i$-th time series represented by the vertex $v_i\in V_t$ belongs to the most correlated time series subset of $G_t$. Thus, the number of the selected time series $n$ is specified by counting the number of all positive components of $\mathbf{a}$. Based on the definition of Pavan and Pelillo~\cite{DBLP:journals/pami/PavanP07}, we can solve the local maximum of $f(\mathbf{a})$ by
\begin{equation}
a_{i}(k+1) =
a_{i}(k)\frac{(\mathbf{C}_t\mathbf{a}(k))_{i}}{\mathbf{a}(k)^{T}\mathbf{C}_t\mathbf{a}(k)}~.\label{iteration update}
\end{equation}
where $a_i(k)$ corresponds to the $i$-th time series represented by $v_i\in V_t$ at
iteration $k$. Based on the element value of $\mathbf{a}$, all time series represented by the vertices $v_{1},\ldots, v_{|V_t|}$ fall into two disjoint subsets, i.e., $$\bold S_{1}(a) = \{v_{i}\in V_t\ |\ a_{i}>0\}$$ and $$\bold S_{2}(a) = \{v_{j}\in V_t\ |\  a_{j}=0\}.$$ Clearly, the set $\bold S_1$ with nonzero values indicates the set of dominant correlated time series, i.e., the set of the most mutually correlated time series. Finally, note that, the solution vector $\mathbf{a}$ also corresponds to a probability distribution of the time series belonging to the dominant set $\bold S_{1}$, i.e., each element $a_i$ corresponds to the probability of the $i$-th time series belonging to $\bold S_{1}$.

\subsection{The Entropic Dynamic Time Warping Kernel}\label{s3.2}

In this subsection, we develop a new kernel method for analyzing time-varying financial networks based on the classical dynamic time warping framework. To this end, we commence by representing the complex networks as discrete dominant entropy time series using the Shannon entropy through the most mutually correlated time series set introduced in Section \ref{s3.1}. \textbf{The reason of characterizing the network using the entropy measure is that the Shannon entropy is an effective way of measuring the uncertainty in the corresponding financial system, associated with the probability distribution of the stocks belonging to the correlated set}. Specifically, for each sample network $G_t(V_t, E_t)$ from $\mathbf{G}$ at time $t$, we first compute the associated commute time matrix $\mathrm{C}_t$. Moreover, by solving the quadratic program problem~\cite{DBLP:journals/pami/PavanP07} on the commute time matrix $\mathrm{C}_t$, we identify the set of dominant correlated time series $\bold S_{1}$ and compute the associated probability distribution $\mathbf{a}$ of the time series belonging to $\bold S_{1}$. Based on Section \ref{s3.1}, the remaining non-dominant correlated time series are included in the set $\bold S_{2}$. With the probability distribution $\mathbf{a}$ to hand, the dominant Shannon entropy of $G_t$ is computed as
\begin{equation}\label{shannon}
H_S(\mathbf{a})=-\sum_{i=1}^{|V_t|} a_{i} \log a_{i},
\end{equation}
where $a_i$ is the probability of the $i$-th time series represented by vertex $v_i\in V_t$. Eq.(\ref{shannon}) indicates that the dominant Shannon entropy is computed by the sum of elements $-a_{i} \log a_{i}$, thus each element $-a_{i} \log a_{i}$ can be seen as a \textbf{dominant sub-entropy} $\bar{H}_S(v_i)$ of the $i$-th time series represented by vertex $v_i$, i.e.,
\begin{equation}\label{sub_shannon}
\bar{H}_S(v_i)=-a_{i} \log a_{i}.
\end{equation}
Note that, if $a_i=0$, we say that the $i$-th time series does not belong to $\bold S_{1}$ and we set $-a_{i} \log a_{i}=0$. With the sub-entropies of all vertices to hand, we compute the dominant entropy characteristics for each network $G_t$ at time $t$ as
\begin{equation}
E_t=\{\bar{H}_S(v_1),\ldots,\bar{H}_S(v_i),\ldots,\bar{H}_S(v_{V})\}^\top,\label{QET}
\end{equation}
where
\begin{equation}
\bar{H}_S(v_i)=\left\{
\begin{array}{l}
-a_{i} \log a_{i} , \ \ \  a_i>0,\ \mathrm{i.e.,}\ v_i\in \bold S_{1}\\
0  , \ \ \ \ \ \ \ \ \ \ \ \ \ \ \ \ \ a_i=0,\ \mathrm{i.e.,}\ v_i\in \bold S_{2}
\end{array}\right.\label{sub_entropy1}
\end{equation}
Eq.(\ref{sub_entropy1}) indicates that we only compute the sub-entropies for the dominant correlated time series in $\bold S_{1}$, and do not consider the non-dominant correlated time series in $\bold S_{2}$.

With the dominant entropy characteristics to hand, we further characterize each network $G_t$ as entropy time series. Let a time window be denoted as a period of $w$ time steps. We shift this window along the whole time steps of the complex system $\mathbf{S}$ to construct the time-varying dominant entropy time series for each network $G_t$ at time $t$. Specifically, for each time window of the network $G_t$, we compute the dominant entropy time series of $G_t$ as
\begin{equation}
\mathcal{S}_t=\{E_{t-w+1}|E_{t-w+2}|\ldots |E_{s}|\ldots |E_{t}\},\label{ETS}
\end{equation}
where $s\in\{t-w+1,t-w+2,\ldots,t\}$, and each column $E_{s}$ of $\mathcal{S}_t$ is the entropy characteristics vector of each network $G_s\in \mathbf{G}$ at time $s$ and is defined by Eq.(\ref{QET}). Clearly, the dominant entropy time series $\mathcal{S}_t$ of the network $G_t$ encapsulates the $w$ time-varying entropy characteristics vectors of the networks $G_{t-w+1}$ at time $t-w+1$ to $G_{t}$ at time $t$.

Assume $G_p\in \mathbf{G}$ and $G_q\in \mathbf{G}$ are a pair of time-varying networks at time $p$ and $q$ respectively, and their associated entropy time series are $$\mathcal{S}_p=\{E_{p-w+1}|E_{p-w+2}|\ldots | E_{p}\}$$ and $$\mathcal{S}_q=\{E_{q-w+1}|E_{q-w+2}|\ldots | E_{q}\}.$$ We define the Entropic Dynamic Warping Kernel (EDTWK) between $G_p$ and $G_q$ as
\begin{align}
k_{\mathbf{DCTE}}(G_p,G_q)& =k_{\mathbf{GA}}(\mathcal{S}_p,\mathcal{S}_q) \nonumber \\
 &=\sum_{\pi\in \mathcal{A}(w,w)} e^{-{D}_{\mathbf{p},\mathbf{q}}(\pi)},
\end{align}
where $k_{\mathbf{GA}}$ is the dynamic time warping inspired Global Alignment Kernel (GAK) defined in Eq.(\ref{GAK}), $\pi$ is the warping alignment between the entropy time series of $G_p$ and $G_q$, $\mathcal{A}(w,w)$ is all possible alignments and ${{D}_{\mathbf{p},\mathbf{q}}(\pi)}$ refers to the alignment cost obtained via Eq.(\ref{DTWD}).

\noindent\textbf{Remarks:} Although the proposed EDTWK kernel is related to the general principles of the GAK kernel. However, the proposed kernel has two distinct theoretical differences. First, the original GAK kernel is only designed for vectorial time series and cannot capture intrinsic relationships between time series. In contrast, our proposed kernel is explicitly designed for time-varying financial networks that reflect correlations between pairs of time series. Second, only the proposed EDTWK kernel can identify the dominant correlated time series through the analysis over the commute time matrix. Based on financial risk theory~\cite{finance}, financial crises are usually caused by a set of most correlated stock time series having less uncertainties. Therefore, only the proposed kernel is able to capture more reliable financial information. In summary, the proposed kernel provides an effective way of incorporating the structural correlations between time series into the process of multiple co-evolving time series analysis.

\subsection{Time Complexity}

For a pair of networks, the proposed kernel $k$ requires time complexity $O(n^3+w^2)$. The reasons are as follows. Assume a family of time-varying networks and each network has $n$ vertices. Computing the dominant commute time entropy kernel $k$ between a pair of networks associated with a time window of $w$ steps requires time complexity $O(n^3 +w^2)$. Because computing the required entropy time series is based on the computation of the commute time. This computation relies on the spectral decomposition of the Laplician matrix and thus requires time complexity $O(n^3)$. Moreover, computing all possible warping alignments over $w$ time steps requires time complexity $O(w^2 )$. Thus, the whole time complexity of the proposed kernel $k$ is $O(n^3+w^2)$.
\subsection{Related works to the Proposed Kernel}\label{s3:TAA}
Comparing to some state-of-the-art approaches, the proposed EDTWK kernel has a number of advantages.

\textbf{First}, unlike the dynamic time warping inspired GAK kernel~\cite{DBLP:conf/icml/Cuturi11}, the proposed kernel is developed for time-varying complex networks. Since the network encapsulates rich co-relationship between pairwise co-evolving time series, the proposed kernel can reflect richer correlated information than the classical dynamic time warping framework for original vectorial time series.

\textbf{Second}, the proposed kernel is based on the new dominant entropy time series that is computed through a quadratic programming method on the commute matrix to identify the most correlated time series subset. As a result, unlike the existing threshold-based approaches~\cite{silva2015modular,ye2015thermodynamic,wang2018PRL} that roughly select pairs of relatively more correlated time series, the proposed kernel can reflect reliable dominant correlated information between time series through the dominant entropy time series. Furthermore, the commute time encapsulates the integrated effectiveness of all possible paths between a pair of vertices. As a result, the dominant entropy time series computed through the commute time matrix can potentially encapsulate the weighted information over all edges, and overcome the shortcoming of information loss arising in the threshold-based approaches.

\textbf{Third}, as we have stated, the time-varying networks are usually complete weighted networks. Most existing graph kernels cannot directly accommodate such network structures and need to transform them into sparse structures. Unfortunately, these sparse structures discard many weighted edges and certainly lead to information loss. By contrast, the commute time is computed through the Laplacian matrix that can directly accommodate complete weighted graphs. Thus, the proposed kernel encapsulates the whole structural information residing on all weighted edges.

In summary, the proposed kernel bridges the gap between state-of-the-art graph kernels and the classical dynamic time warping framework for time-varying networks, providing a new alternative way for analyzing time series more effectively.

\section{Experiments of Time Series Analysis}\label{s4}

We empirically validate the effectiveness of the proposed kernel approach on a family of time-varying financial networks extracted from the New York Stock Exchange (NYSE) dataset~\cite{silva2015modular,ye2015thermodynamic}. The NYSE dataset consists of 347 stocks associated with their daily closing prices over 6004 transaction days starting from January 1986 to February 2011. These prices are all collected from the public financial dataset on Yahoo (http://finance.yahoo.com). To abstract the time-varying financial network structures, we employ a time window of fixed size (i.e., 28 days). We slide this fixed sized window along time to derive a sequence from the 29th trading day to the 6004th trading day, where each temporal time window encapsulates a set of 347 co-evolving daily stock price time series of the 347 stocks over 28 days. We characterize the trades between various stocks as a network structure with each stock as the vertex. Specifically, for each time window we calculate the absolute value of the Pearson correlation between the time series for pairwise stocks as their edge weight. This in turn generates a family of time-varying financial network with a fixed number of 347 vertices and varying edge weights for the 5976 trading days. The aim of this study is to investigate whether the proposed kernel approach can be used to detect fluctuations in trading network structure due to global political
or economic events.

\subsection{Evaluation of The Entropy Time Series}\label{exp:s1}
\begin{figure}
\centering
\subfigure[Black Monday]{\includegraphics[width=0.49\linewidth]{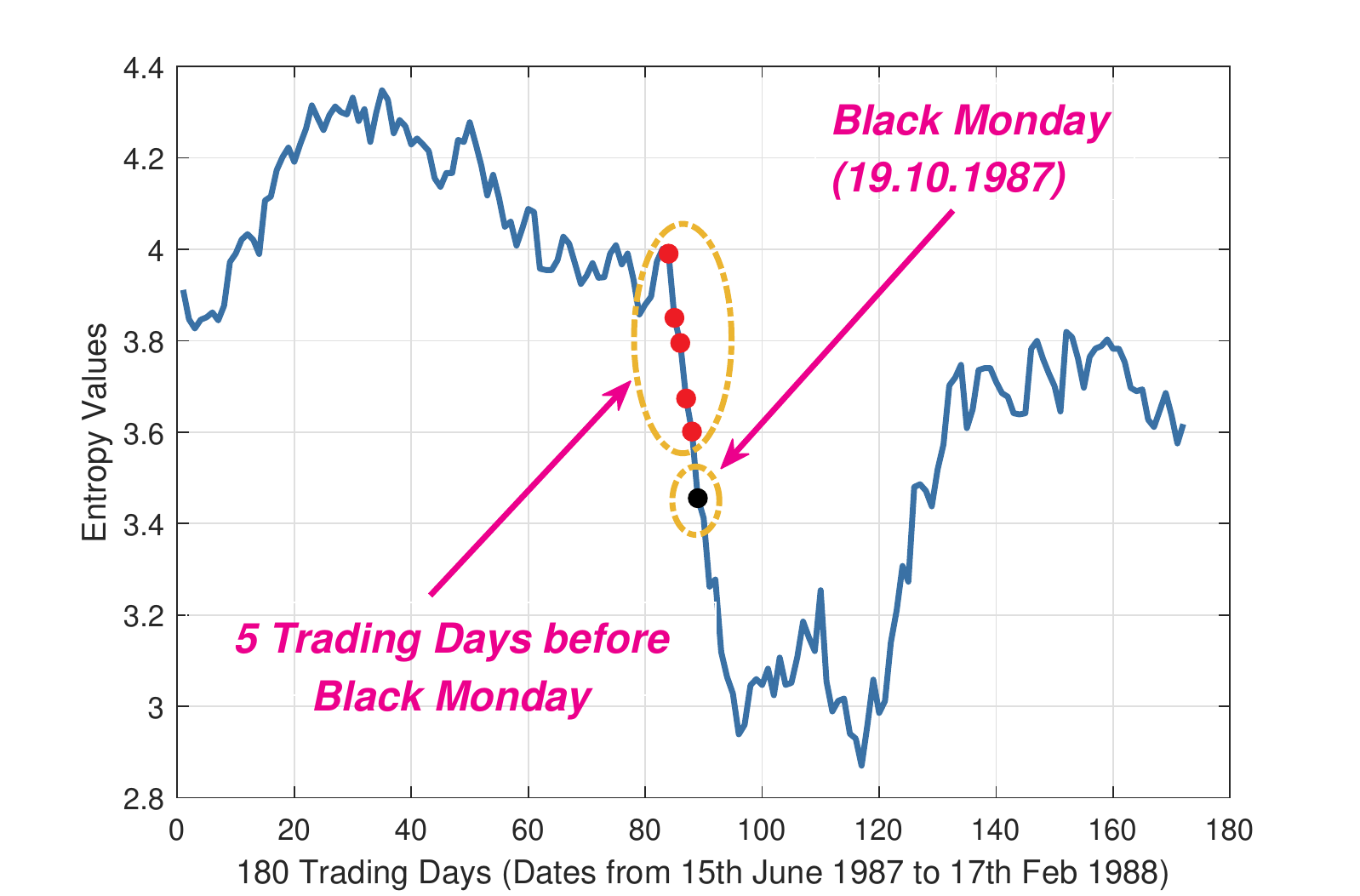}}
\subfigure[Dot-com Bubble Burst]{\includegraphics[width=0.49\linewidth]{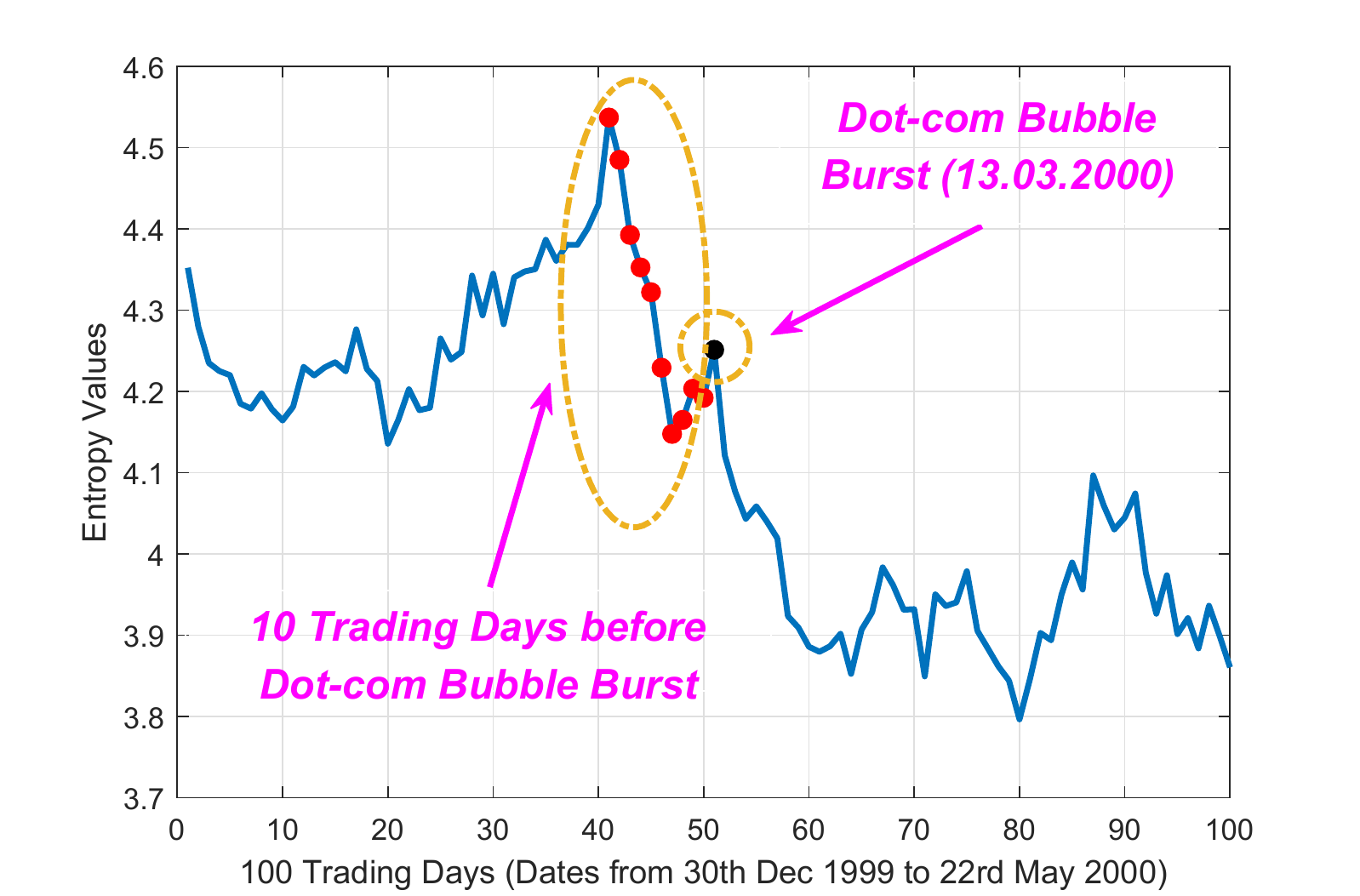}}
\subfigure[Bankrupt of New Centry Financial]{\includegraphics[width=0.49\linewidth]{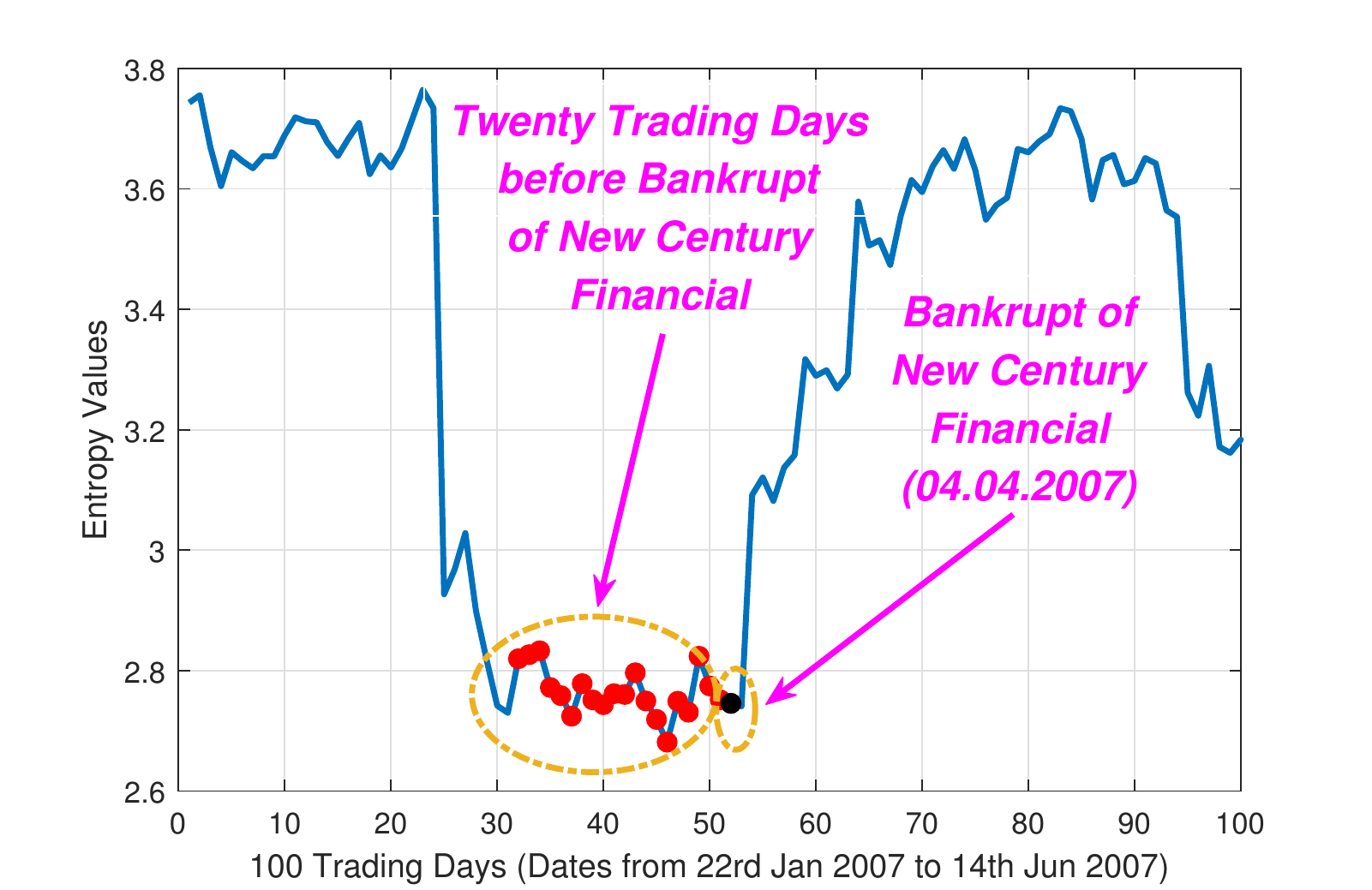}}
\subfigure[Lehman Crisis]{\includegraphics[width=0.49\linewidth]{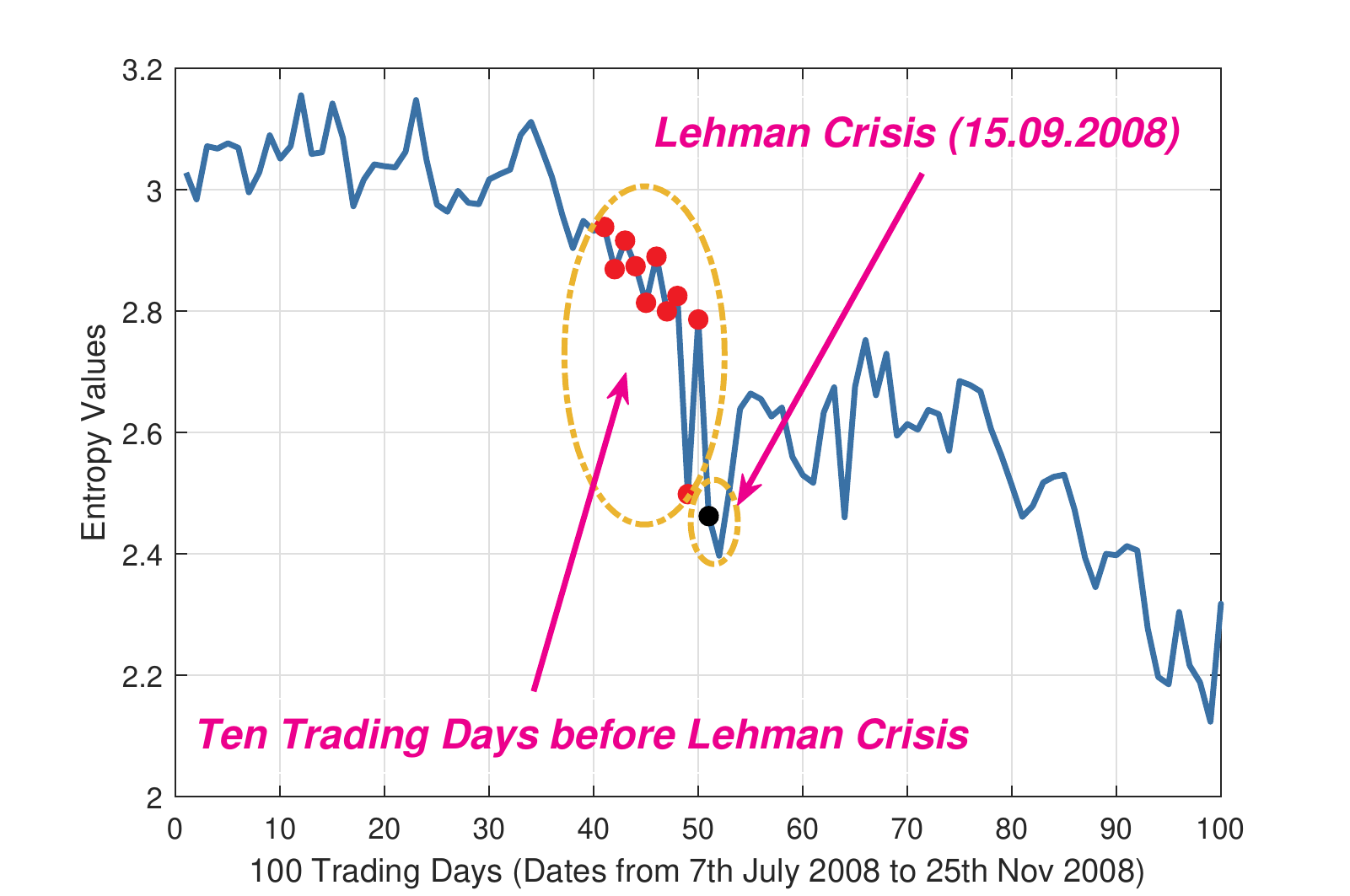}}
\subfigure[Enron Crisis]{\includegraphics[width=0.49\linewidth]{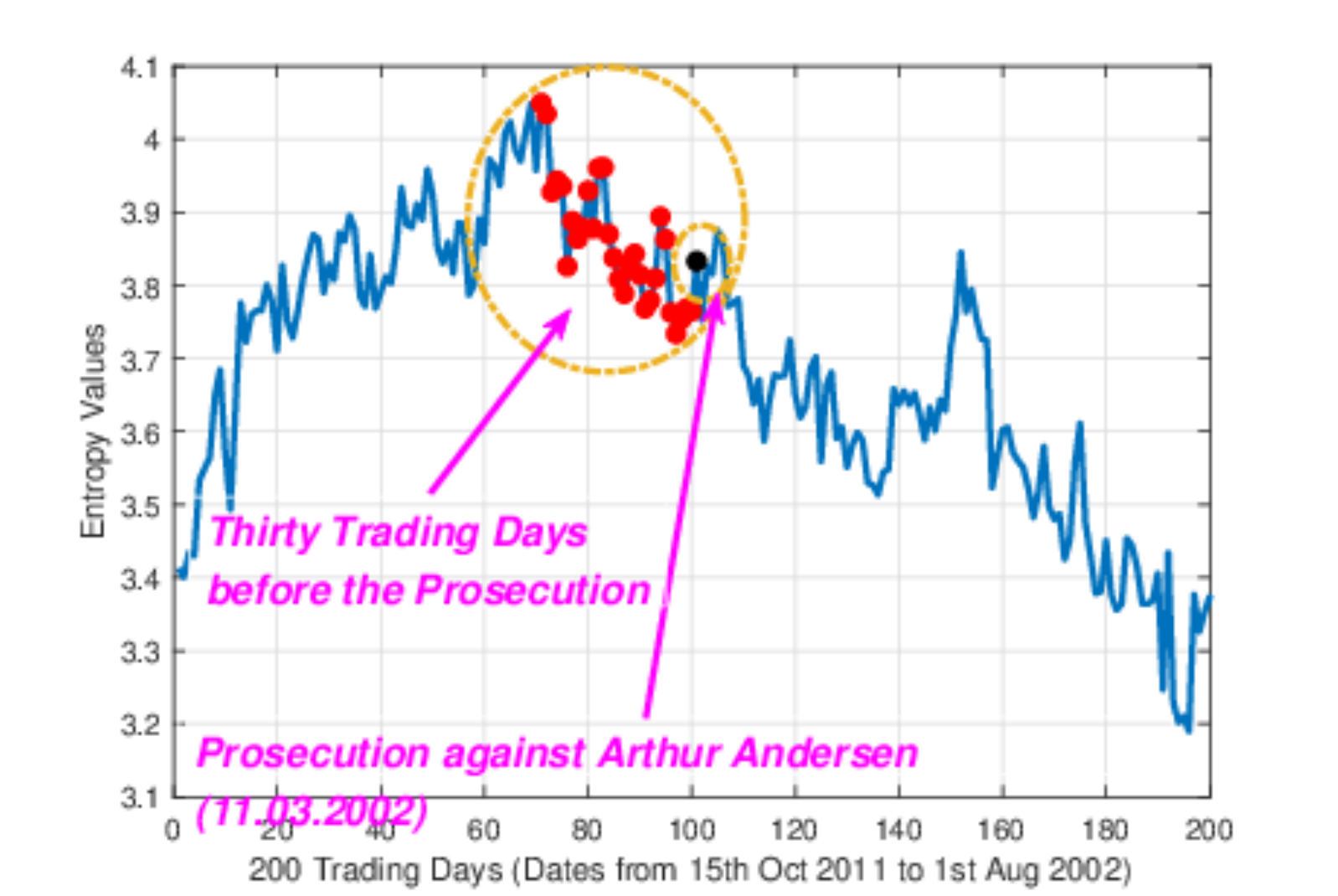}}
\subfigure[1997 Asian Financial Crisis]{\includegraphics[width=0.49\linewidth]{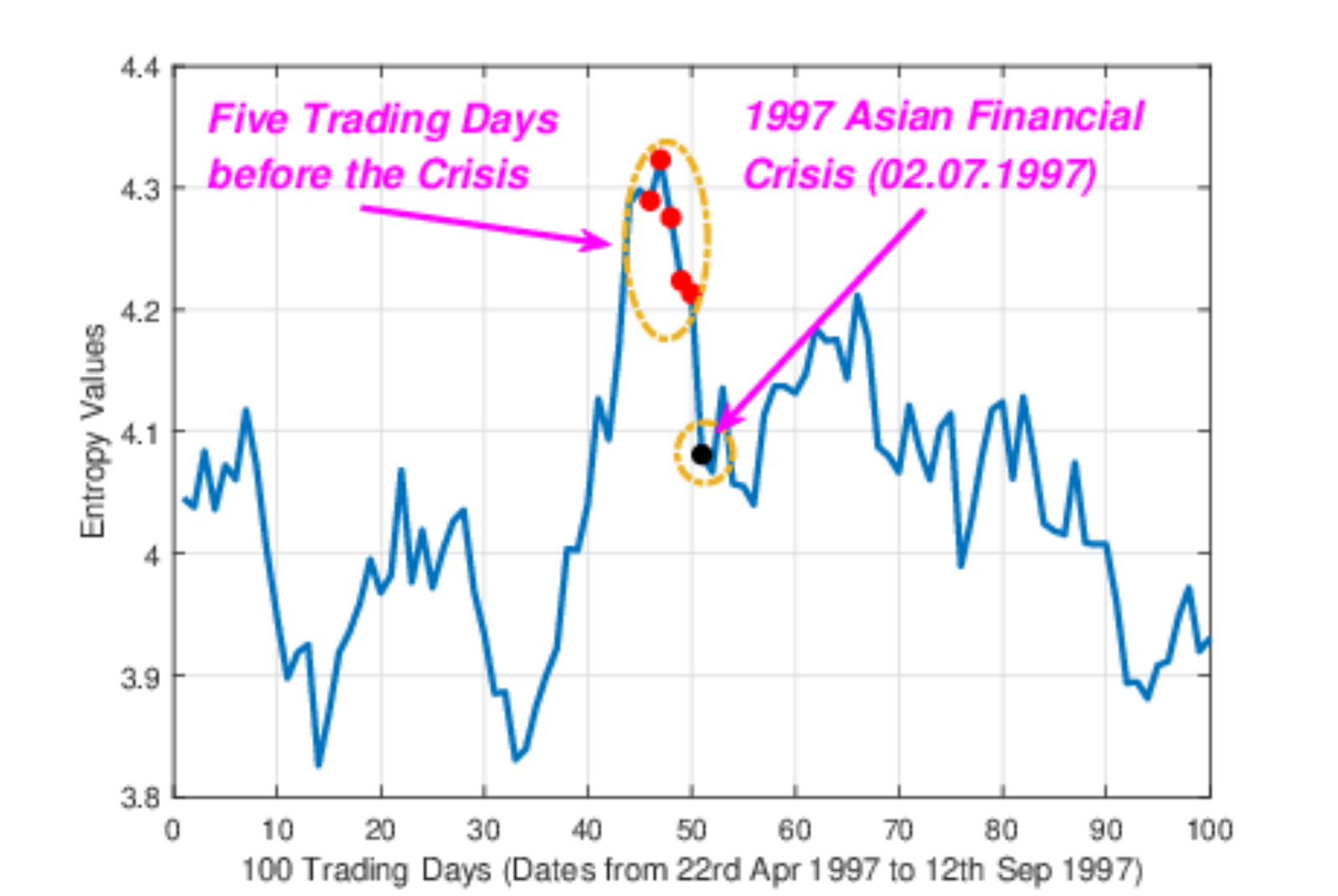}}
\vspace{-10pt}
\caption{Evaluation of Dominant Shannon Entropy.} \label{EtropyEvaluation}
\vspace{-10pt}
\end{figure}

We commence by exploring whether the dominant entropy time series can significantly characterize the time-varying financial networks, since these new developed time series play an important role for the proposed kernel. Specially, we investigate the evolutionary behavior of the NYSE stock market through calculating the dominant Shannon entropy on the time-varying financial networks at each time step, i.e., we investigate how the sum of the dominant sub-entropies of the network varies with increasing time $t$. We exhibit the results in Fig.\ref{EtropyEvaluation}, where the x-axis corresponds to the date (time) and the y-axis corresponds to the dominant Shannon entropy values. Fig.\ref{EtropyEvaluation} shows that the dominant Shannon entropy is sensitive to different financial crises (i.e., Black Monday~\cite{blackmonday}, Dot-com Bubble Burst~\cite{anderson2010speculative}, Bankrupt of New Centry Financial,  Lehman Crisis in Sub-prime Crisis Period~\cite{subprime}, Enron Crisis, and 1997 Asian Financial Crisis), and the entropy values usually lead to a rapid decrease even many days before the significant financial event. In other words, each significant fluctuation of the dominant Shannon entropy values corresponds to a financial crisis, and provides early warning before the crisis occurs. The reason for the effectiveness is that the financial networks are constructed by computing the correlation between pairwise stock time series and the dominant Shannon entropy is computed based on the dominant correlated time series subset that is identified through the commute time matrix. Based on the financial risk theory stated by~\cite{finance}, the financial crisis is usually caused by a set of most correlated stocks having less uncertainties. Thus, the dominant Shannon entropy, that characterizes the the dominant correlated stocks, tends to significantly drop down before a financial crisis. The experiments demonstrates that the proposed dominant entropy time series through the commute time can capture significant financial information, satisfying financial theory.

Note that, although Fig.\ref{EtropyEvaluation} indicates that the dominant Shannon entropy is effective for identifying the extreme financial events in the evolution of the time-varying financial networks, the entropy measure can only represent the network characteristics in an one-dimensional pattern space and thus ignores the information regarding specific changes in the network structure. In contrast, our proposed kernel associated with the dominant entropy is able to map the network structures into a high dimensional Hilbert space via kernelizing the entropies. In other words, our kernels should better preserve structural information contained in the network structures.

\subsection{Kernel Embeddings of Financial Networks from kPCA}\label{exp:s2}


\begin{figure}
\centering
\subfigure[Path for EDTWK]{\includegraphics[width=0.49\linewidth]{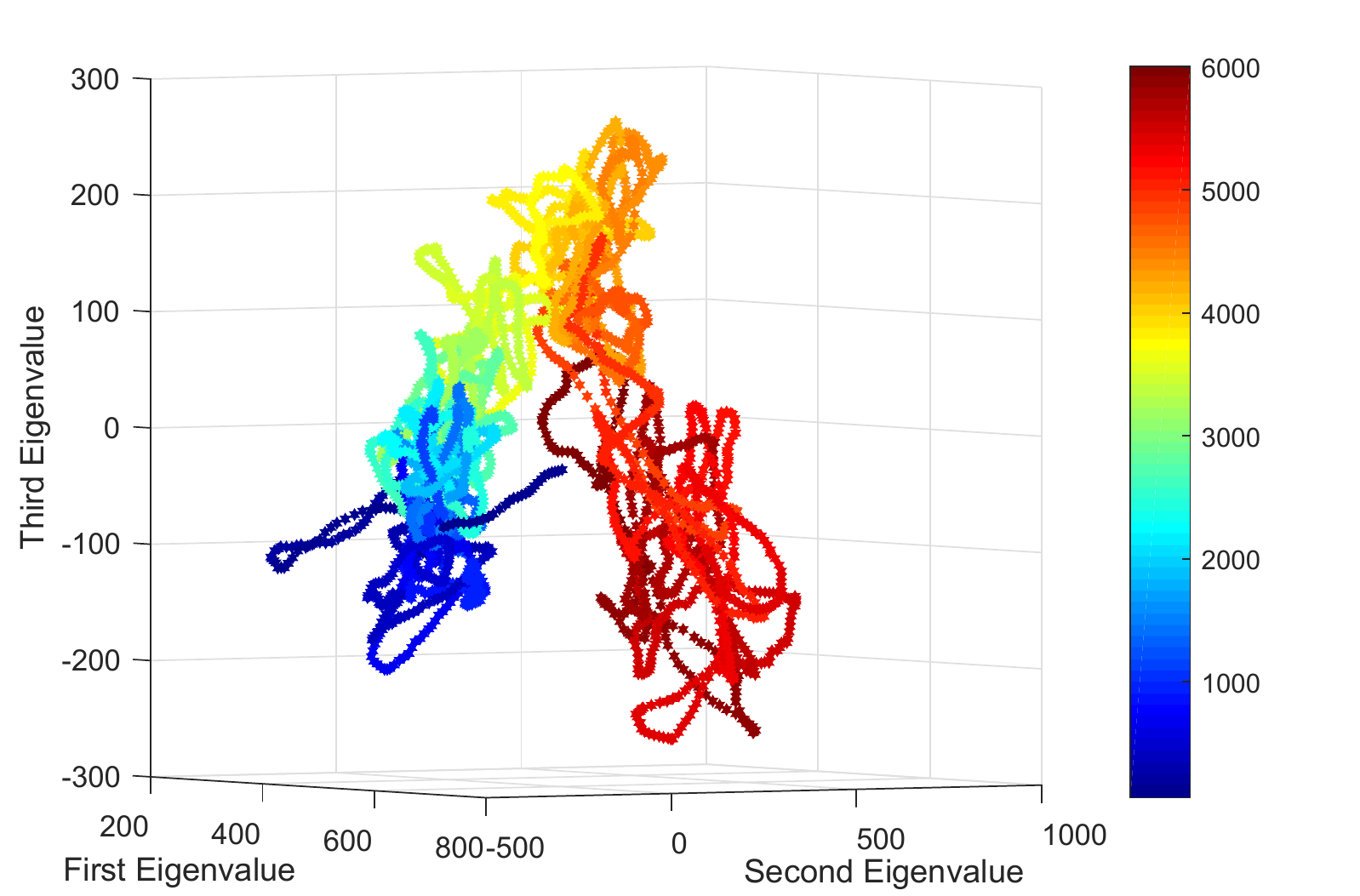}}
\subfigure[Path for GC]{\includegraphics[width=0.49\linewidth]{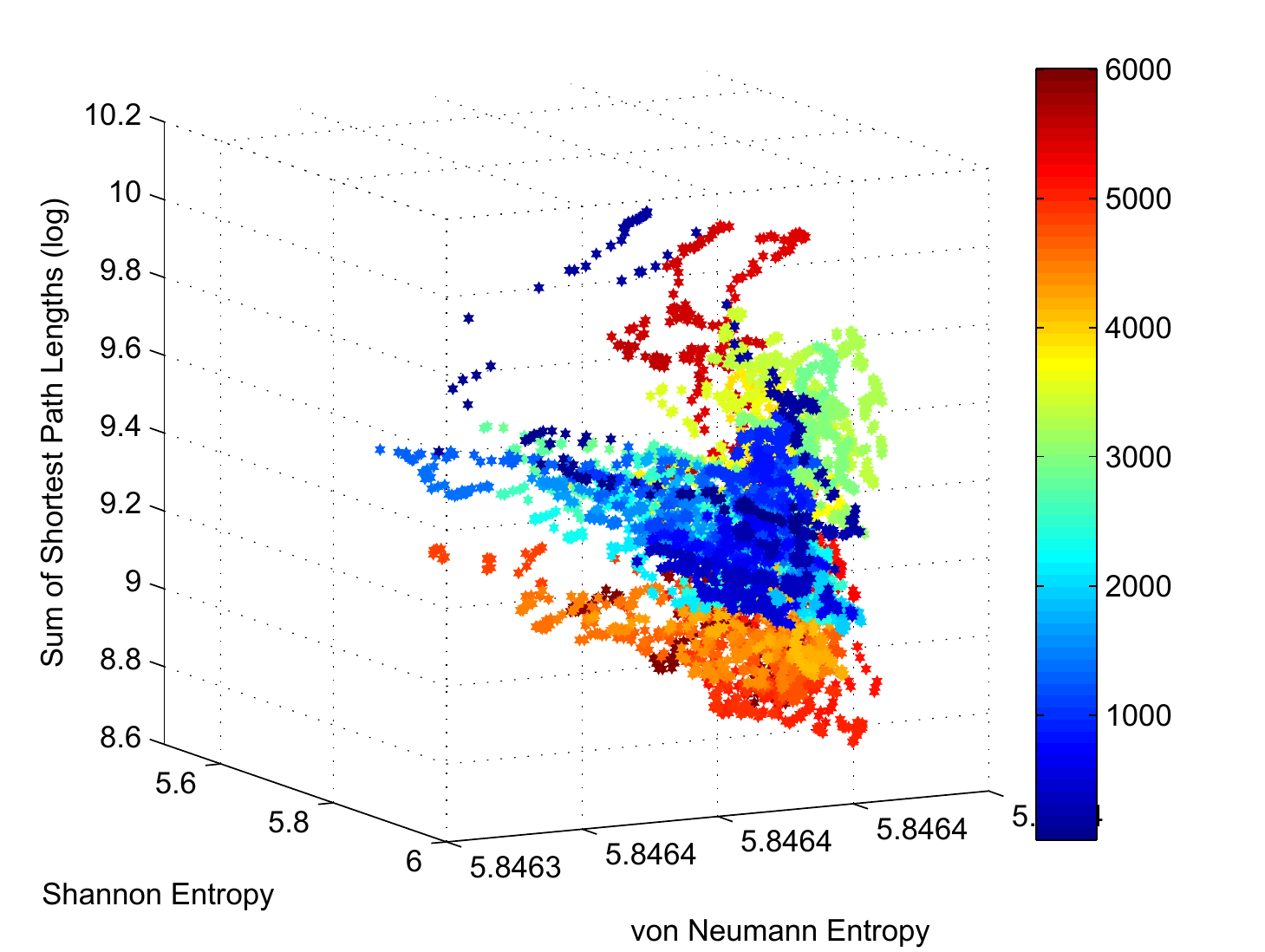}}
\subfigure[Path for GAK]{\includegraphics[width=0.49\linewidth]{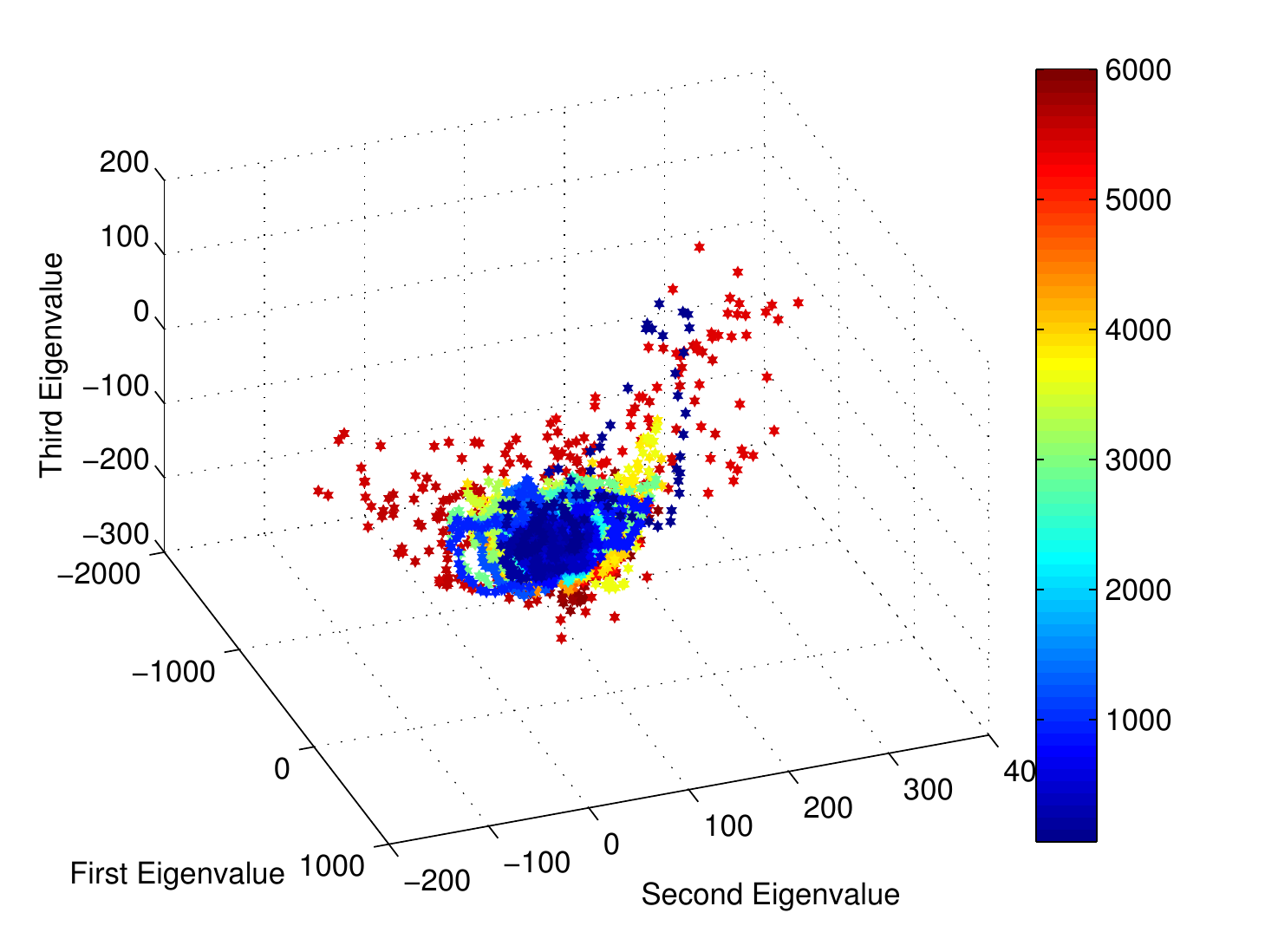}}
\subfigure[Path for WLSK]{\includegraphics[width=0.49\linewidth]{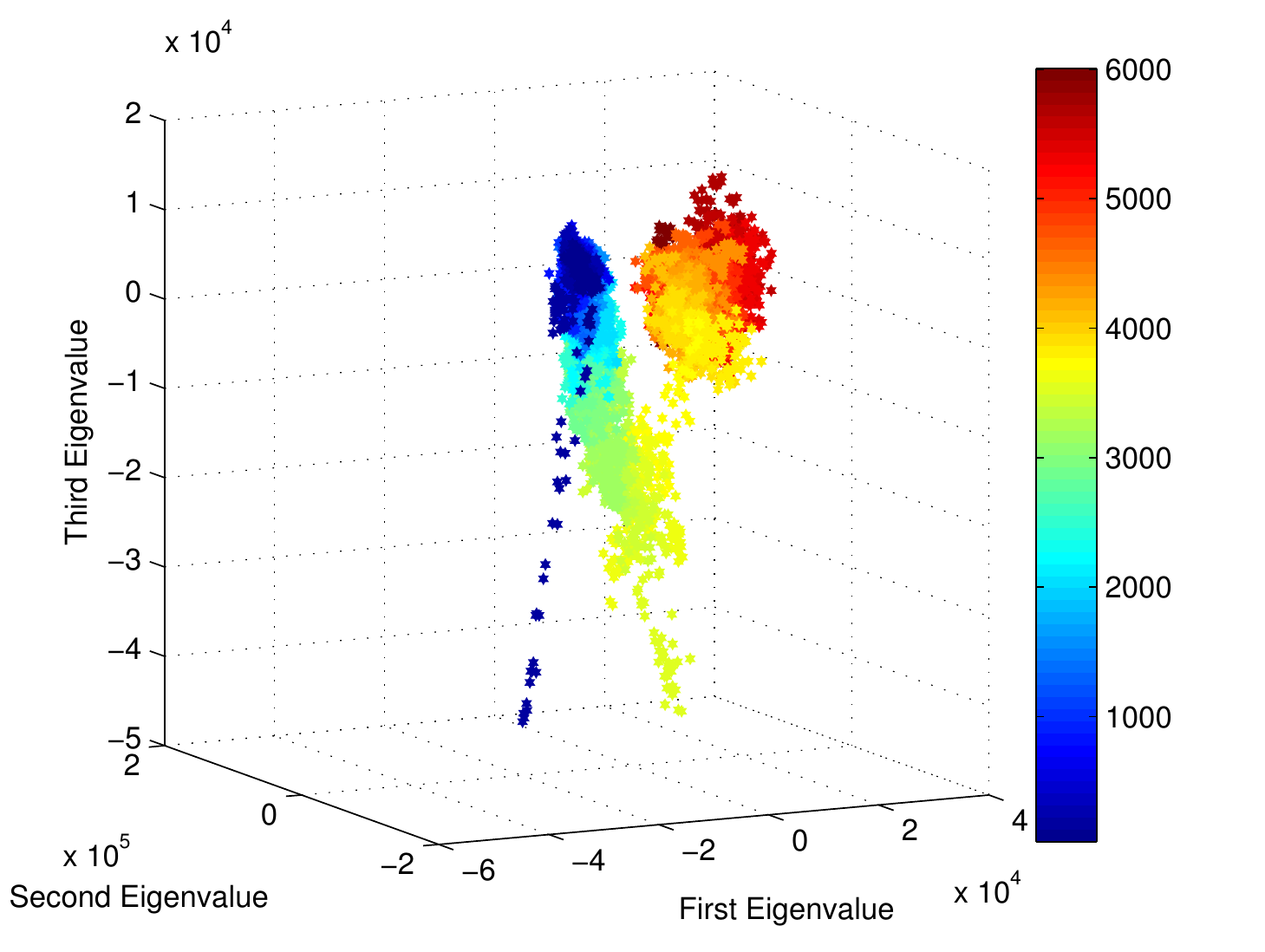}}
\subfigure[Path for QJSK]{\includegraphics[width=0.49\linewidth]{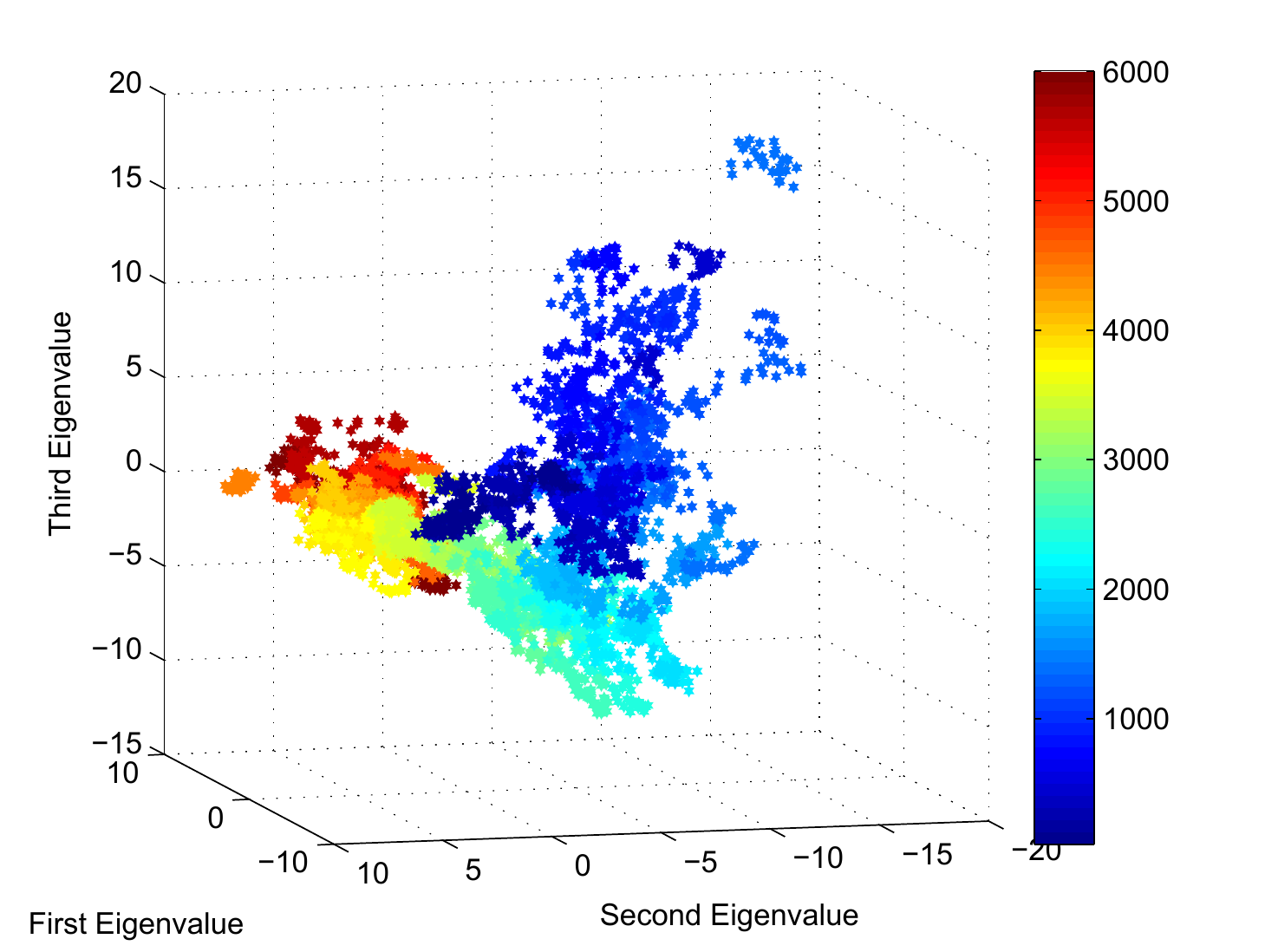}}
\subfigure[Path for FLK]{\includegraphics[width=0.49\linewidth]{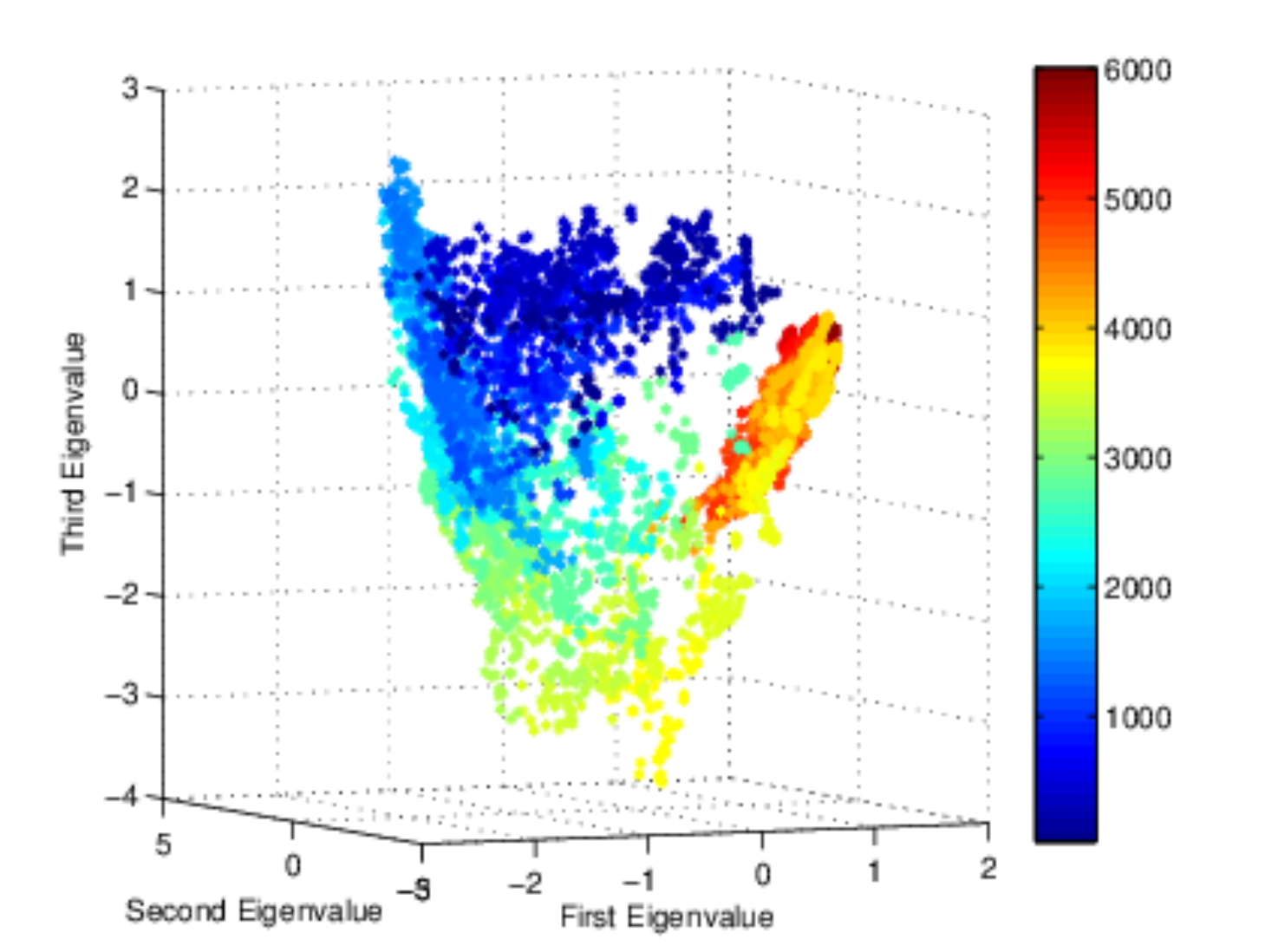}}
\subfigure[Path for EDTWKO]{\includegraphics[width=0.49\linewidth]{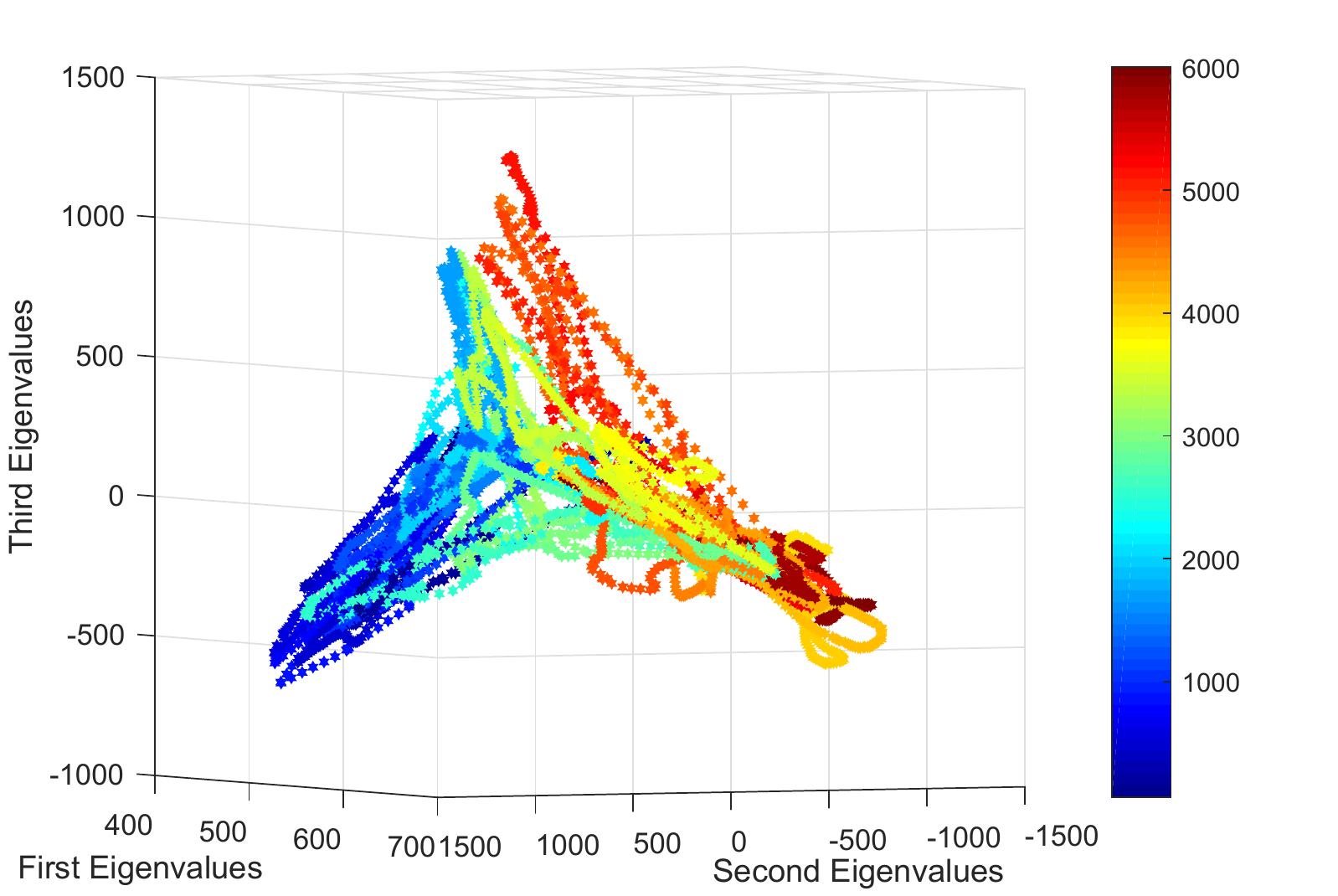}}
\subfigure[Path for QK]{\includegraphics[width=0.49\linewidth]{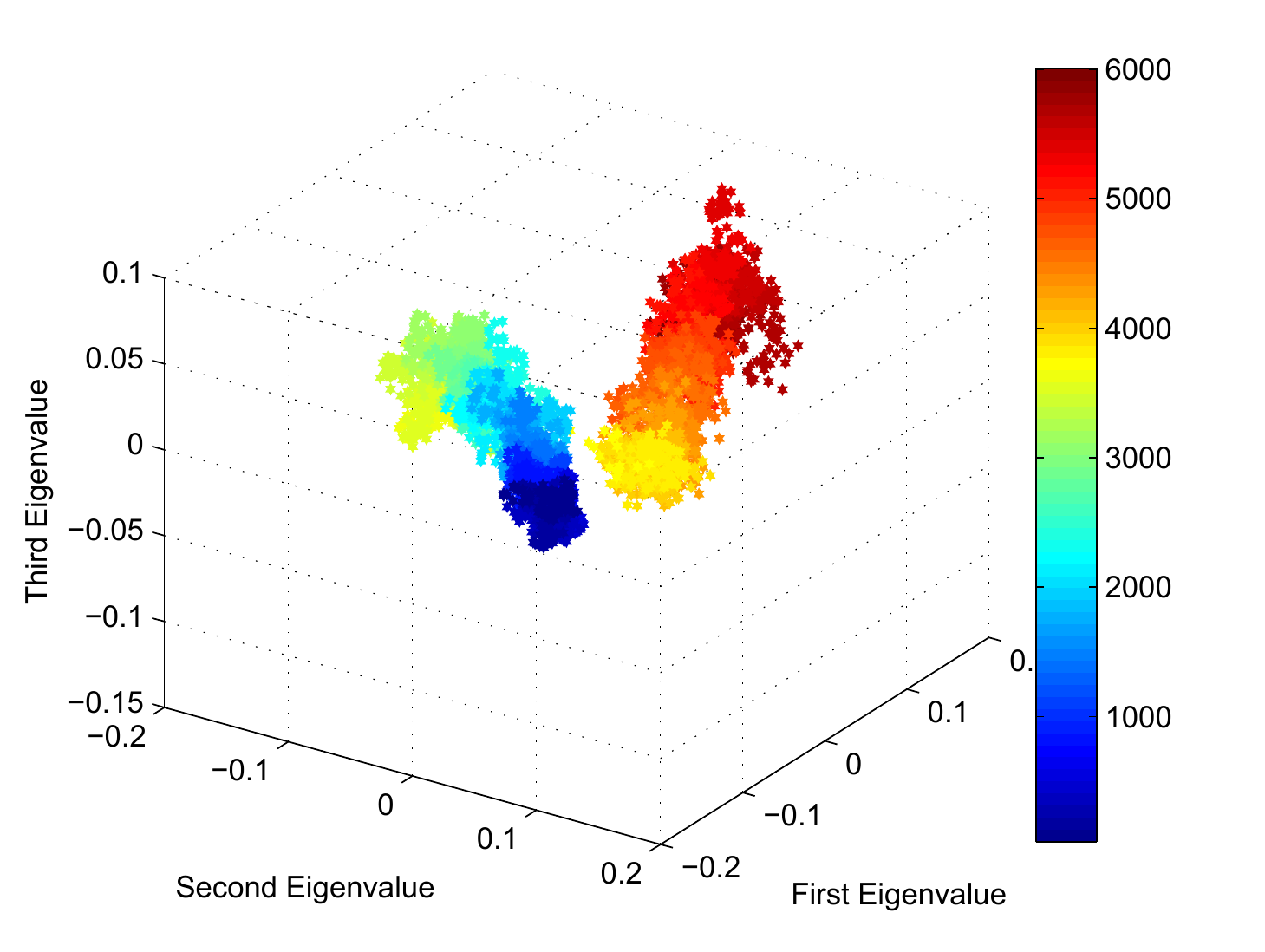}}
\vspace{-10pt}
\caption{Path of financial networks over all trading days.}\label{allembeddings}
\vspace{-20pt}
\end{figure}

In this subsection, we evaluate the performance of the proposed EDTWK kernel on time-varying networks of the NYSE dataset, and explore whether the proposed kernel can distinguish the network evolution with time. Specifically, associated with the proposed kernel, we perform kernel Principle Component Analysis (kPCA)~\cite{witten2011data} on the kernel matrix of the financial networks and embed the networks in a vector space. We visualize the embedding results through the first three principal components in Fig.\ref{allembeddings}(a). In addition, we compare the proposed kernel to three classical graph characterization methods (GC), that is, the Shannon entropy associated with the classical steady state random walk~\cite{DBLP:journals/pr/Bai0TH15}, the von Neumann entropy associated with the normalized Laplician matrix~\cite{DBLP:journals/isci/DehmerM11}, and the average length of the shortest path over all pairs of vertices~\cite{silva2015modular}. All the three network characterization methods can accommodate complete weighted graphs. The visualization spanned by the three graph characterizations are shown in Fig.\ref{allembeddings}(b). Finally, we also compare the proposed kernel with four state-of-the-art kernel methods, that is, the dynamic time warping inspired global alignment kernel (GAK) for original time series~\cite{DBLP:conf/icml/Cuturi11} and three graph kernels for time-varying financial networks. Specifically, the graph kernels include the Weisfeiler-Lehman subtree kernel (WLSK)~\cite{DBLP:journals/jmlr/ShervashidzeVPMB09}, the quantum Jensen-Shannon kernel (QJSK)~\cite{DBLP:journals/pr/Bai0TH15}, and the feature space Laplacian graph kernel (FLGK)~\cite{DBLP:conf/nips/KondorP16}. For the GAK kernel, we also adopt a time window of 28 days for each trading day. For the WLSK kernel, since it cannot accommodate a complete weighted graph and the edge weight, we transform each original network into a minimum spanning tree and ignore the edge weights. For the QJSK and FLGK kernels, since they can accommodate edge weight, we directly apply these kernels to the original financial networks. For the four kernel methods, we also perform kPCA on the resulting kernel matrices and embed the original time series or the time-varying networks into a vector space, and the embedding results are exhibited in Fig.\ref{allembeddings}(c), \ref{allembeddings}(d), \ref{allembeddings}(e) and \ref{allembeddings}(f), respectively.

\begin{table}
\vspace{-0pt}
\centering {
\footnotesize
\caption{The Distance Stress of the Network Embeddings}\label{T:DSE} \vspace{-10pt}
\begin{tabular}{|c||c||c||c||c|}
\hline
~Methods ~       &~EDTWK  ~ &~GC~       &~GAK~     & ~WLSK~    \\ \hline \hline
~Distance Stress~&~$1.1931$~& ~$5.5215$~&~$2.9677$~& ~$4.0045$~\\ \hline \hline
~Methods ~       &~QJSK~    &~FLK~      & ~EDTWKO~ & ~QK~\\ \hline \hline
~Distance Stress~&~$2.7220$~&~$2.9297$~ & ~$2.133$~& ~$2.9167$~\\ \hline
\end{tabular}
} \vspace{-10pt}
\end{table}

As we previously stated, the commute time matrix computed from the original absolute Pearson correlation based adjacency matrix of each financial network can be seen as the enhanced Pearson correlation matrix for the proposed DETWK kernel. Thus, the commute time matrix plays a significant role to determine the effectiveness of the proposed kernel. To further validate its effectiveness, we also compare the proposed DETWK kernel associated with commute time matrices with that associated with the original weighted adjacency matrices (DETWKO), and the result is displayed in Fig~\ref{allembeddings}(g). In addition, we compare the proposed kernel with the quantum-inspired kernel (QK)~\cite{DBLP:journals/ijon/CuiBZWH19}, since the QK kernel can also accommodate the original Pearson correlation based adjacency matrices of financial networks through the commute time matrices. The result of the QK kernel is exhibited in Fig~\ref{allembeddings}(h).

Fig.\ref{allembeddings} exhibits the traces of the time-varying financial networks (or the original time series) in the different kernel spaces together with the classical graph characterization pattern space over all trading days. The color bar of each subfigure indicates the specific date over time. We observe that the embedding from the proposed EDTWK kernel exhibits a better manifold structure. Moreover, the embedding resulting from the EDTWK, WLSK and QK kernels are better distributed than those defined by the remaining approaches for comparisons. Another interesting phenomenon exhibited in Fig.\ref{allembeddings} is that only the proposed EDTWK kernel produces a clear time-varying trajectory associated with the financial networks of consecutive time steps, i.e., the embedding of each time-varying financial network on the current time step is closely near to that on the last time step in the embedding space. By contrast, the alternative methods hardly result in such a trajectory and the associated embeddings tend to distribute as clusters. To further demonstrate this effectiveness of the proposed EDTWK kernel, we compute the distance stress of the financial network embeddings based on different methods. Specifically, the distance stress $\mathrm{DS}$ is defined as
\begin{equation}
\mathrm{DS}= \frac{\sum_t \parallel x_t-x_{t-1}\parallel^2}{ \sum_t \parallel x_t-x_{\mathrm{tn}}\parallel^2},
\end{equation}
where $t=2,3,\ldots,n$, $x_t$ is the network embedding vector at time $t$, and $x_{\mathrm{tn}}$ is the nearest network embedding vector of $x_t$. For each embedding vector $x_t$ at time $t$, if the nearest embedding vector is always the embedding vector at last time step (i.e., $x_{t-1}$), the value of the distance stress $\textbf{DS}$ will be $1$. In other words, the distance stress value nearer to $1$ indicates the better performance of the embeddings to form a clear time-varying trajectory. The distance stress values of the financial network embeddings based on different methods are show in Table~\ref{T:DSE}. It is clear that only the distance stress value of the proposed EDTWK kernel is nearer to $1$. This further indicates that \textbf{the proposed method can better preserve the ordinal arrangement of the time-varying financial networks}.

\begin{figure}
\vspace{-10pt}
\centering
\subfigure[For Commute Time]{\includegraphics[width=0.49\linewidth]{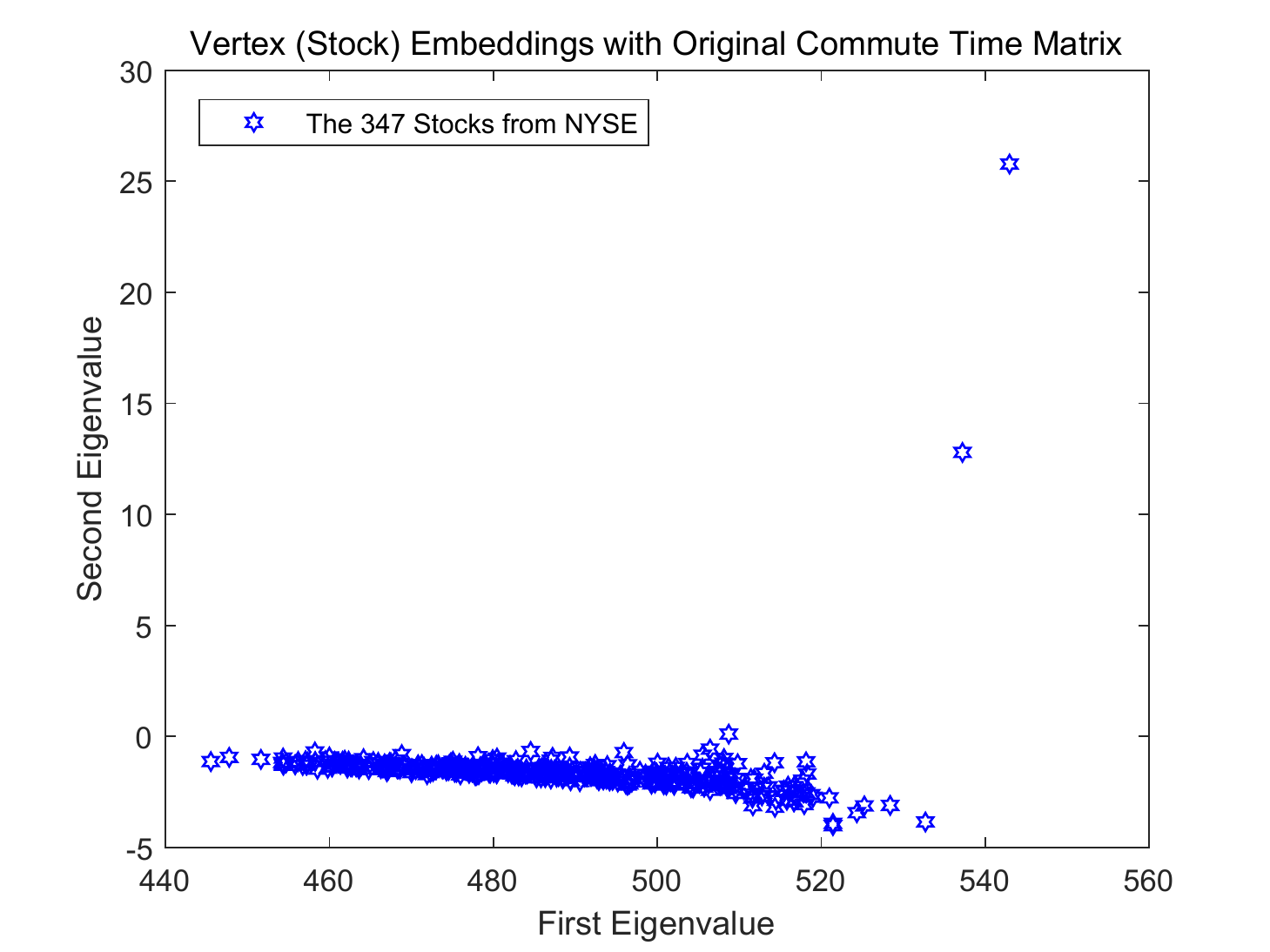}}
\subfigure[For Pearson Correlation]{\includegraphics[width=0.49\linewidth]{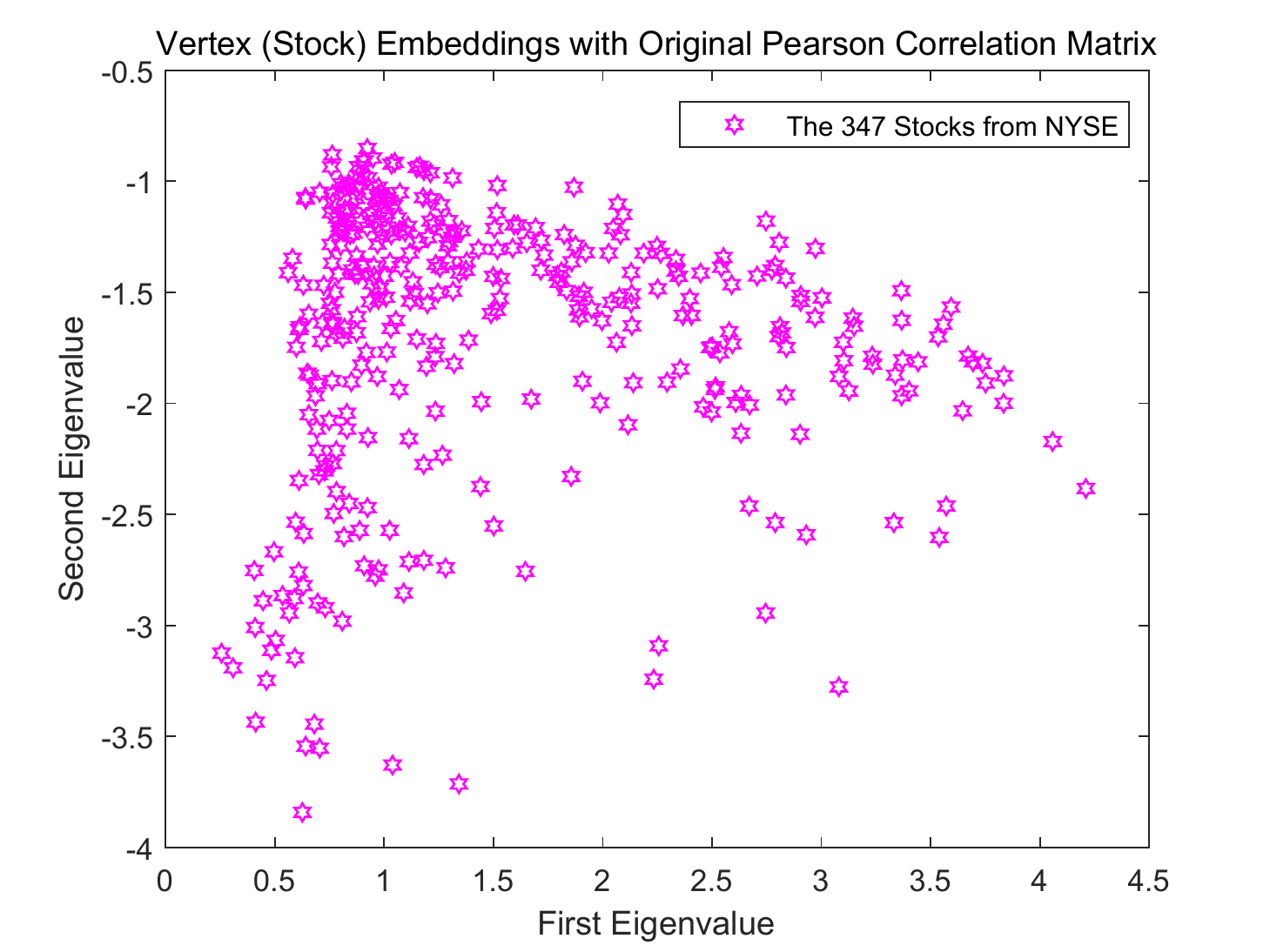}}
\vspace{-10pt}
\caption{Vertex (Stock) Embedding Comparisons.} \label{embeddingsStock}
\vspace{-20pt}
\end{figure} 

Finally, although the EDTWK kernel (i.e., the proposed kernel associated with the original weighted adjacency matrix) can also form a clear time-varying trajectory, its embedding points of the financial networks are not well distributed. \textbf{This reveals that the proposed approach associated with the commute time matrix has better performance than that associated with the original absolute Pearson correlation matrix, demonstrating the effectiveness of the commute time integrated in the proposed approach}. To further reveal the effectiveness of adopting the commute time matrix, we randomly select a financial network. For this network we perform multidimensional on both its commute time matrix and the Pearson correlation matrix, to embed each vertex (i.e., the 347 stocks from NYSE) into a 2-dimensional pattern space. Suppose the affinity matrix in this question is $C$ (i.e., the commute time or the Pearson correlation matrix), then the centred similarity matrix is given by
\begin{equation}
K=-1/2(I-J/n)C(I-J/n),
\end{equation} 
where $n$ is the number of vertices, $I$ is the $n\times n$ identity matrix, and $J$ is the $n\times n$ all-ones matrix. If $K=\Phi \Lambda \Phi^T$ is the eigendecompositon of the kernel matrix in terms of the diagonal matrix of ordered eigenvalues $\Lambda$ and the correspondingly ordered matrix of column eigenevtors $\Phi$, then the matrix with the embedding co-ordinates as column vectors is $X=\sqrt\Lambda$. Here we discard rows corresponding to negative eigenvalues and take the leading two rows. The results are shown in Fig.\ref{embeddingsStock}. It is interesting that the stock embedding points through the commute time matrix distribute well and form an approximately linear manifold structure. By contrast, the stock embedding points through the Pearson correlation  matrix disperse over a larger volume of the space. Note that we will observe similar results for most of our financial networks. This reveals that when compared to the Pearson correlation matrix, the commute time based vertex affinity matrix offers the advantage of capturing more reliable relationships between the stocks which reside in an approximately linear subspace.
  
Overall, the above observations indicates that the proposed EDTWK kernel associated with the commute time matrix can better understand the structural evolutions of the financial networks with time than the remaining methods.


\subsection{Kernel Embeddings for Financial Crisis Analysis}\label{exp:s3}
\begin{figure}
\centering
\subfigure[Black Monday for EDTWK]{\includegraphics[width=0.49\linewidth]{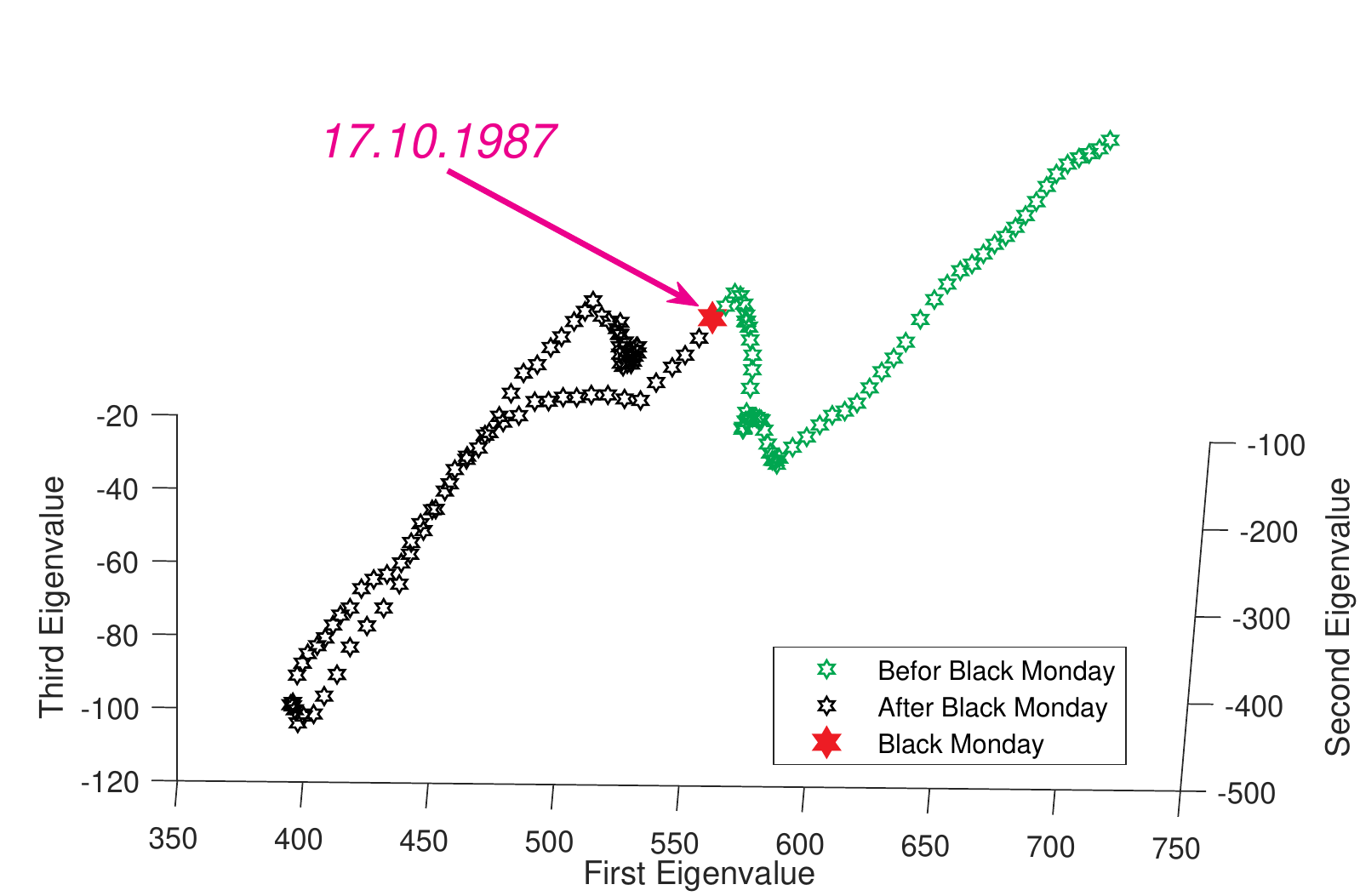}}
\subfigure[Black Monday for GC]{\includegraphics[width=0.49\linewidth]{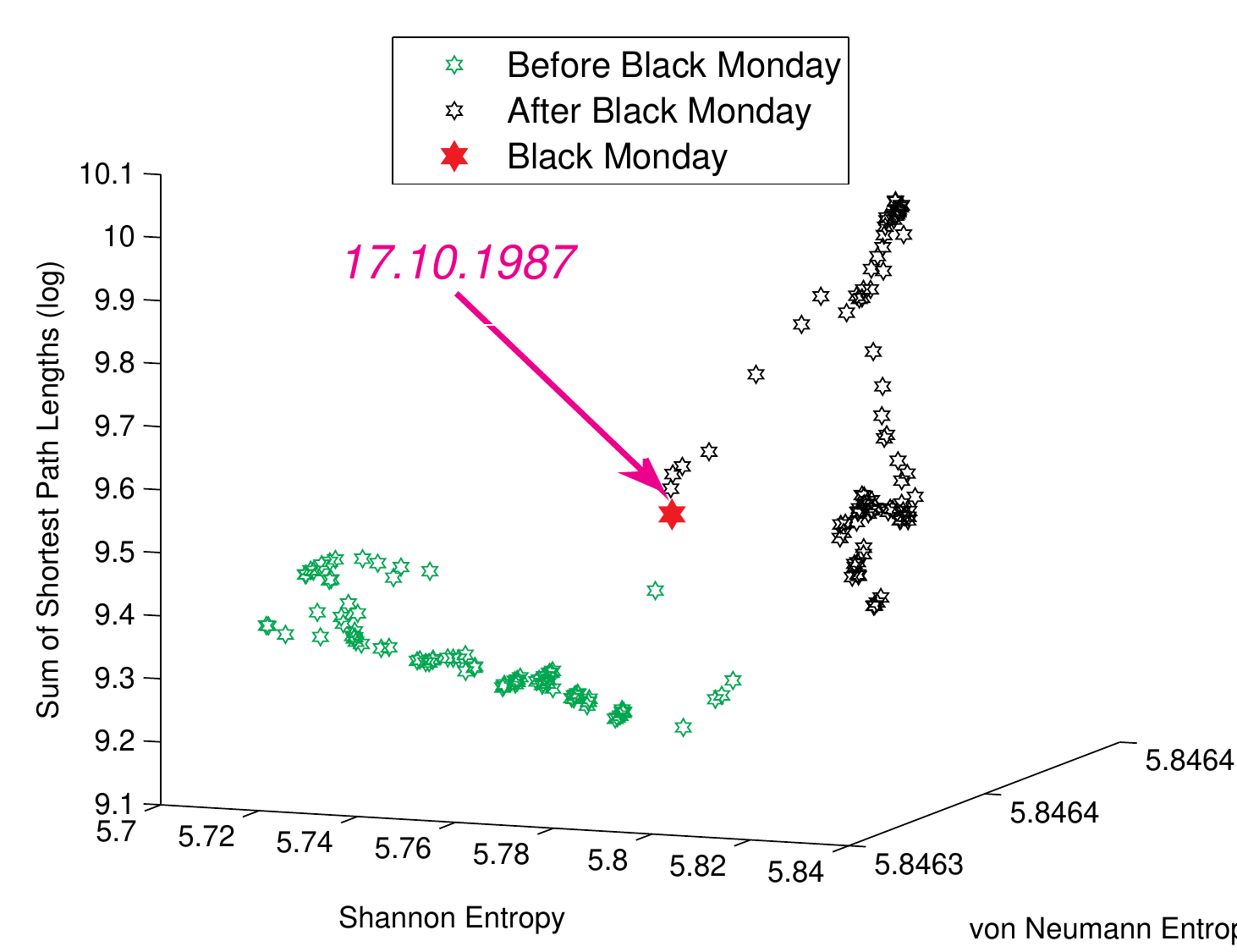}}
\subfigure[Black Monday for GAK]{\includegraphics[width=0.49\linewidth]{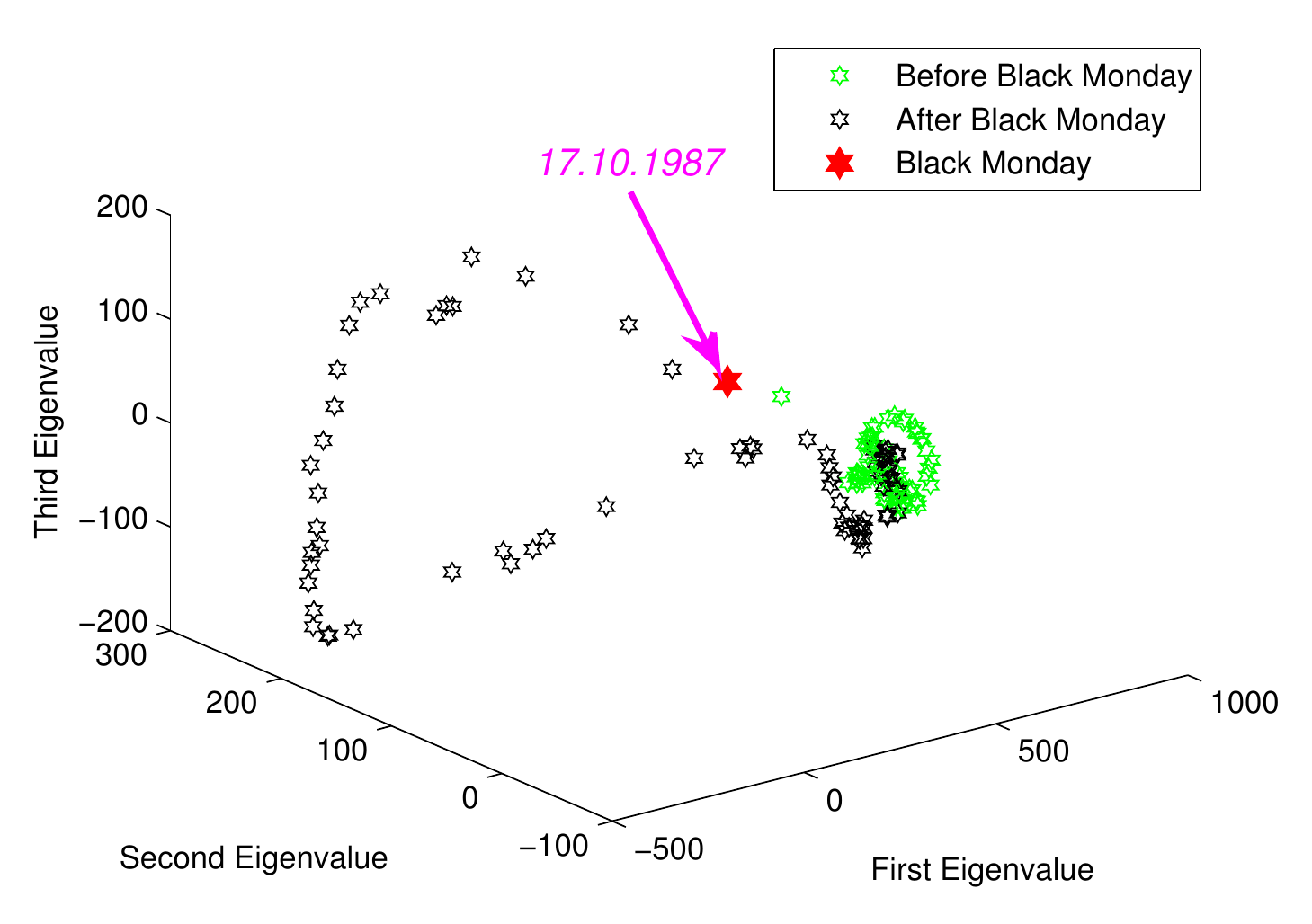}}
\subfigure[Black Monday for WLSK]{\includegraphics[width=0.49\linewidth]{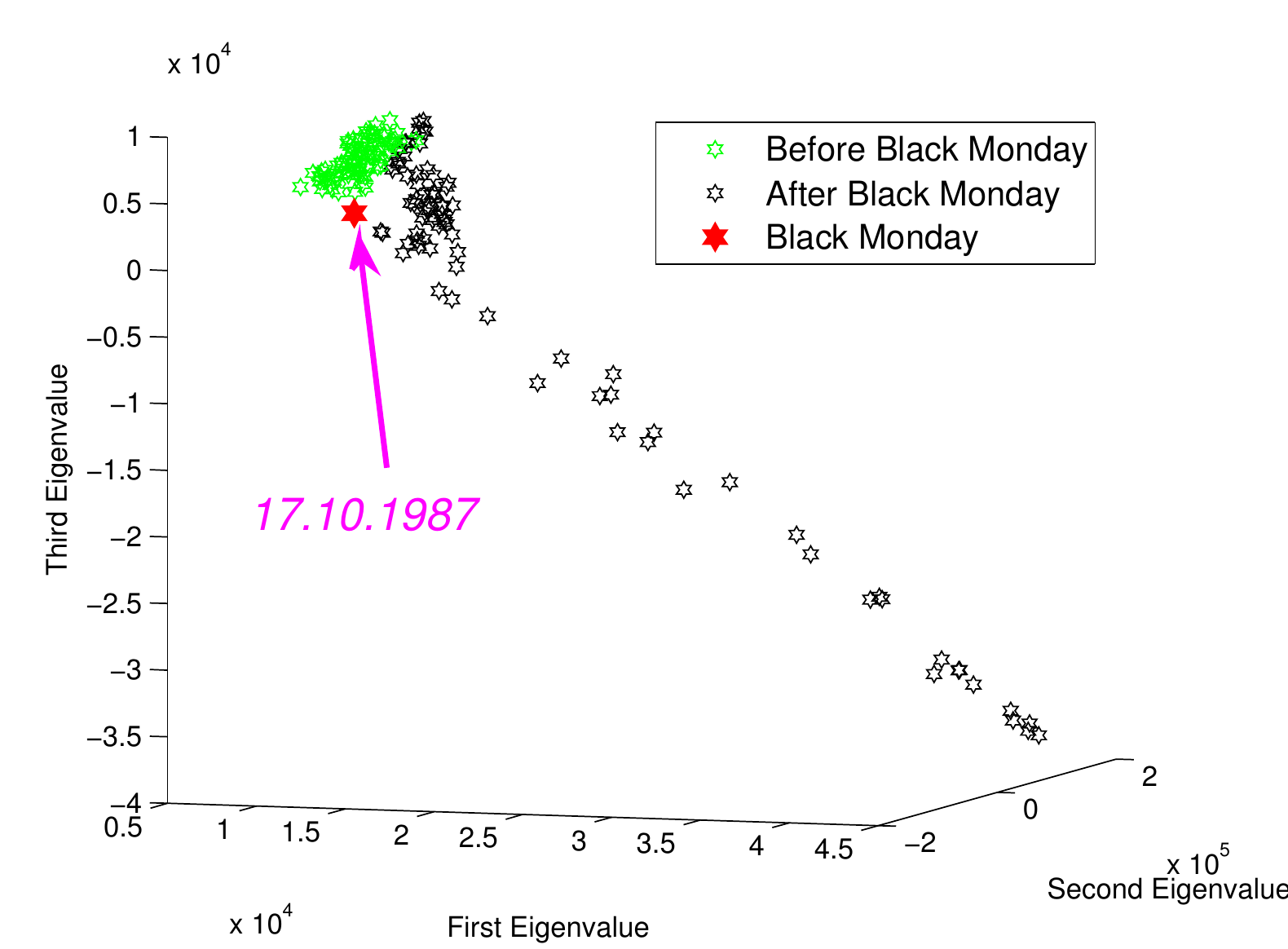}}
\subfigure[Black Monday for QJSK]{\includegraphics[width=0.49\linewidth]{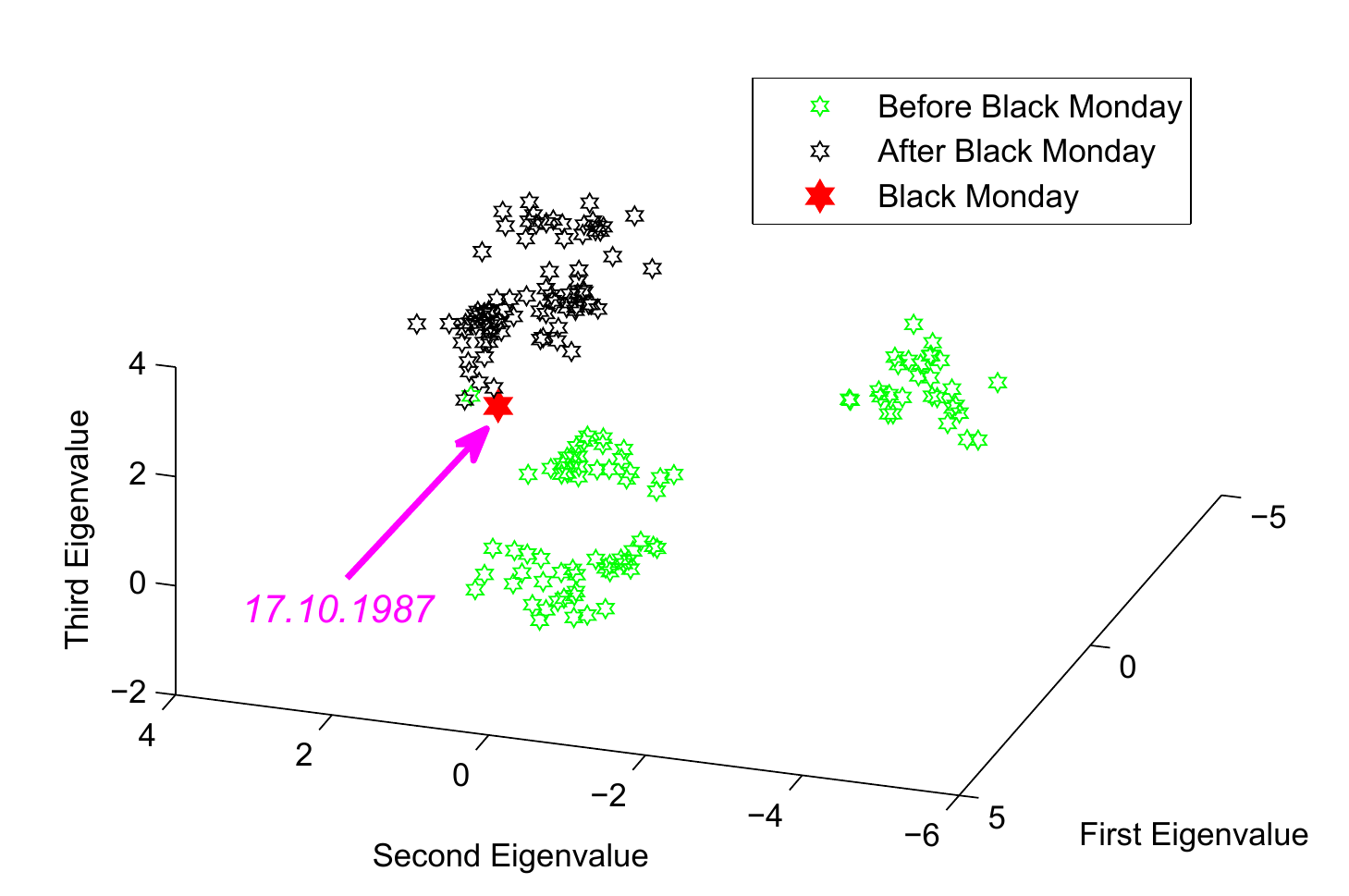}}
\subfigure[Black Monday for FLK]{\includegraphics[width=0.49\linewidth]{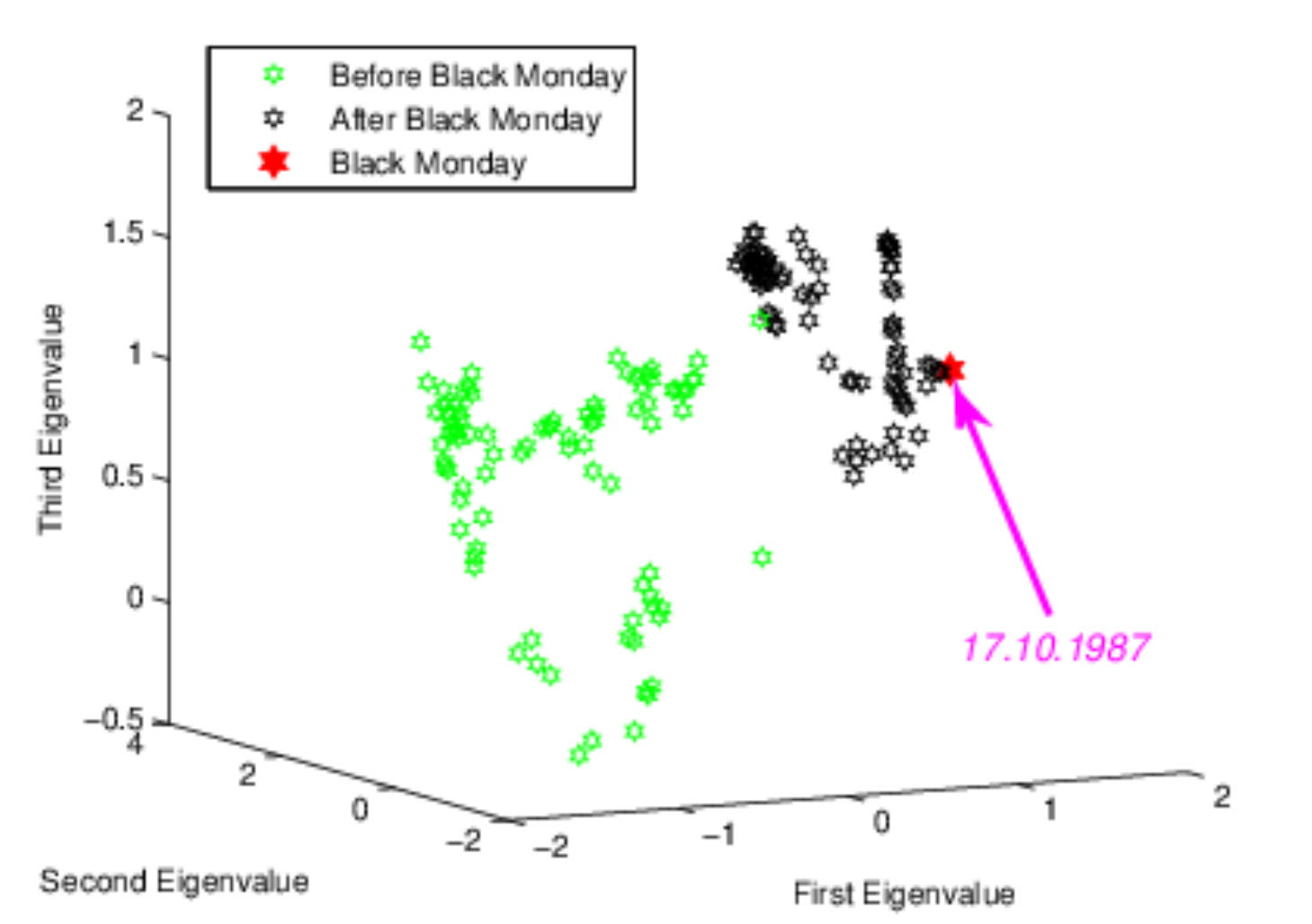}}
\subfigure[Black Monday for EDTWKO]{\includegraphics[width=0.49\linewidth]{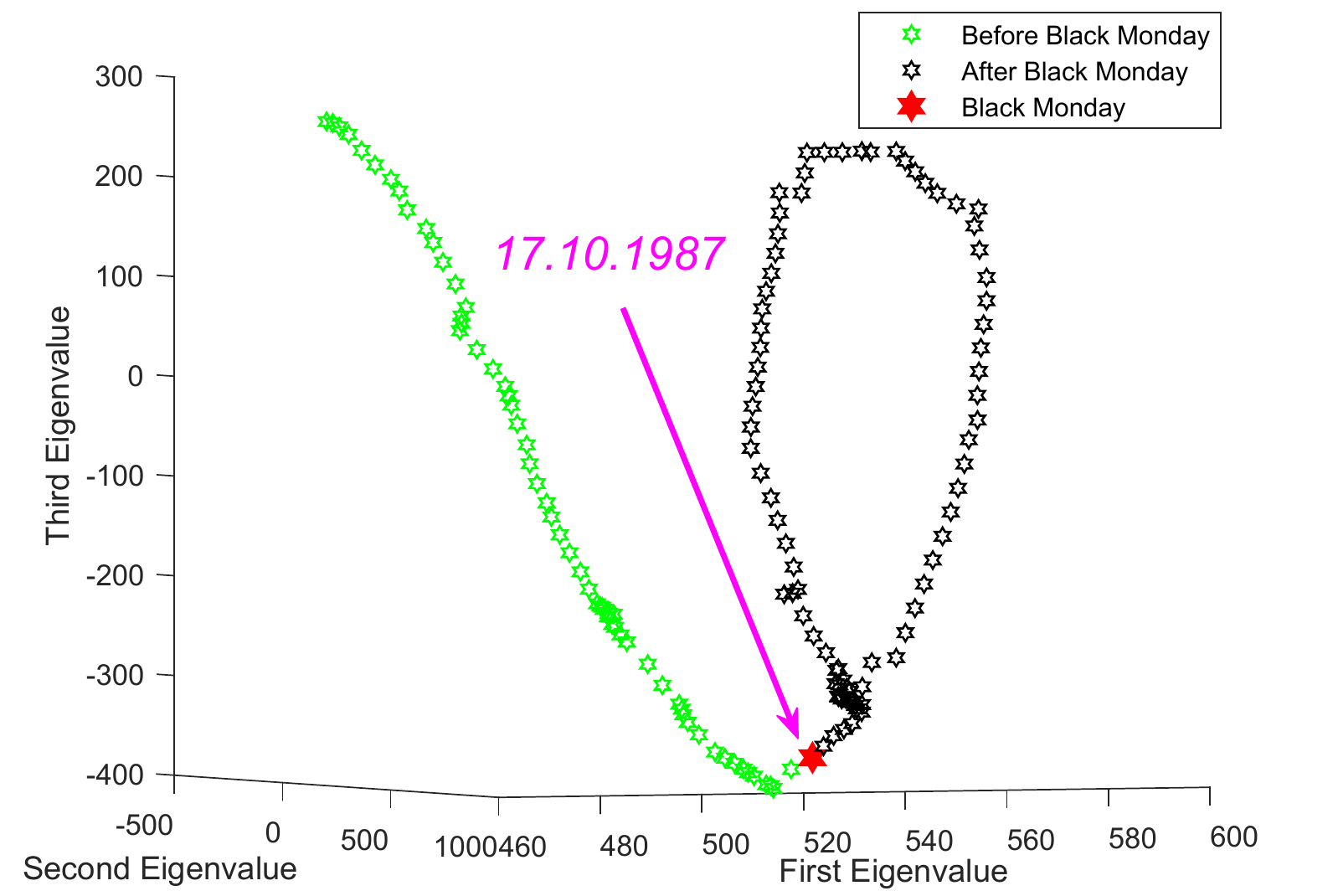}}
\subfigure[Black Monday for QK]{\includegraphics[width=0.49\linewidth]{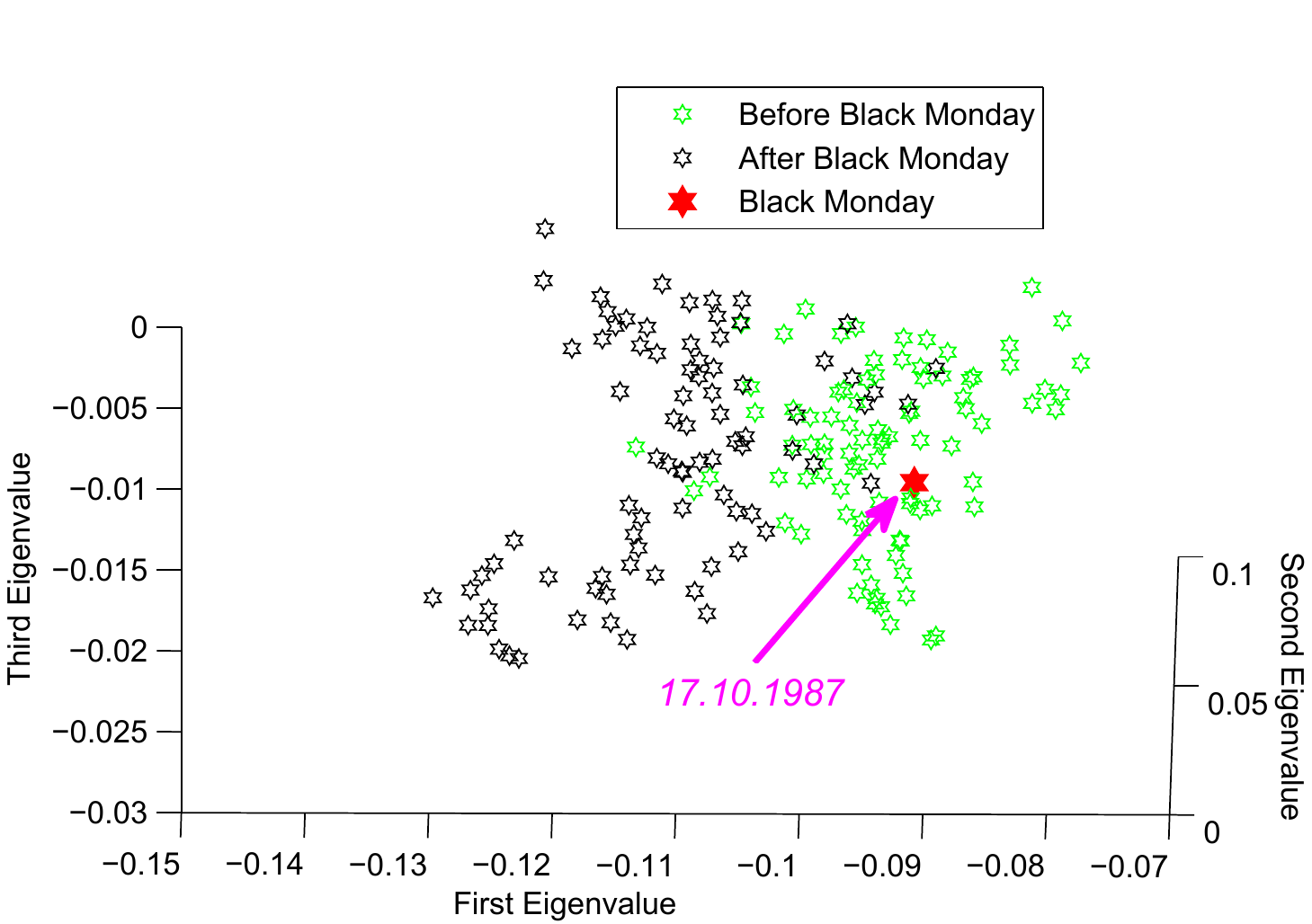}}
\vspace{-10pt}
\caption{The 3D embeddings of Black Monday.} \label{embeddingsB}
\vspace{-10pt}
\end{figure}

\begin{figure}
\vspace{-0pt}
\centering
\subfigure[Dot-com Bubble for EDTWK]{\includegraphics[width=0.49\linewidth]{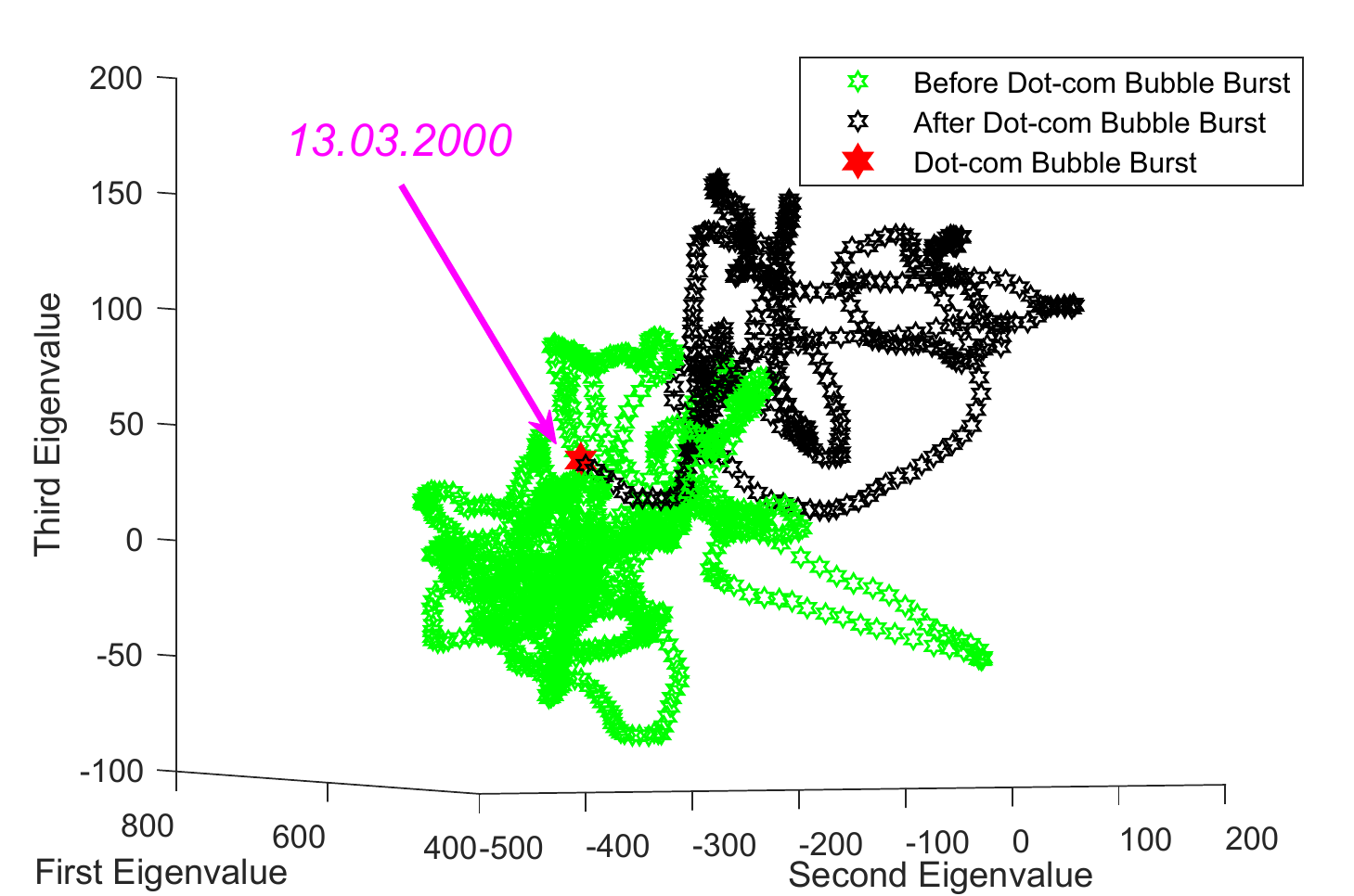}}
\subfigure[Dot-com Bubble for GC]{\includegraphics[width=0.49\linewidth]{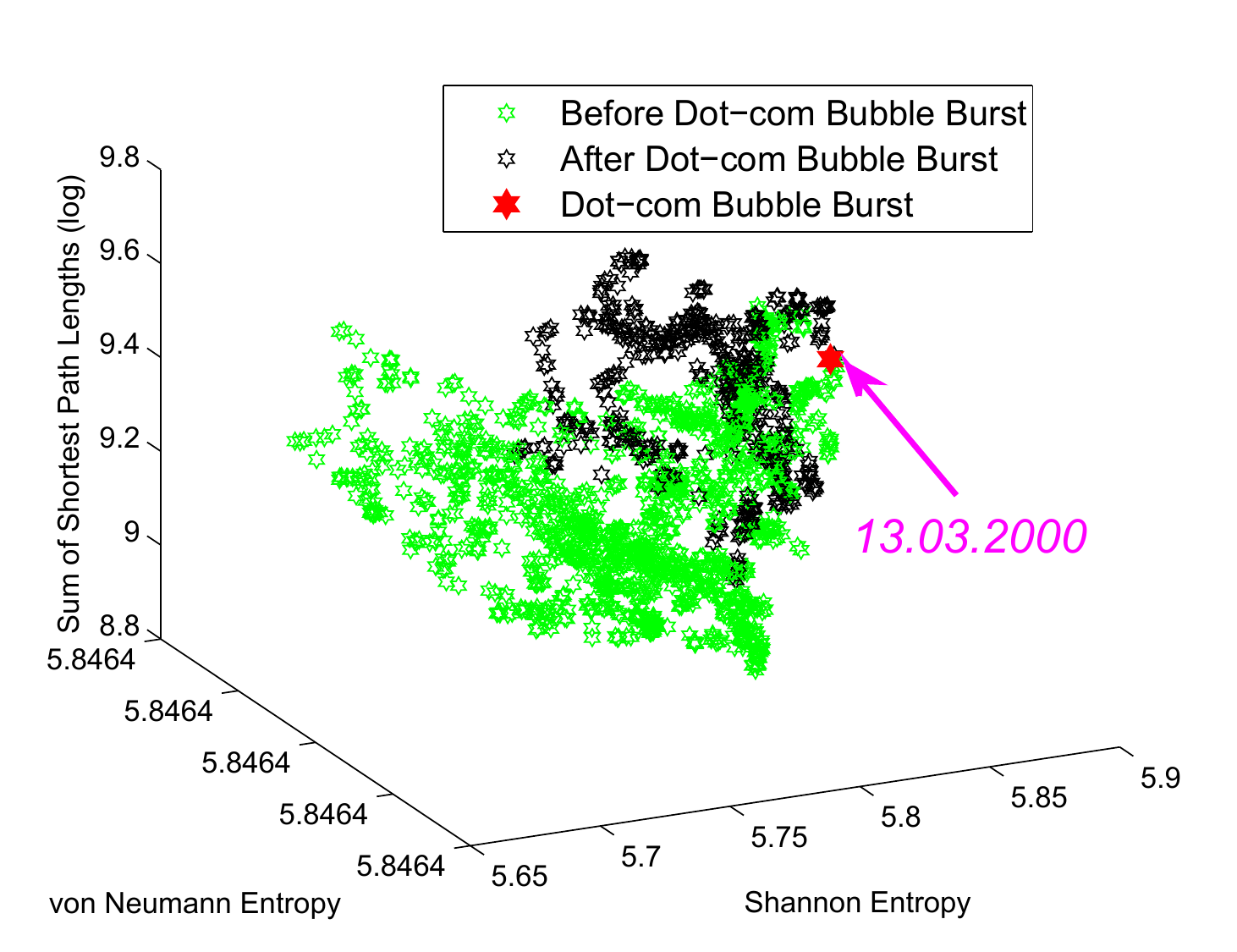}}
\subfigure[Dot-com Bubble for GAK]{\includegraphics[width=0.49\linewidth]{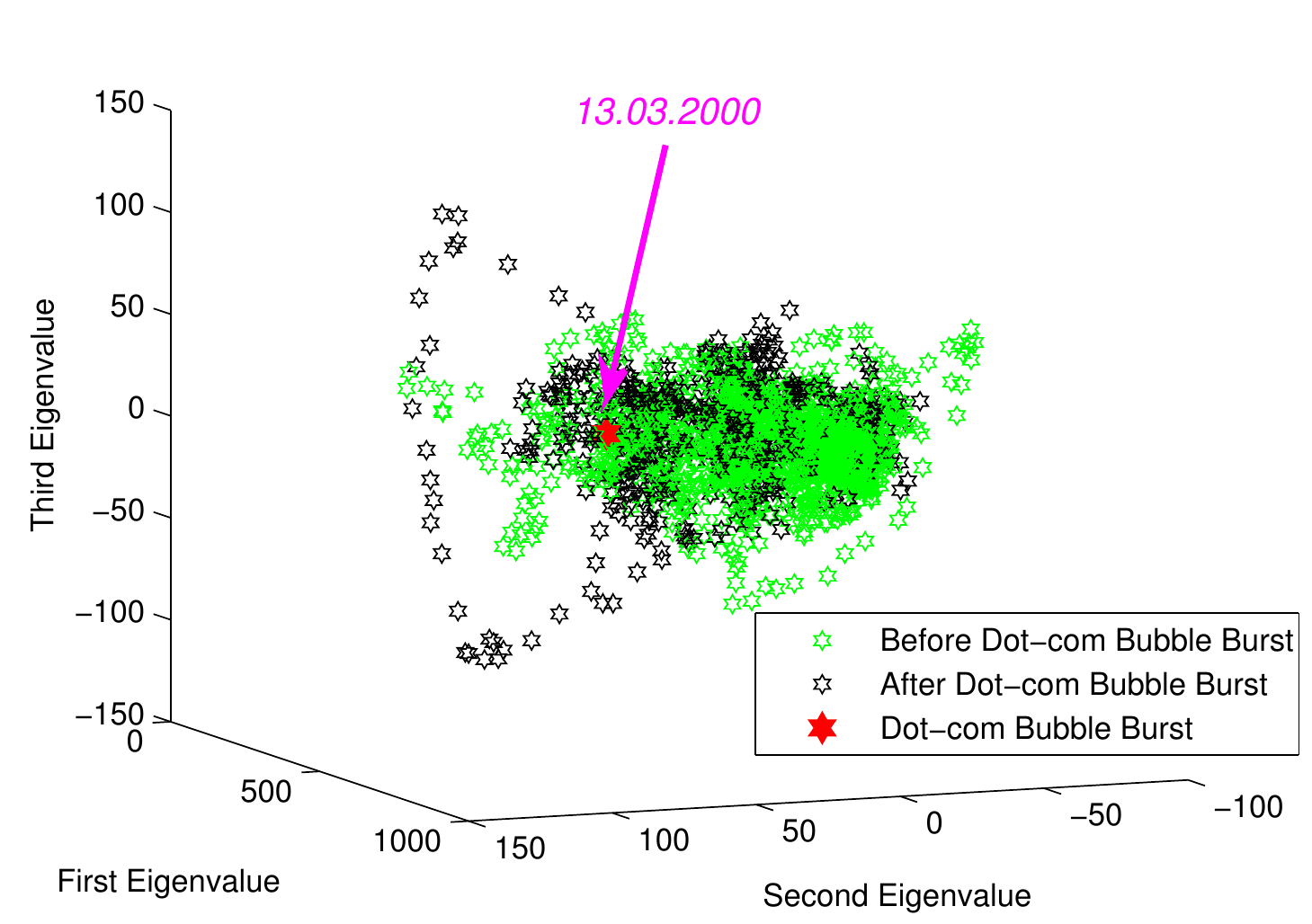}}
\subfigure[Dot-com Bubble for WLSK]{\includegraphics[width=0.49\linewidth]{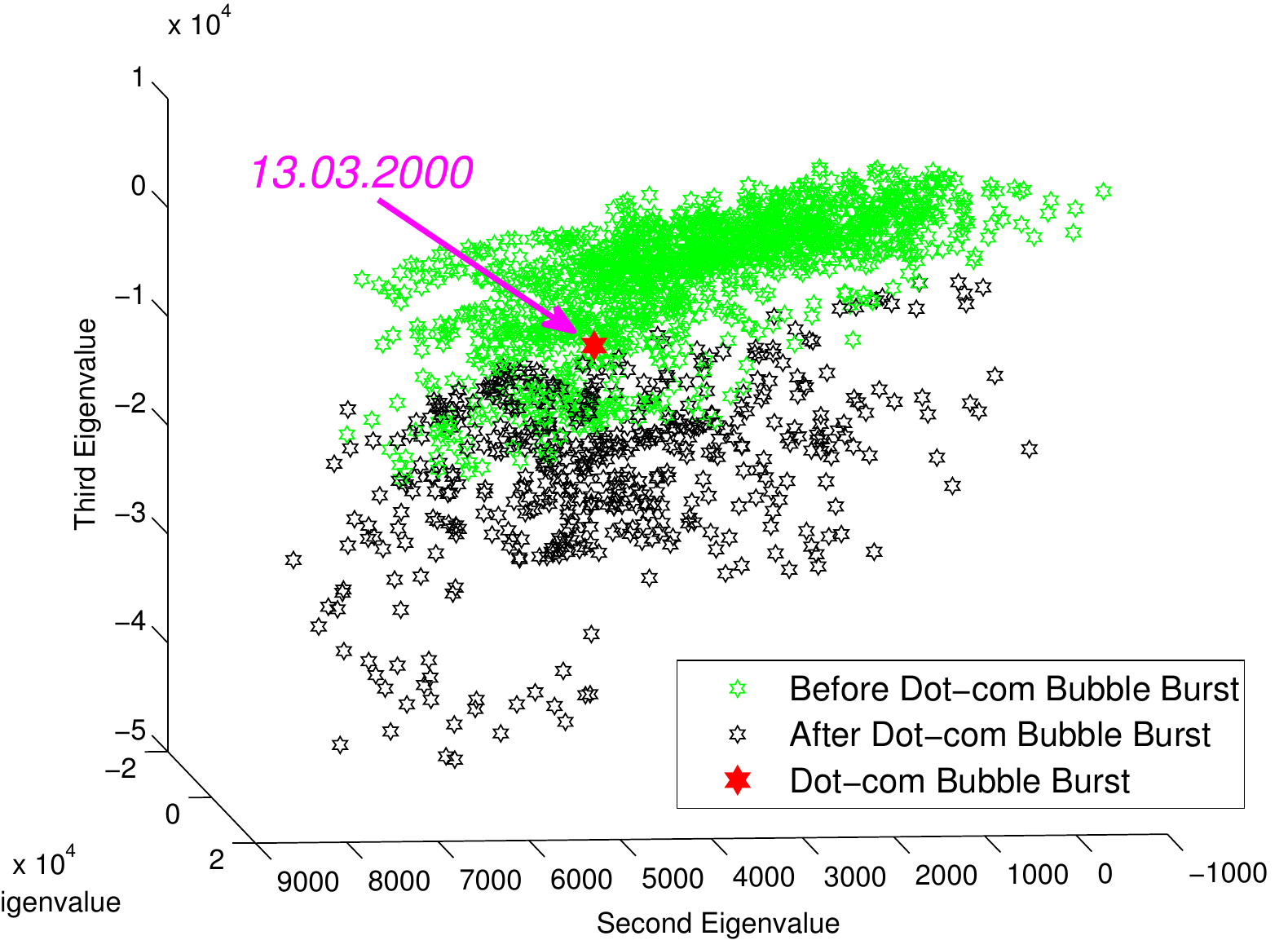}}
\subfigure[Dot-com Bubble for QJSK]{\includegraphics[width=0.49\linewidth]{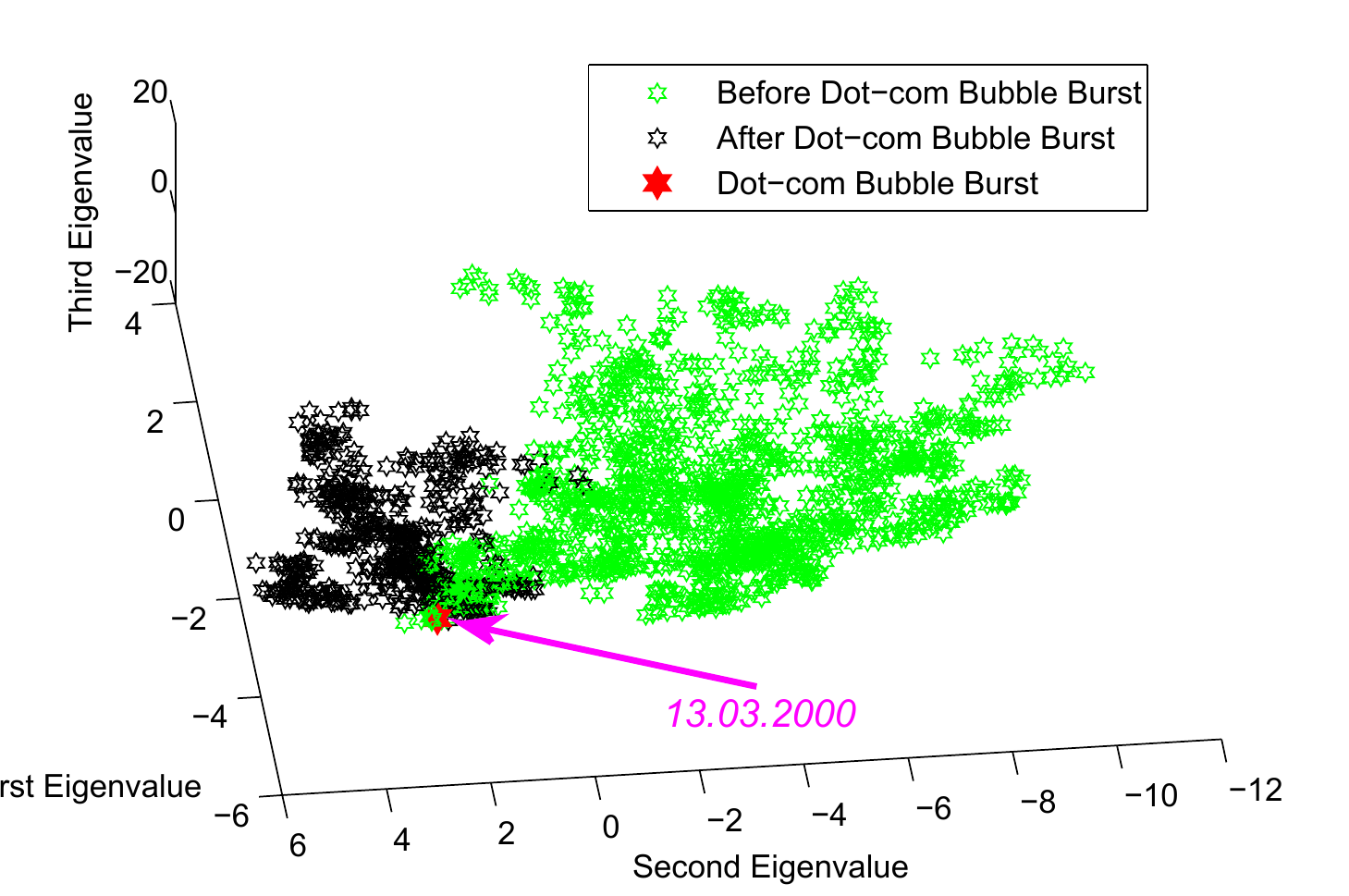}}
\subfigure[Dot-com Bubble for FLK]{\includegraphics[width=0.49\linewidth]{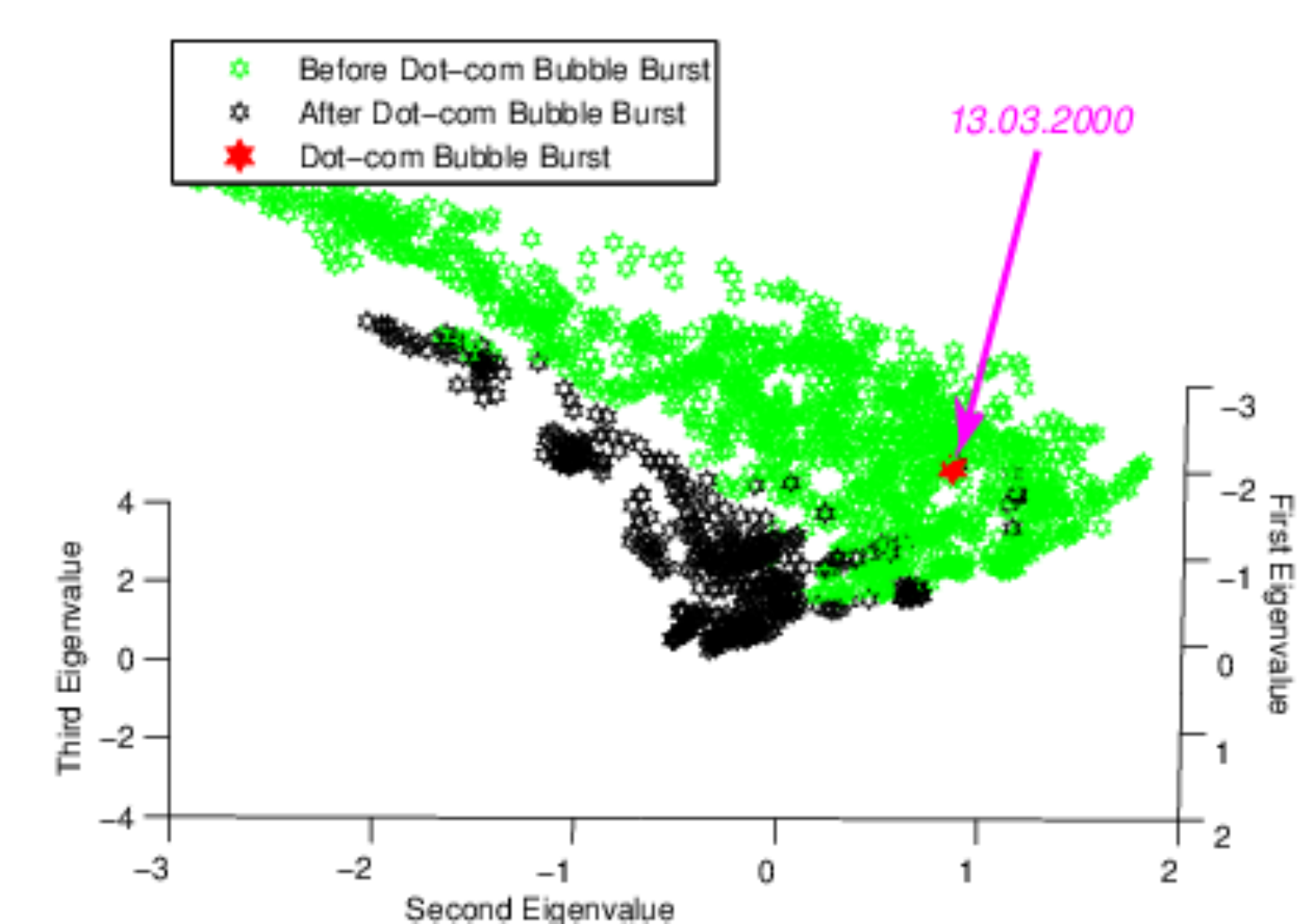}}
\subfigure[Dot-com Bubble for EDTWKO]{\includegraphics[width=0.49\linewidth]{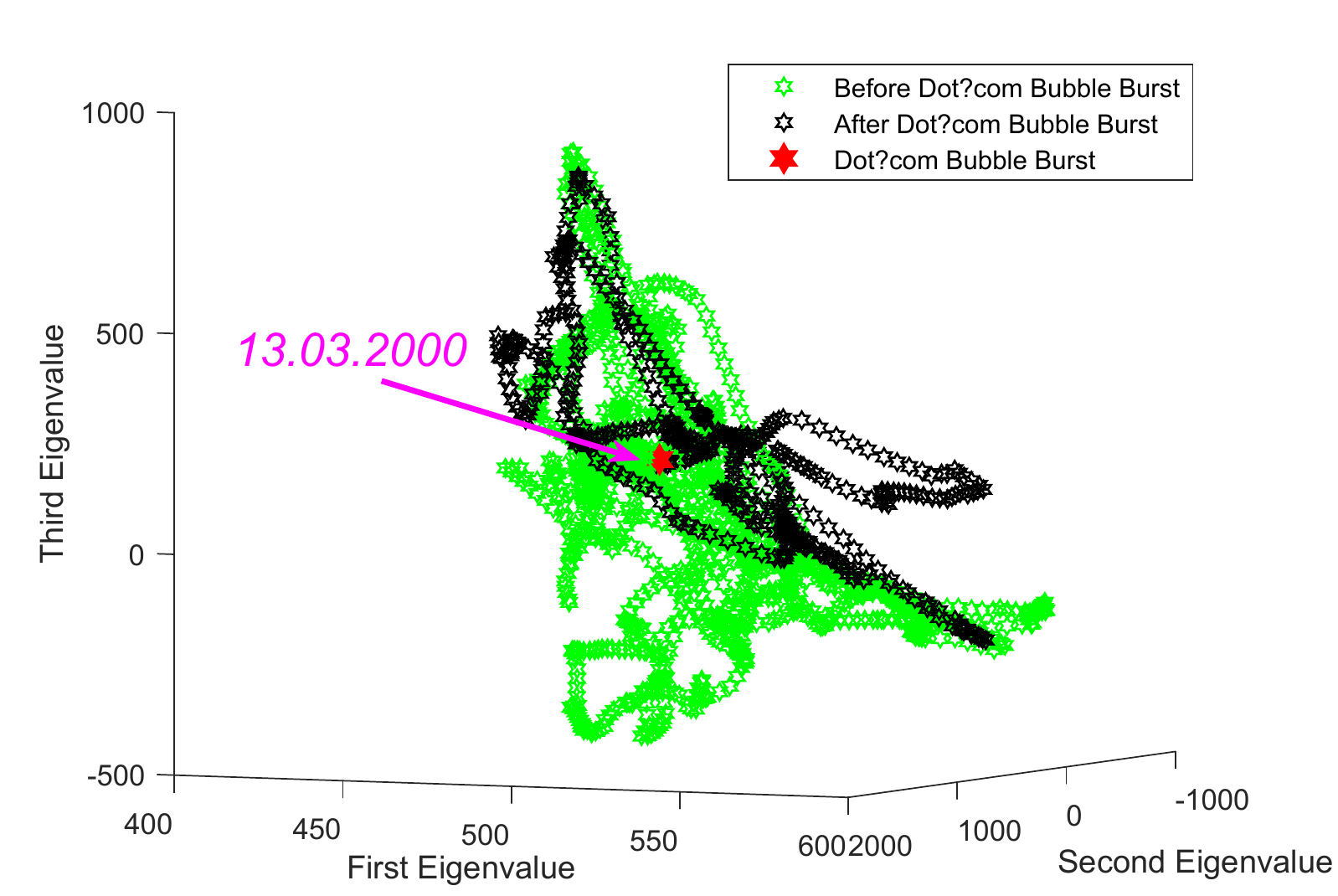}}
\subfigure[Dot-com Bubble for QK]{\includegraphics[width=0.49\linewidth]{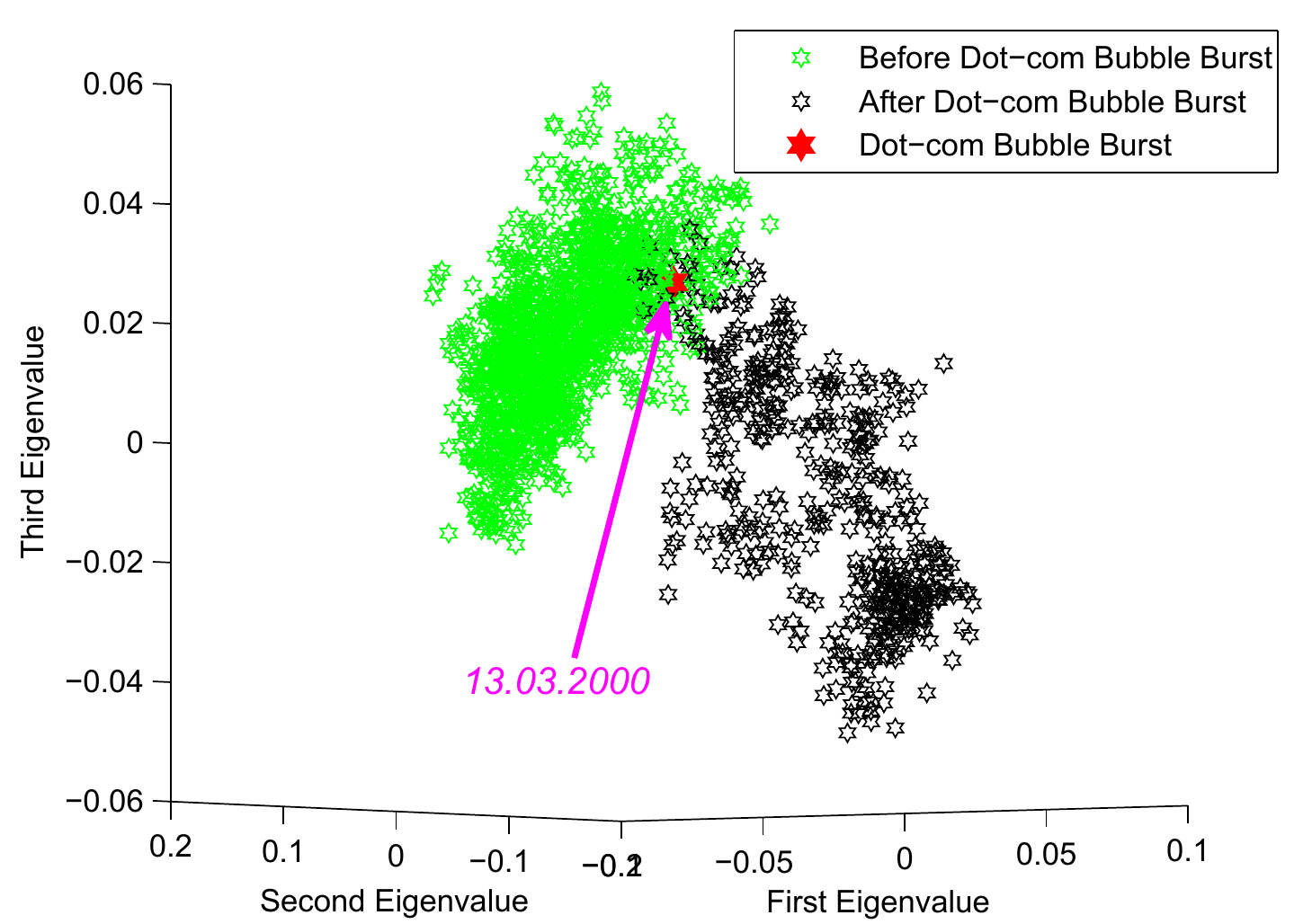}}
\vspace{-10pt}
\caption{The 3D embedding of Dot-com Bubble Burst.} \label{embeddingsD}
\vspace{-20pt}
\end{figure}

\begin{figure}
\vspace{-0pt}
\centering
\subfigure[Enron Incident for EDTWK]{\includegraphics[width=0.49\linewidth]{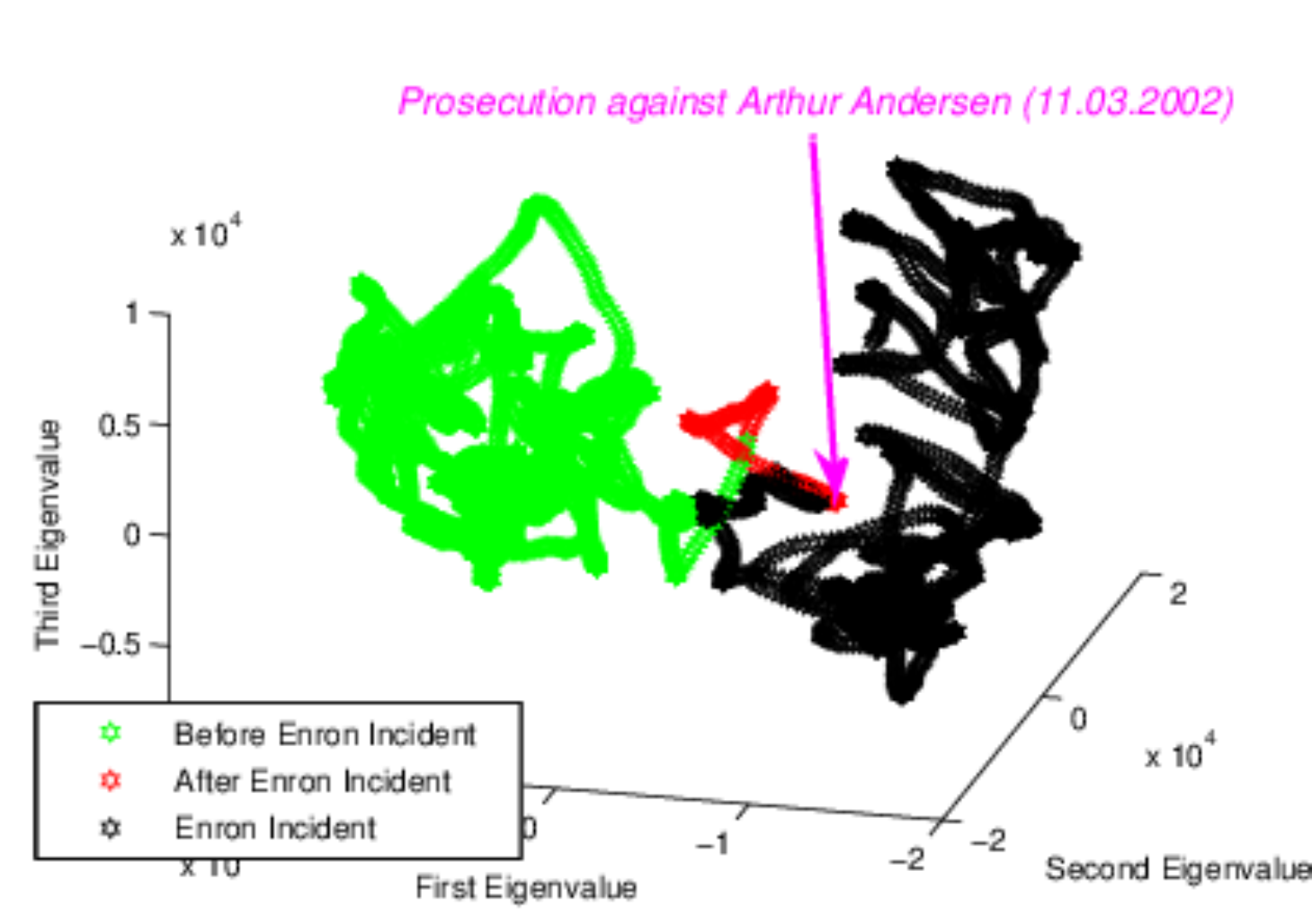}}
\subfigure[Enron Incident for GC]{\includegraphics[width=0.49\linewidth]{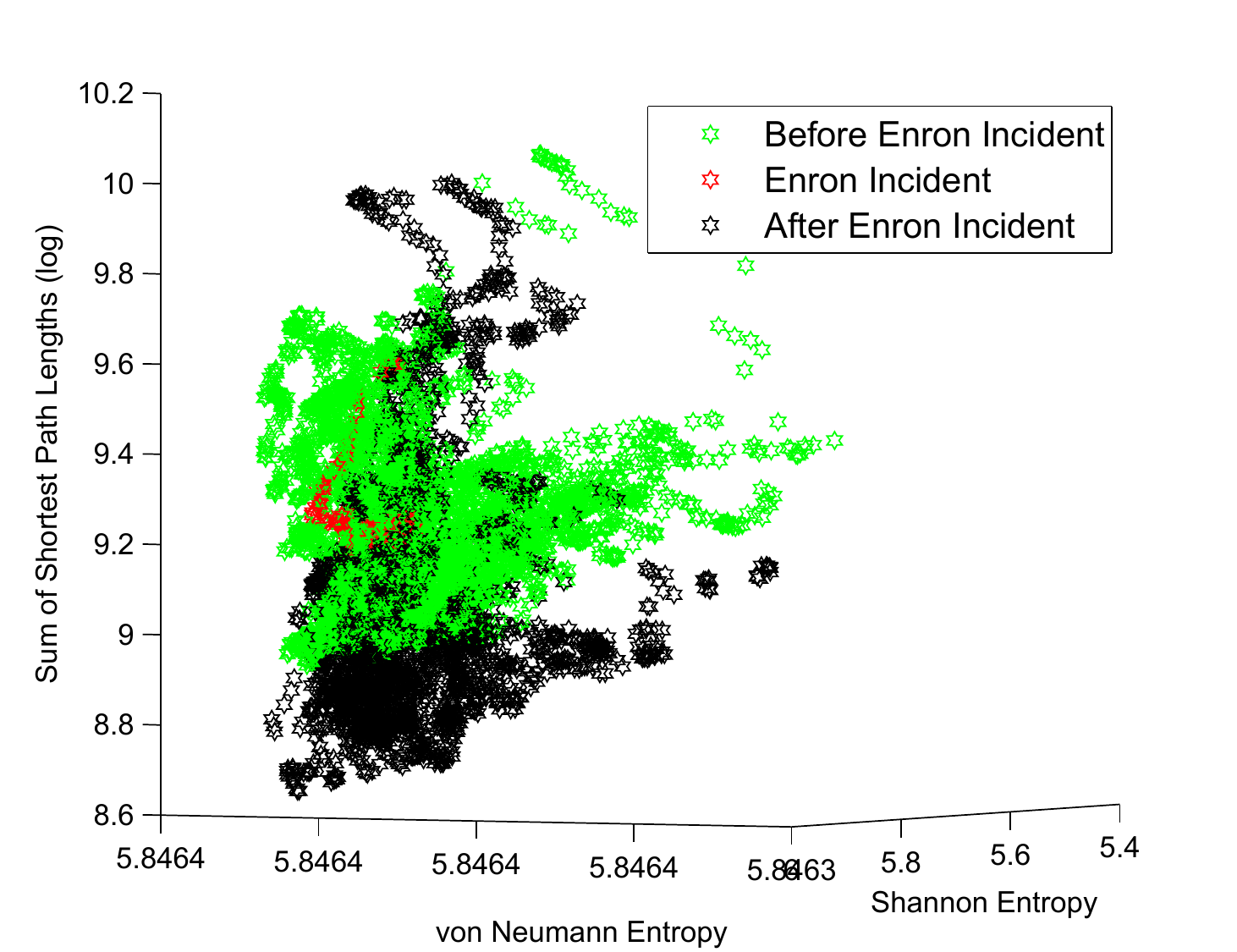}}
\subfigure[Enron Incident for GAK]{\includegraphics[width=0.49\linewidth]{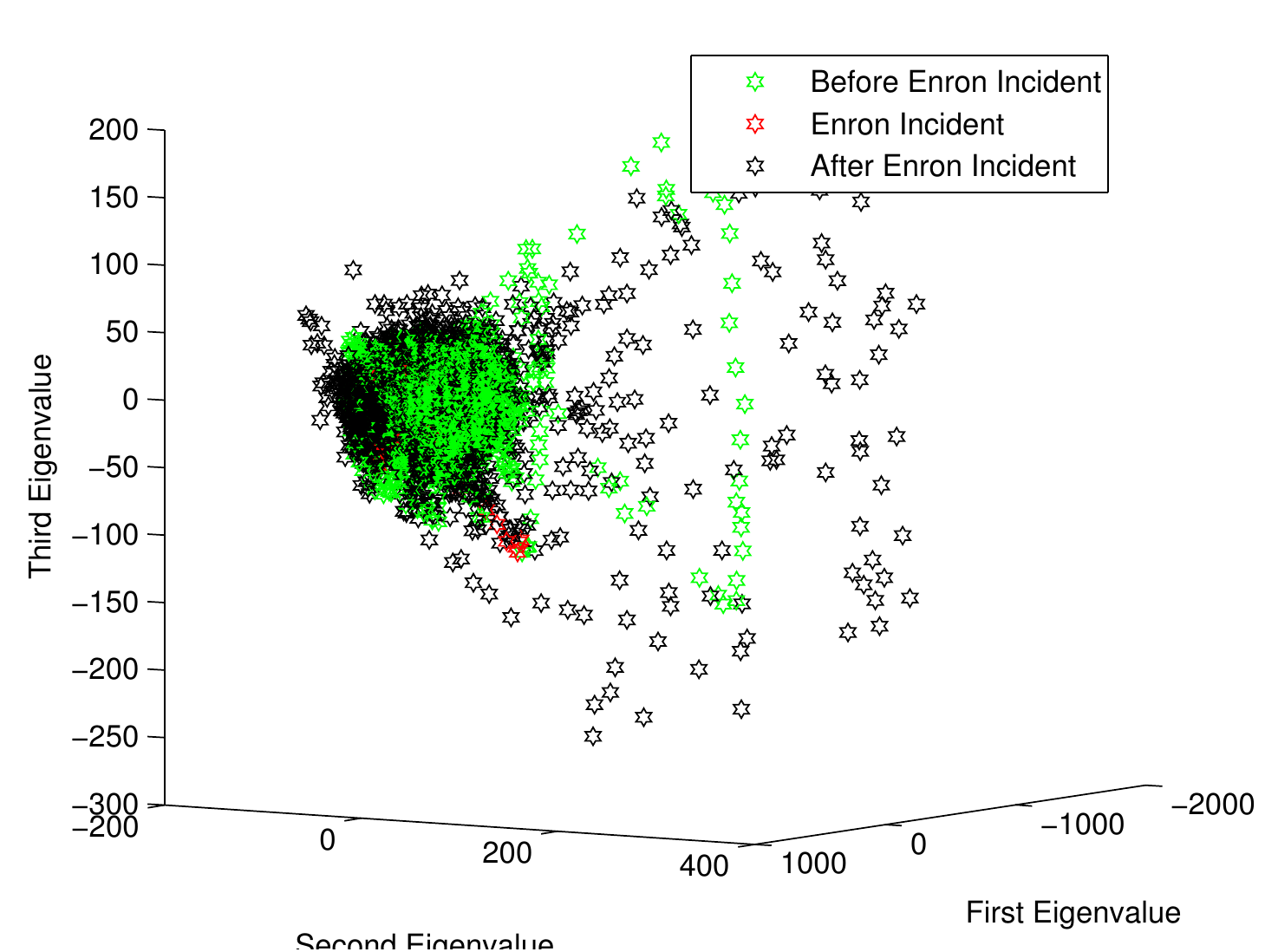}}
\subfigure[Enron Incident for WLSK]{\includegraphics[width=0.49\linewidth]{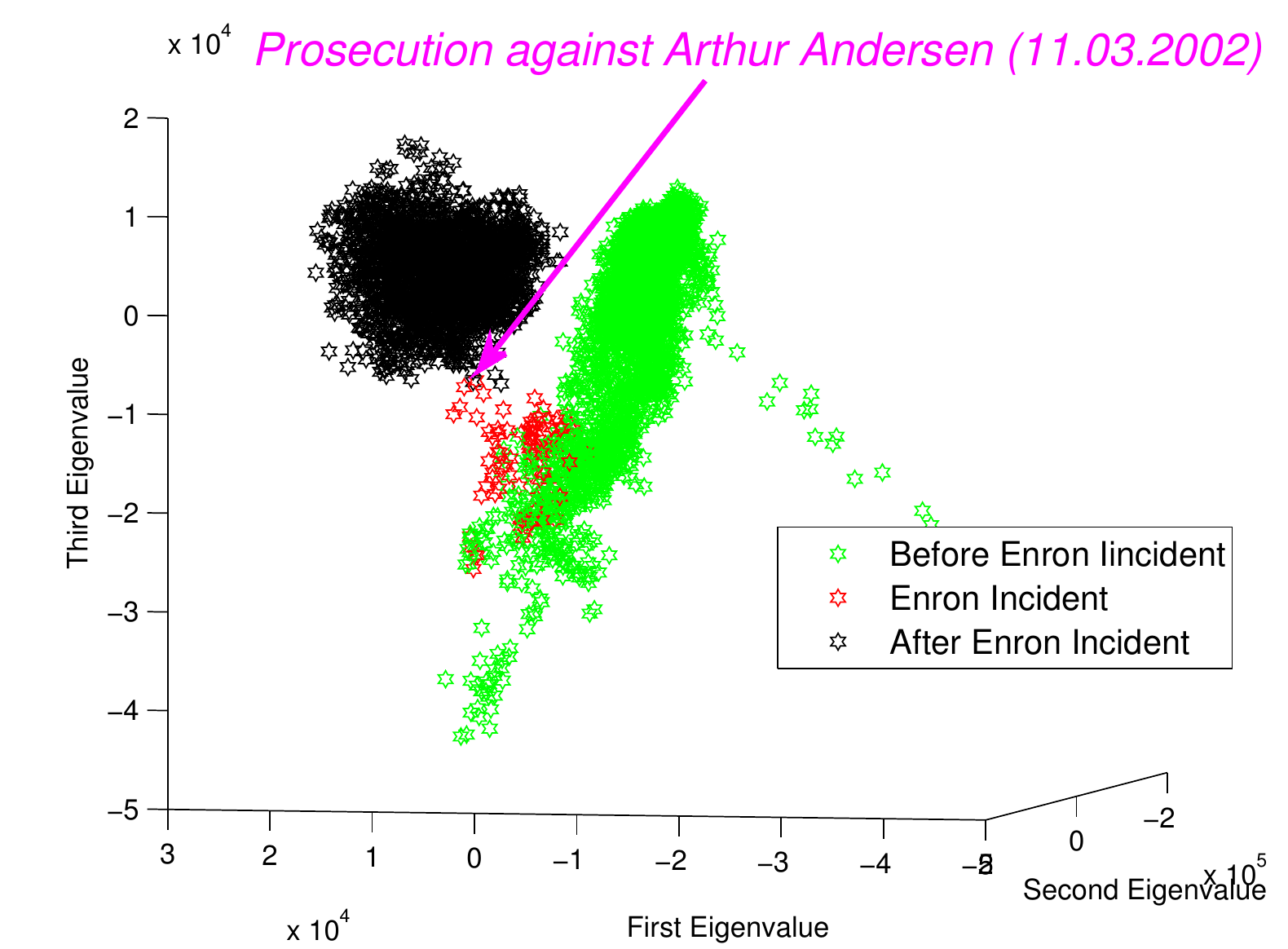}}
\subfigure[Enron Incident for QJSK]{\includegraphics[width=0.49\linewidth]{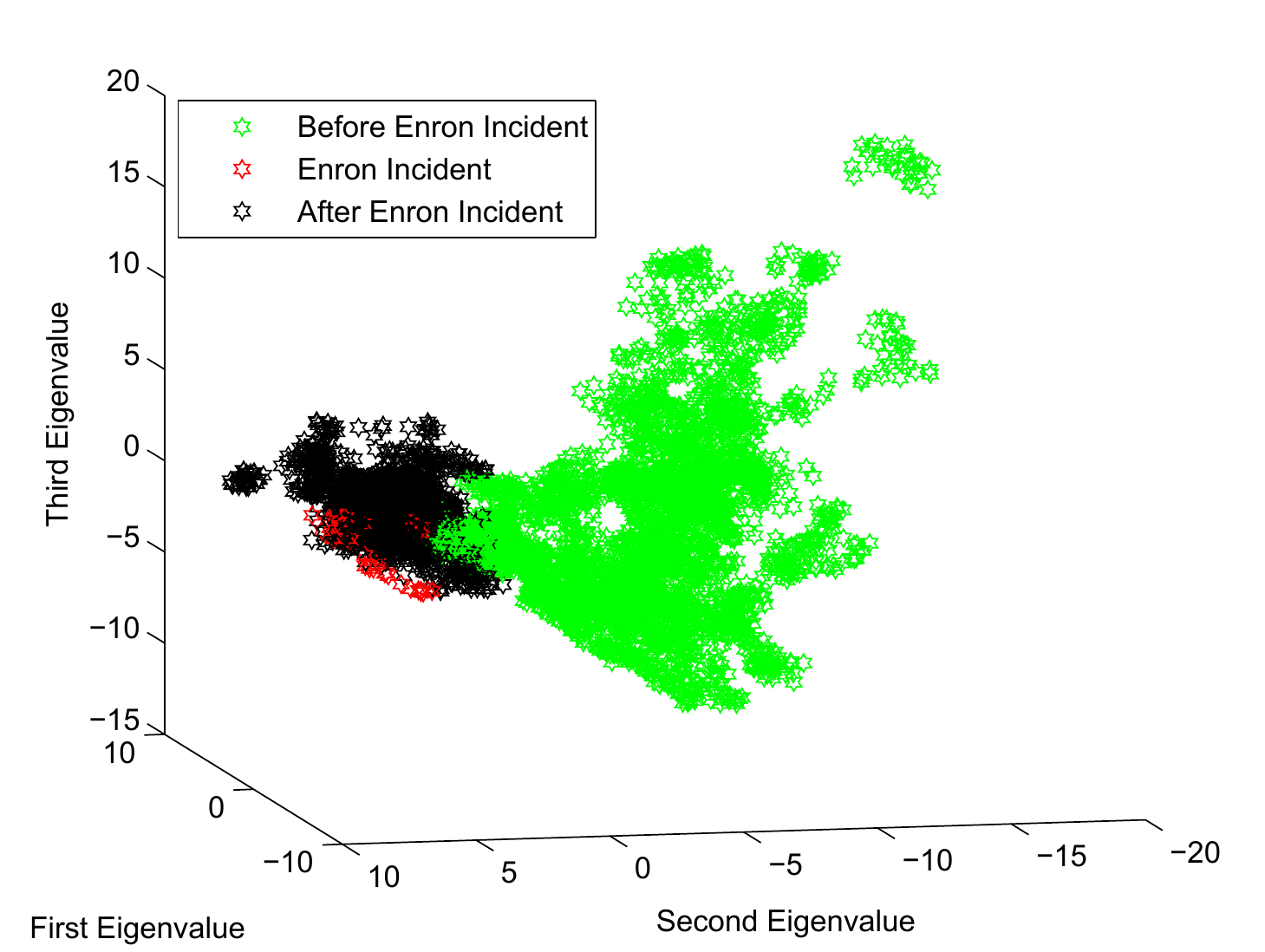}}
\subfigure[Enron Incident for FLK]{\includegraphics[width=0.49\linewidth]{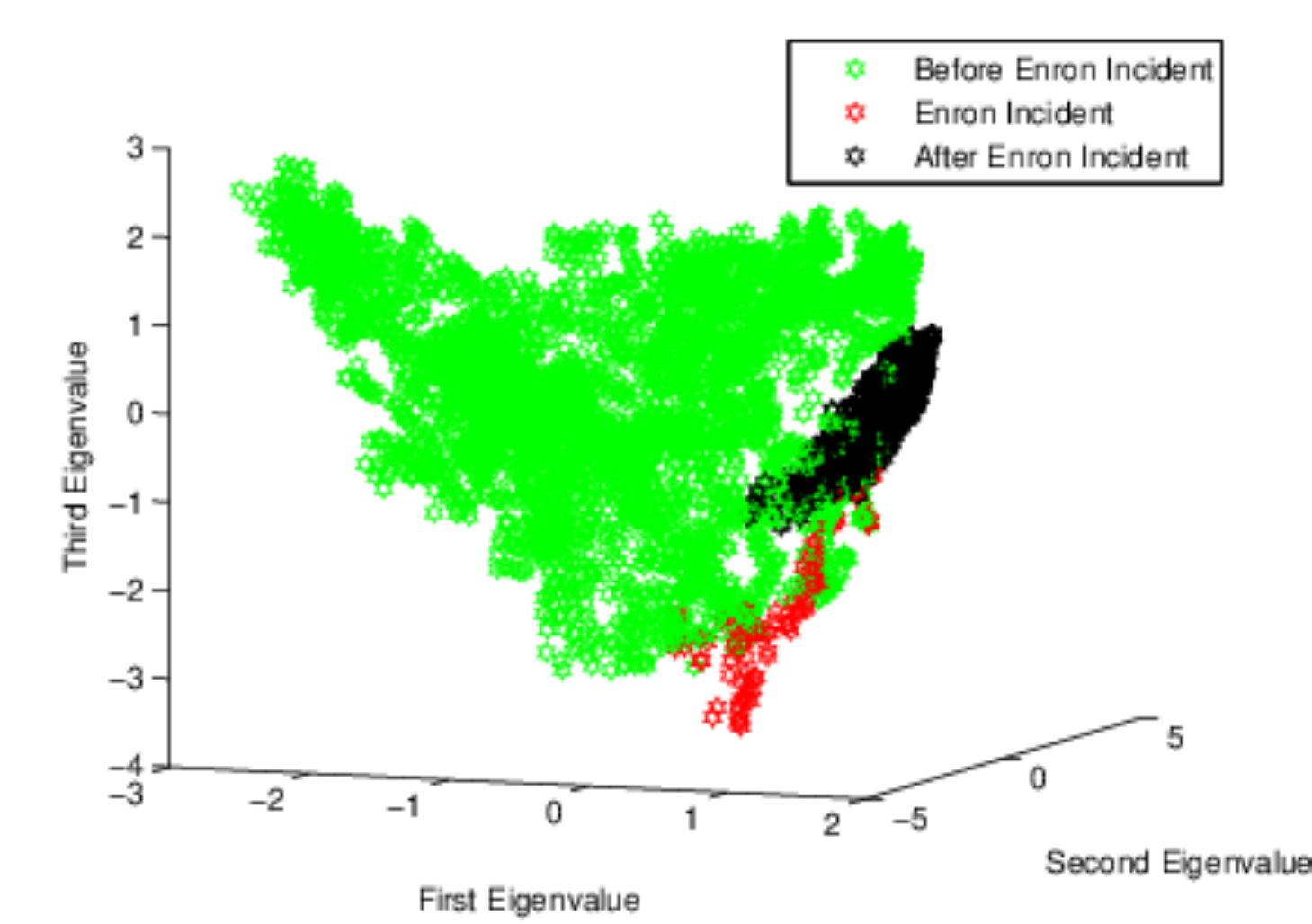}}
\subfigure[Enron Incident for EDTWKO]{\includegraphics[width=0.49\linewidth]{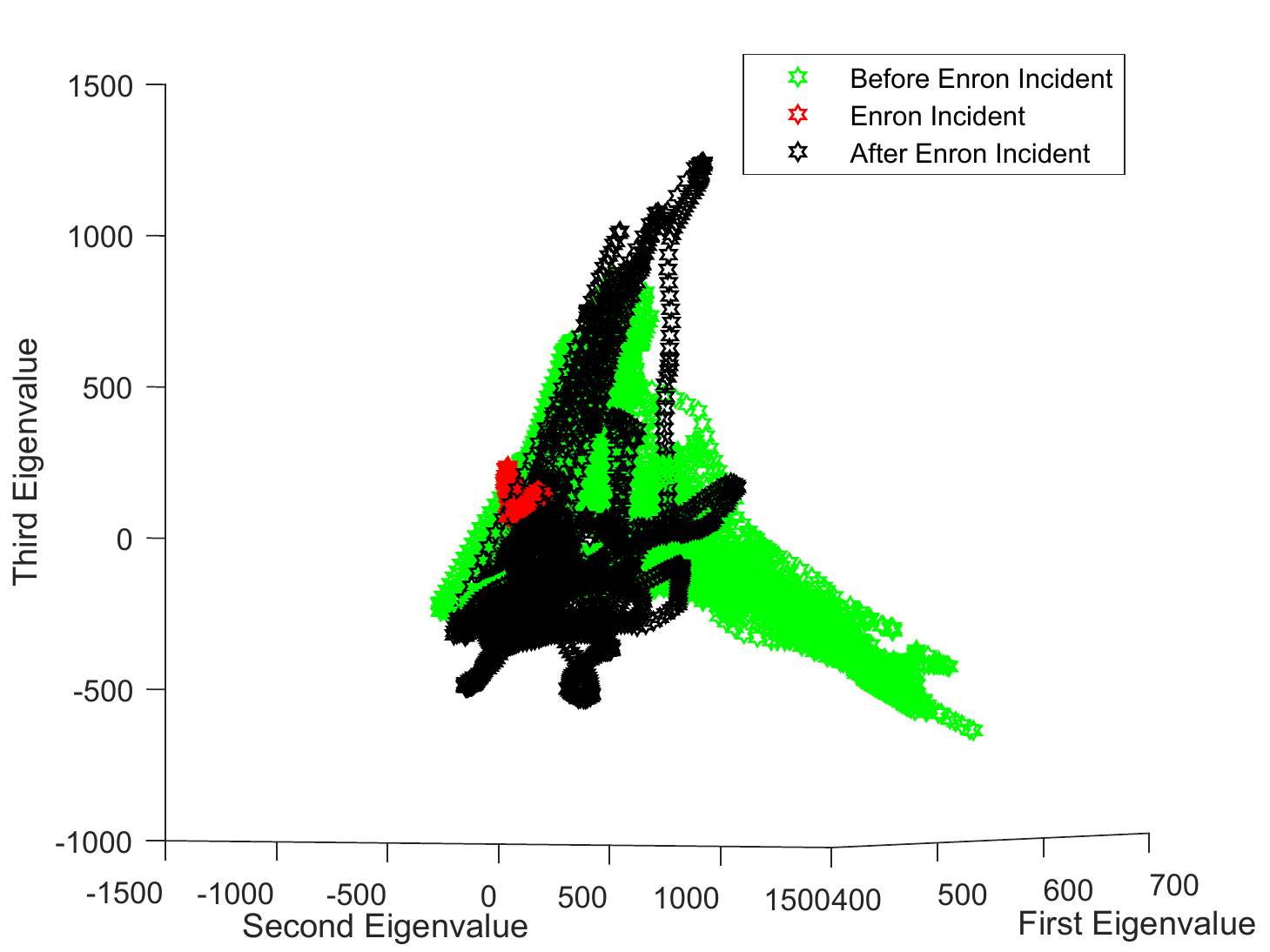}}
\subfigure[Enron Incident for QK]{\includegraphics[width=0.49\linewidth]{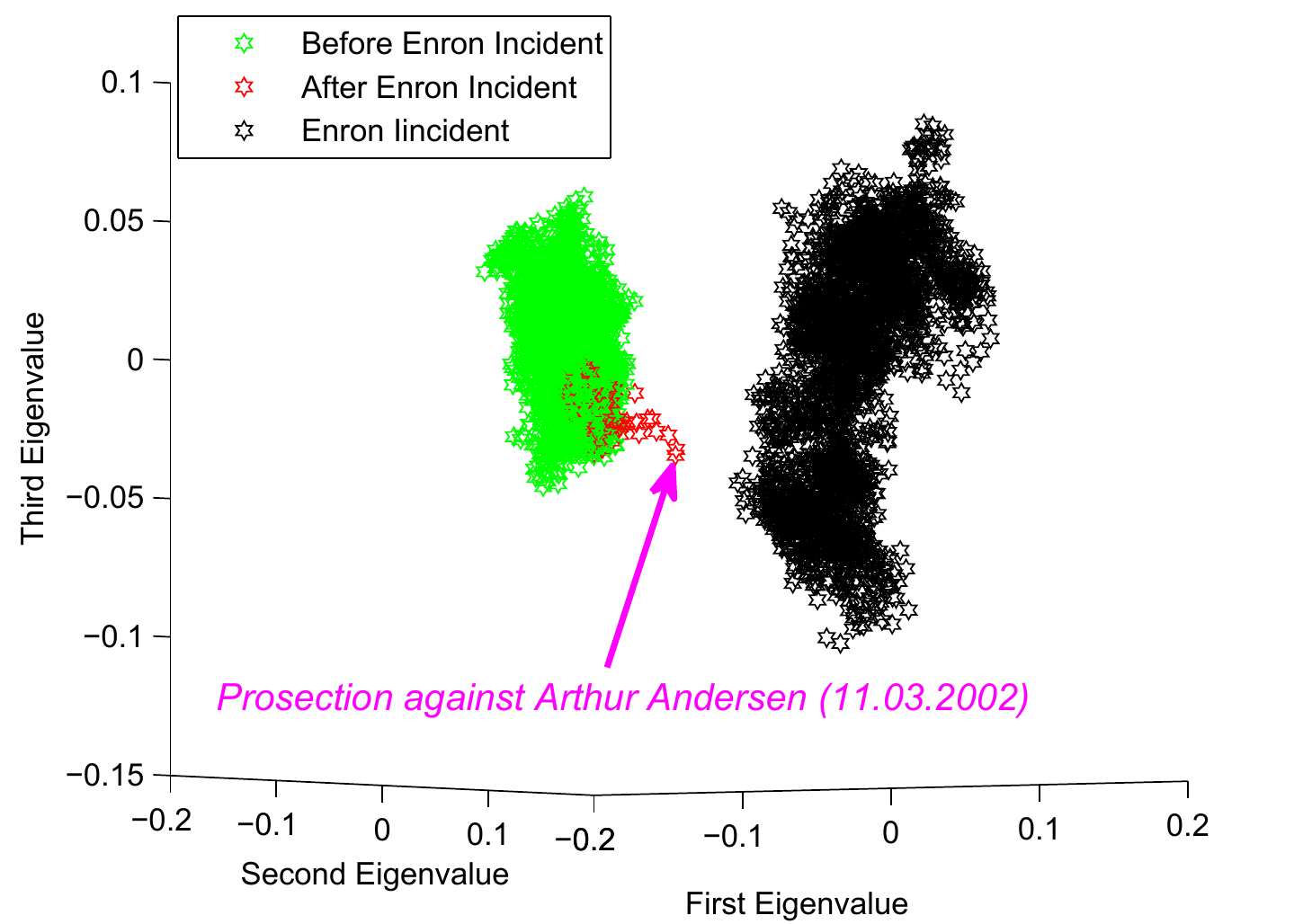}}
\vspace{-10pt}
\caption{The 3D embedding of Enron Incident.} \label{embeddingsE}
\vspace{-20pt}
\end{figure}
To take our study one step further, in this subsection we exhibit more details of the kernel embedding results for three different financial crisis periods that have been preliminarily explored in Section~\ref{exp:s1}. Specifically, for differen methods, Fig.\ref{embeddingsB} corresponds to the Black Monday period (\emph{\textbf{from 15th June 1987 to 17th February 1988}}), Fig.\ref{embeddingsD} corresponds to the Dot-com Bubble period (\emph{\textbf{from 3rd January 1995 to 31st December 2001}}), and Fig.\ref{embeddingsE} corresponds to the Enron Incident period (\emph{\textbf{the red points, from 16th October 2001 to 11th March 2002}}). These figures show that the Black Monday (\emph{\textbf{17th October, 1987}}), the Dot-com Bubble Burst (\emph{\textbf{13rd March, 2000}}) and the Enron Incident period (\emph{\textbf{from 2nd December 2001 to 11th March 2002}}) are all extreme financial events. The embedding points of the proposed EDTWK kernel before and after these events can be better separated into independent clusters, and the points representing the extreme financial events are in the middle of the corresponding clusters. Another interesting phenomenon from Fig.\ref{embeddingsE} is that the embedding points of the networks between year 1986 and year 2011 are distinctly divided by the Prosecution against Arthur Andersen (\emph{\textbf{3rd November, 2002}}). Since the prosecution symbolizes the end of the Enron Incident, the Enron Incident can be considered as a watershed at the beginning of the 21st century, which significantly distinguishes the financial networks of the 21st century from those of the 20th century. These observations again indicate that the proposed kernel can well understand and detect the abrupt financial incident significantly changing the network structures. Although some methods are competitive, only the proposed kernel produces a clearer trajectory in terms of the embedding distributions, i.e., the proposed kernel can better reflect the time-varying transition with time. Note that, we can observe similar results if we explore the proposed kernel and the remaining methods on the alternative financial crisis mentioned in Section~\ref{exp:s1}.

\begin{figure}
\vspace{-0pt}
\centering
\subfigure[Newcentury for EDTWK]{\includegraphics[width=0.49\linewidth]{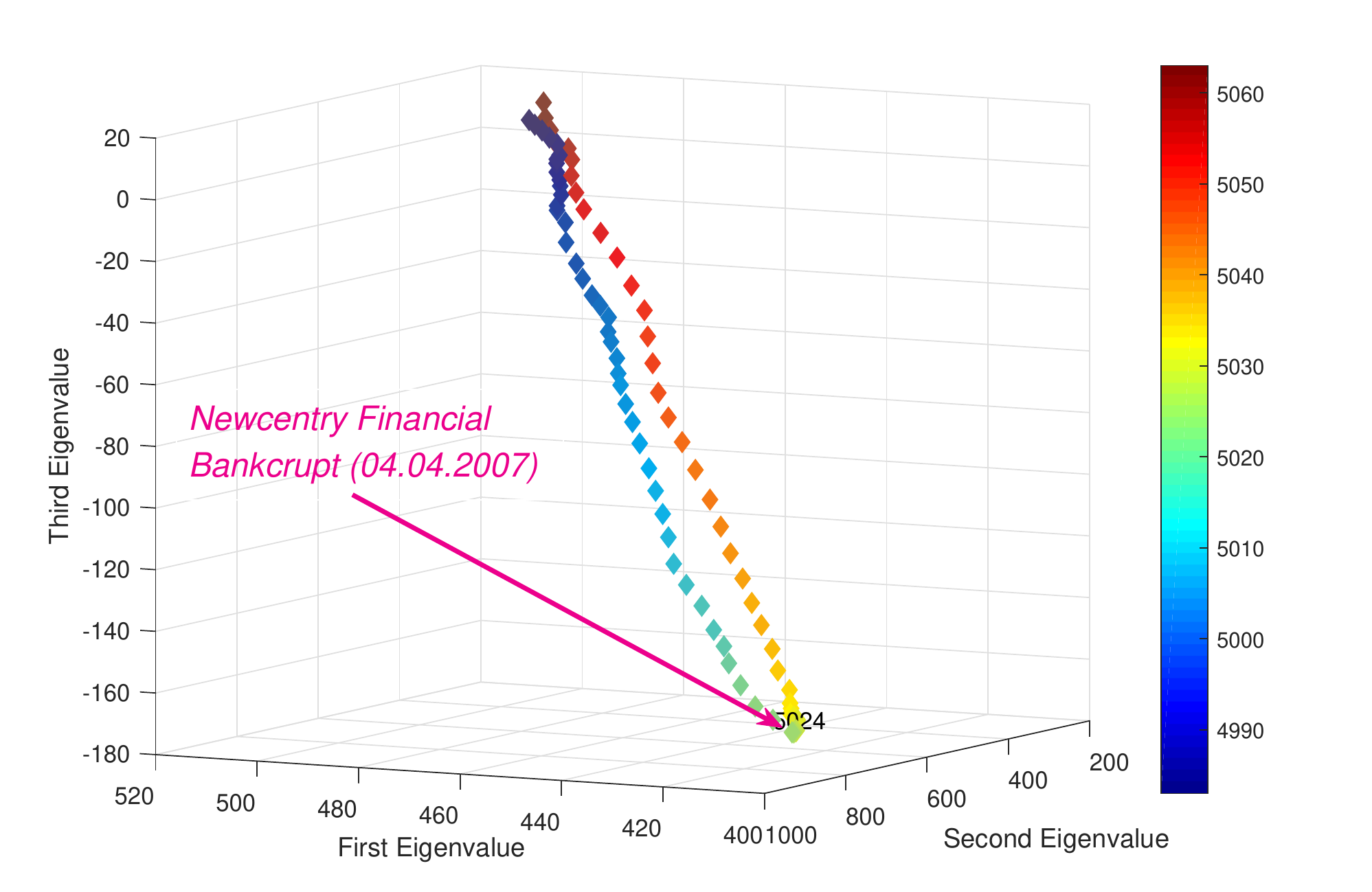}}
\subfigure[Newcentury for WLSK]{\includegraphics[width=0.49\linewidth]{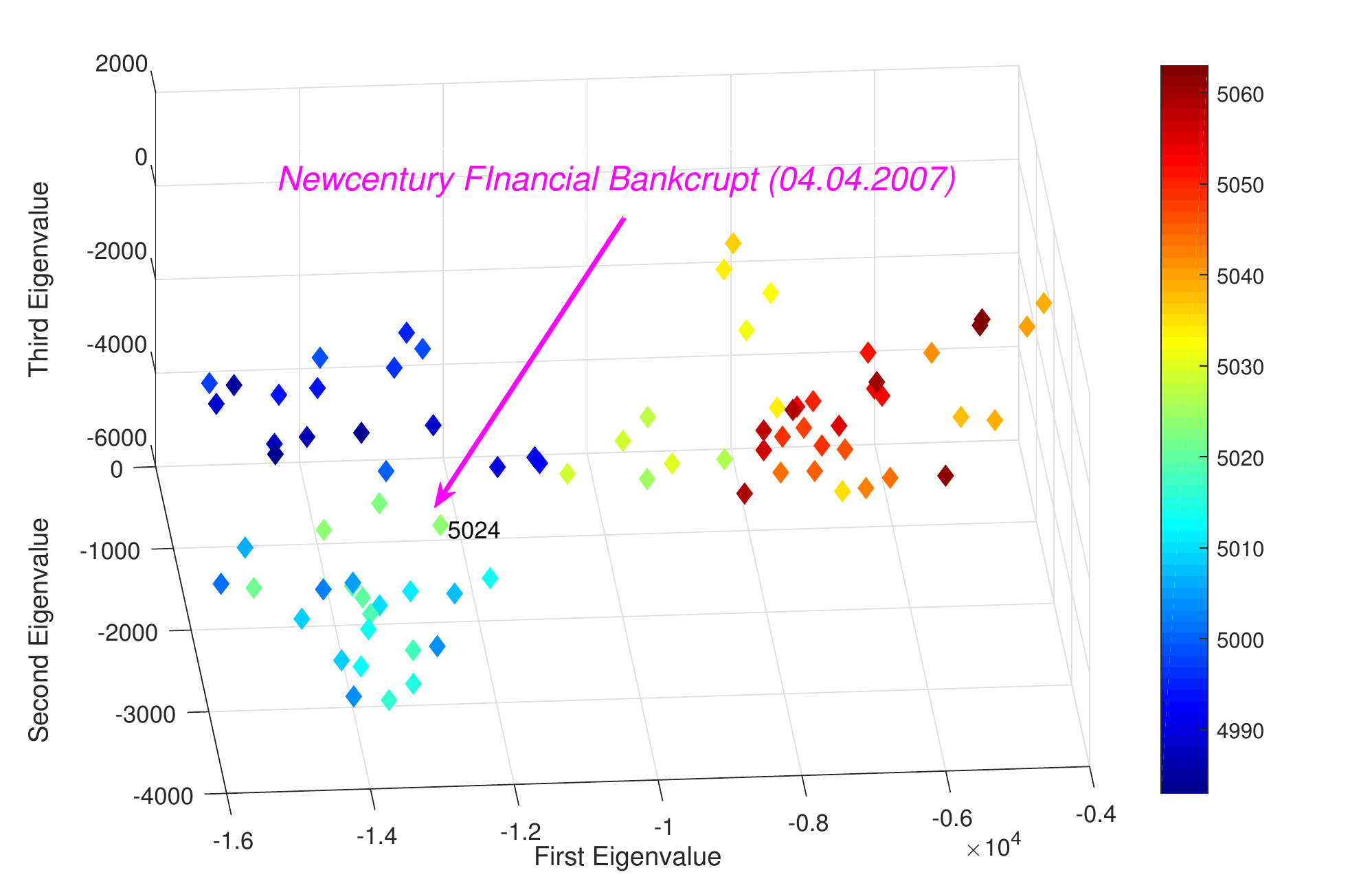}}
\subfigure[Lehman Crisis for EDTWK]{\includegraphics[width=0.49\linewidth]{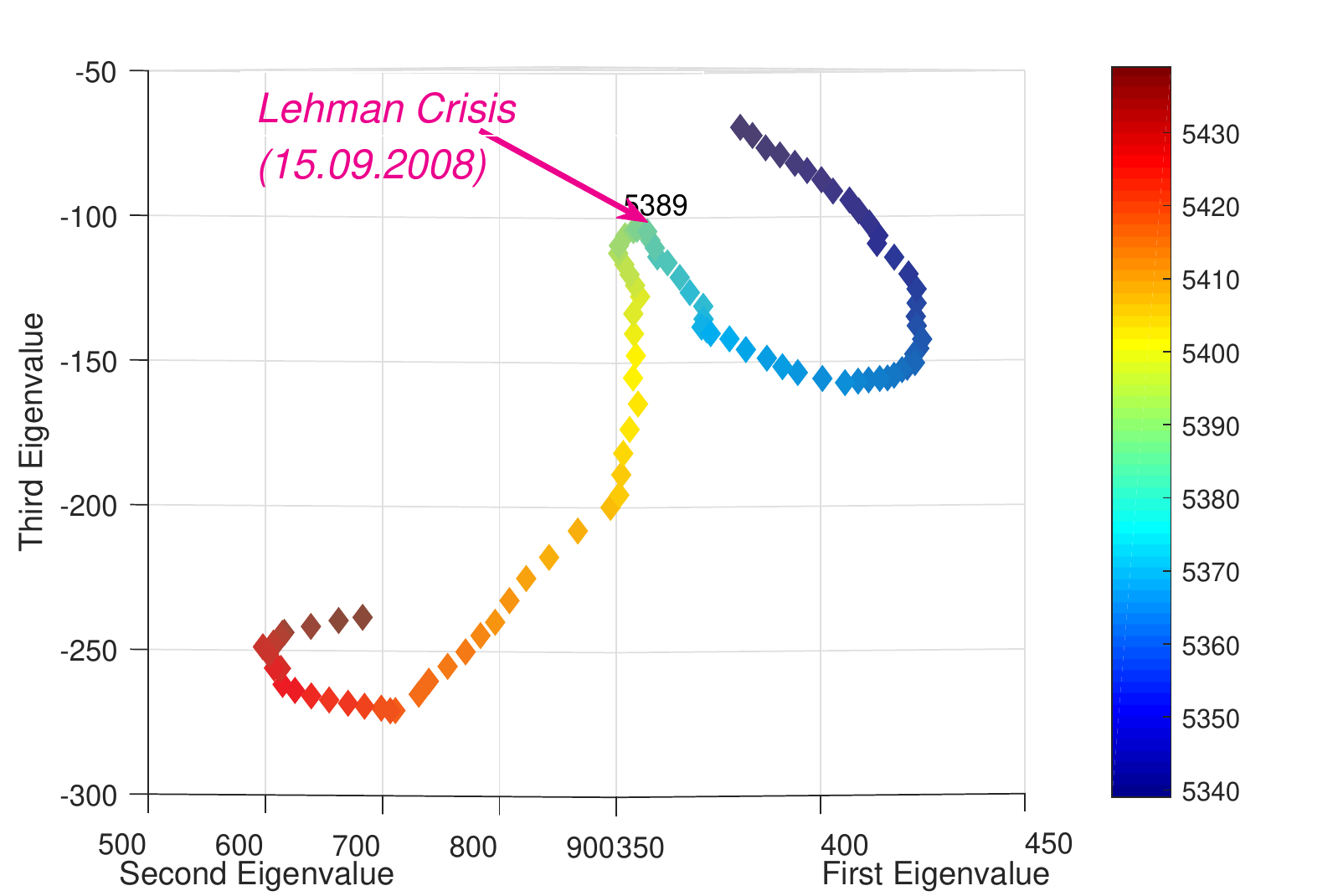}}
\subfigure[Lehman Crisis for WLSK]{\includegraphics[width=0.49\linewidth]{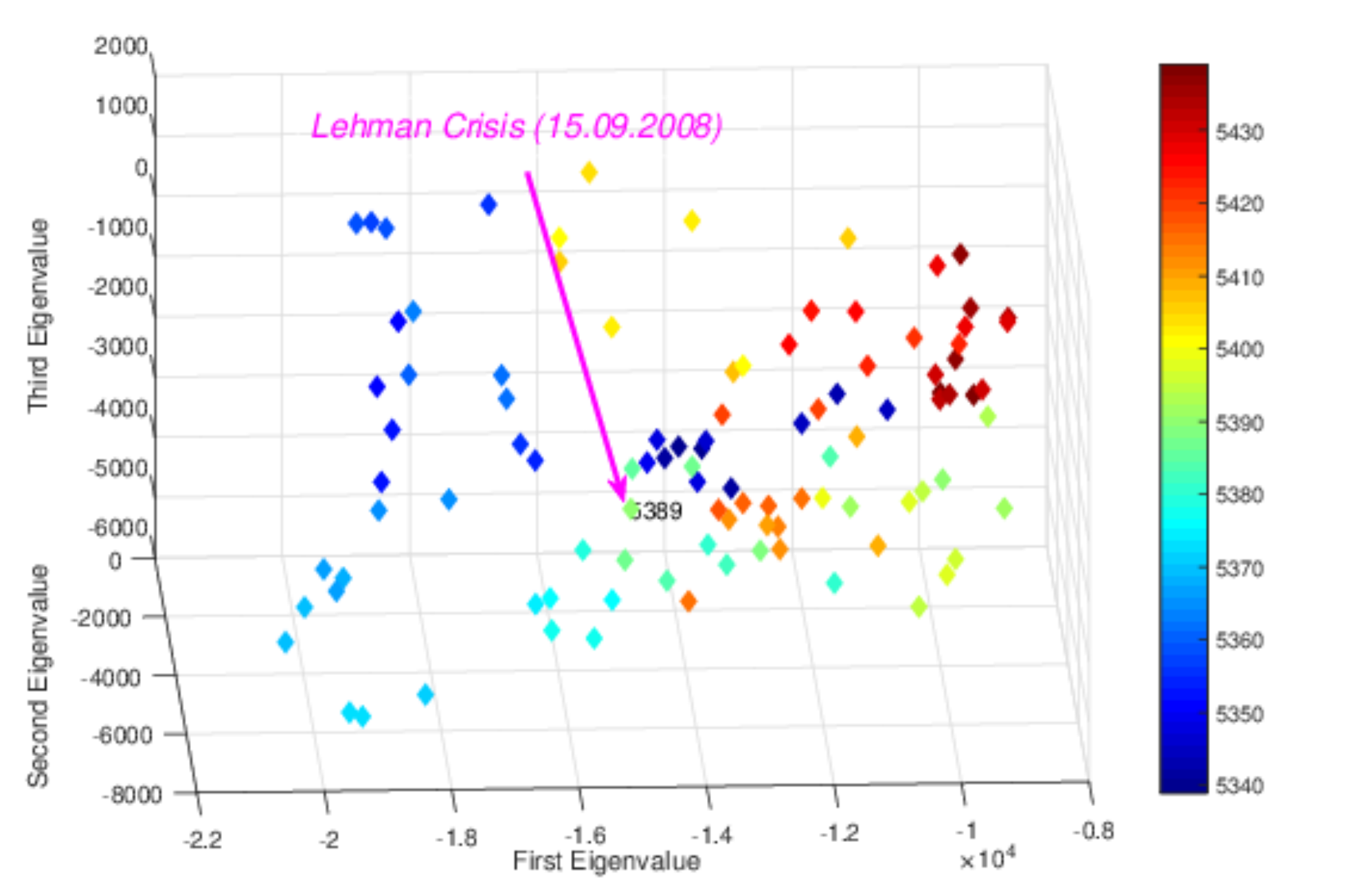}}
\vspace{-10pt}
\caption{Time-varying Trajectory Comparisons of the EDTWK and WLSK Kernels for Sub-prime Crisis.} \label{embeddingsSPC}
\vspace{-20pt}
\end{figure}

The above experimental results indicate that the WLSK kernel as well as the QK kernel are the most competitive alternative methods to the proposed EDTWK kernel. Since the embedding results of these two kernels can also be better divided before and after a financial event. To further demonstrate the effectiveness of the proposed kernel, we compare the three methods on two financial events happened in Sub-Prime Crisis period (\emph{\textbf{from 2nd January 2006 to 1st July 2009}}) in more details. The financial events for comparisons are the Newcentury Financial Bankruptcy (\emph{\textbf{4th April 2007}}) and the Lehman Brothers Bankruptcy (\emph{\textbf{15th September 2008}}), both having significant impacts in the world finance history. Specifically, for each method, we visualize the set of points that indicate the path of the kPCA embeddings with time over about 90 trading days (i.e., 90 points) around each of the two events. The results are shown in Fig.\ref{embeddingsSPC}, and the colour bar beside each subfigure represents the data in time series. Note that, we only show the result from the WLSK kernel, since we will observe
the similar phenomenon from both the WLSK and QK kernel. For the proposed kernel, we observe that the point distribution forms a clear trajectory with time, and the trajectory around each of the financial events usually undergoes significant changes, i.e., the trajectory starting from the financial event will tempestuously change the distribution direction during a short time period. By contrast, the point distributions from the WLSK kernel are chaos and cannot observe any significant phenomenon. This demonstrates that only the proposed kernel has better capability to both characterize and distinguish financial crises.

\subsection{Evaluations of the Kernel Matrix}\label{exp:s4}

\begin{figure}
\vspace{-0pt}
\centering
\subfigure[For EDTWK on All Days]{\includegraphics[width=0.49\linewidth]{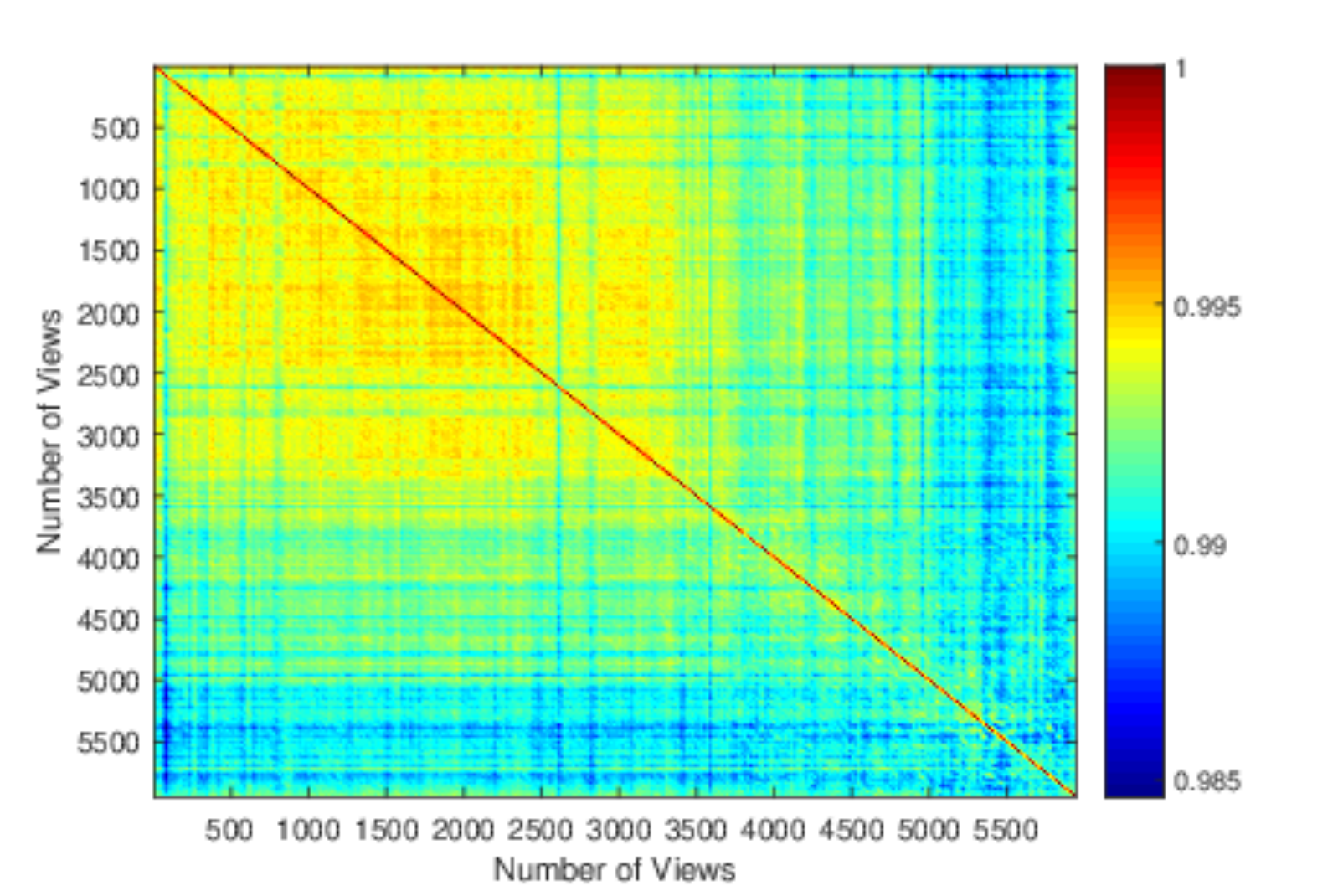}}
\subfigure[For WLSK on All Days]{\includegraphics[width=0.49\linewidth]{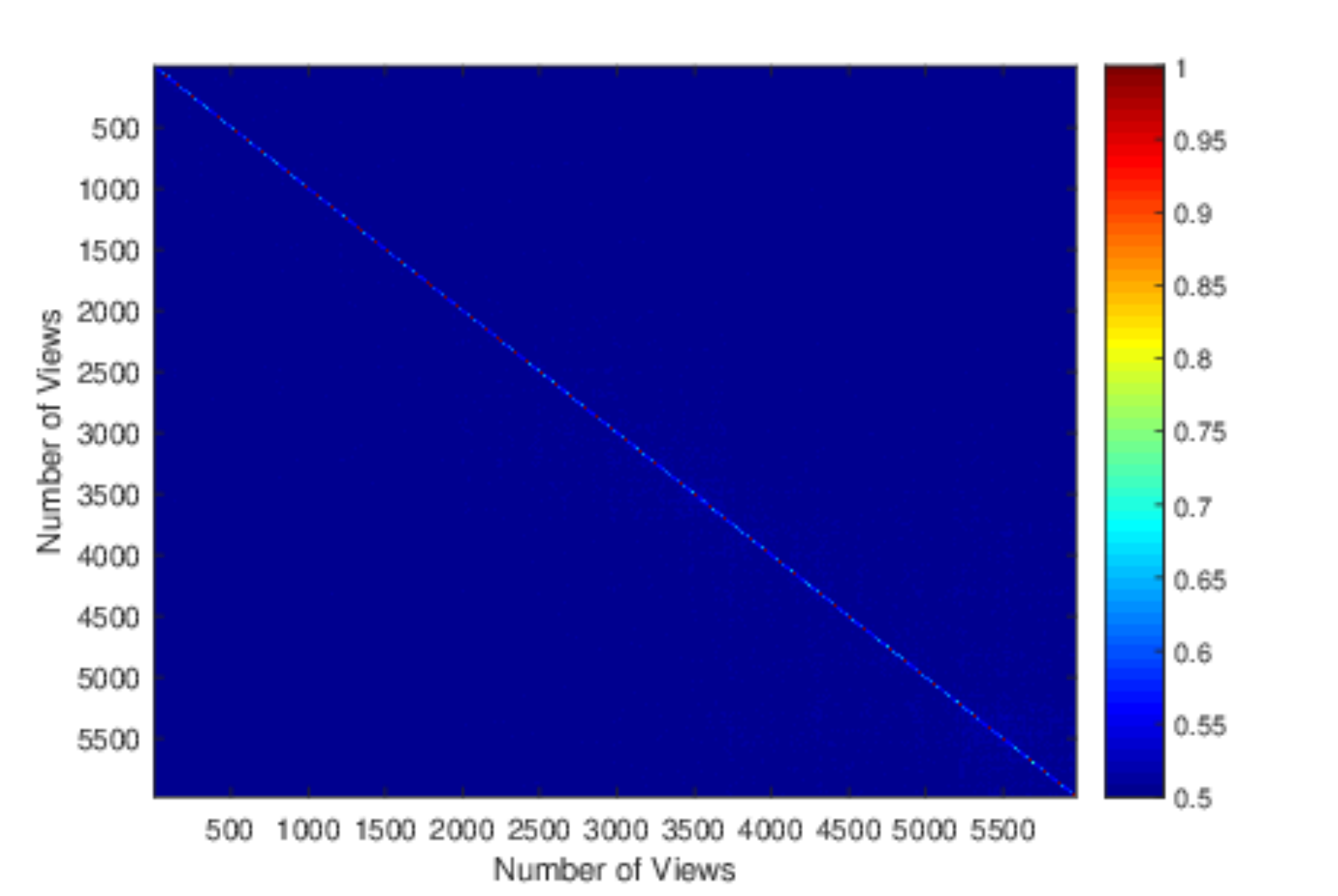}}
\subfigure[For EDTWK on Newcentry]{\includegraphics[width=0.49\linewidth]{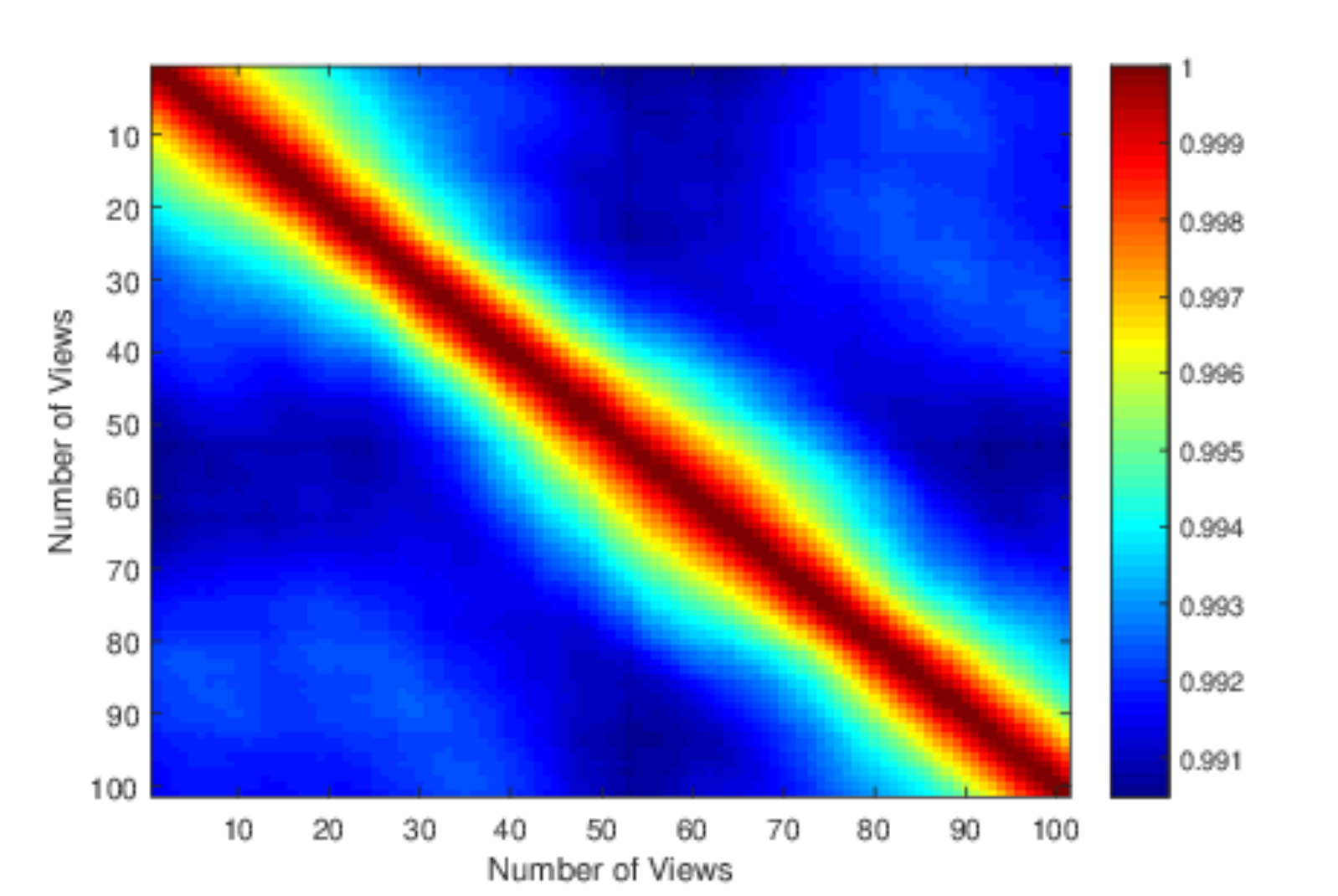}}
\subfigure[For WLSK on Newcentry]{\includegraphics[width=0.49\linewidth]{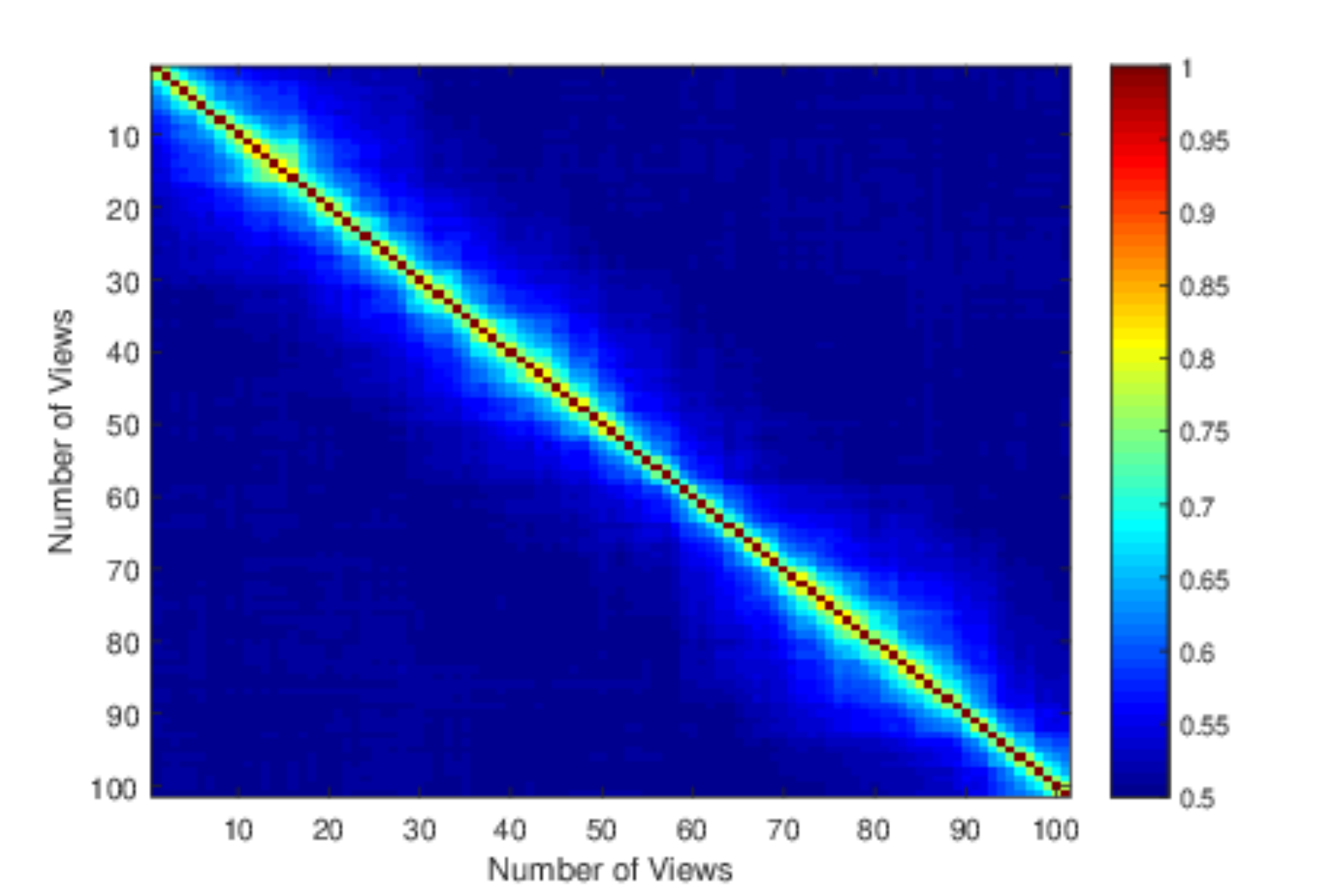}}
\subfigure[For EDTWK on Lehman]{\includegraphics[width=0.49\linewidth]{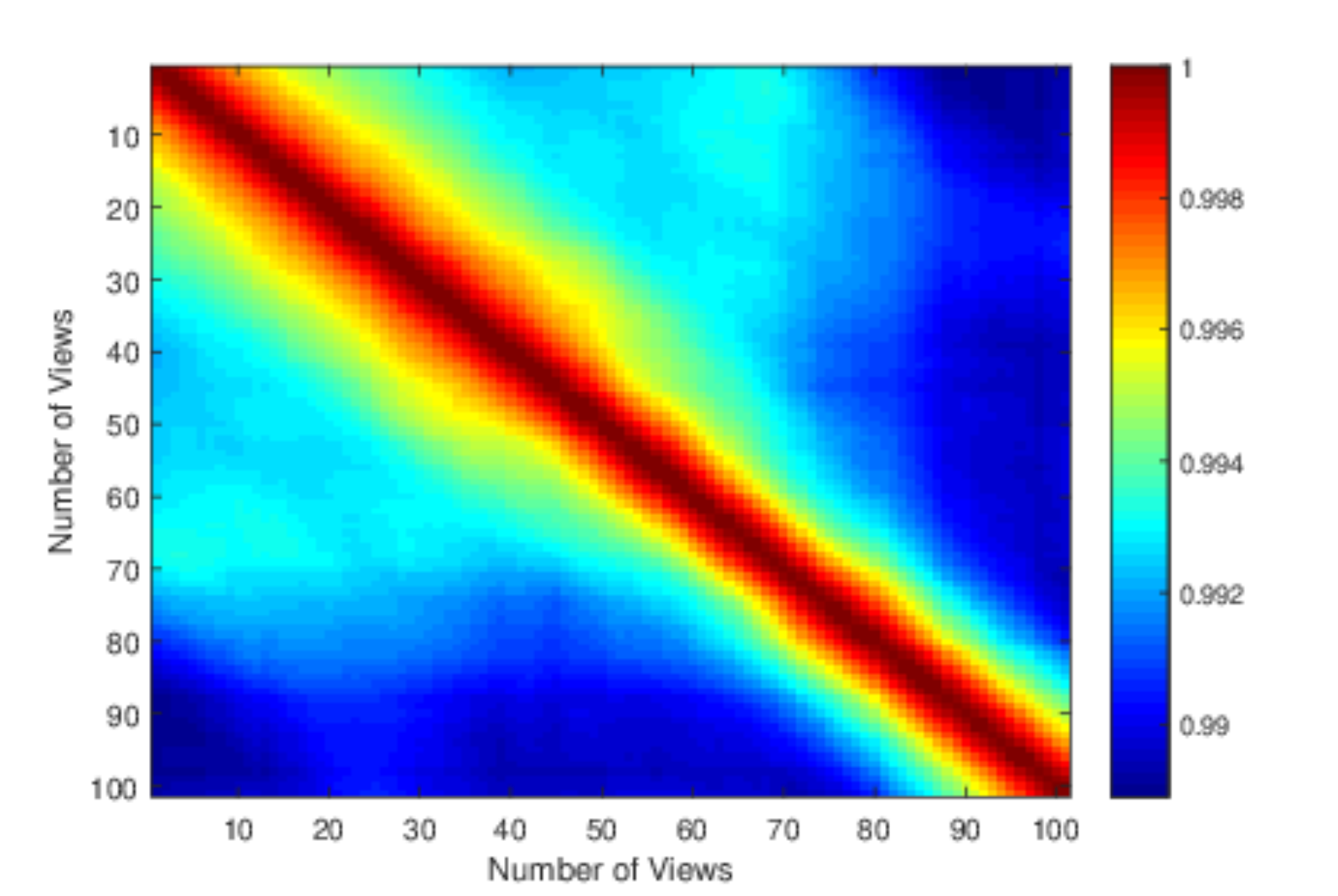}}
\subfigure[For WLSK on Lehman]{\includegraphics[width=0.49\linewidth]{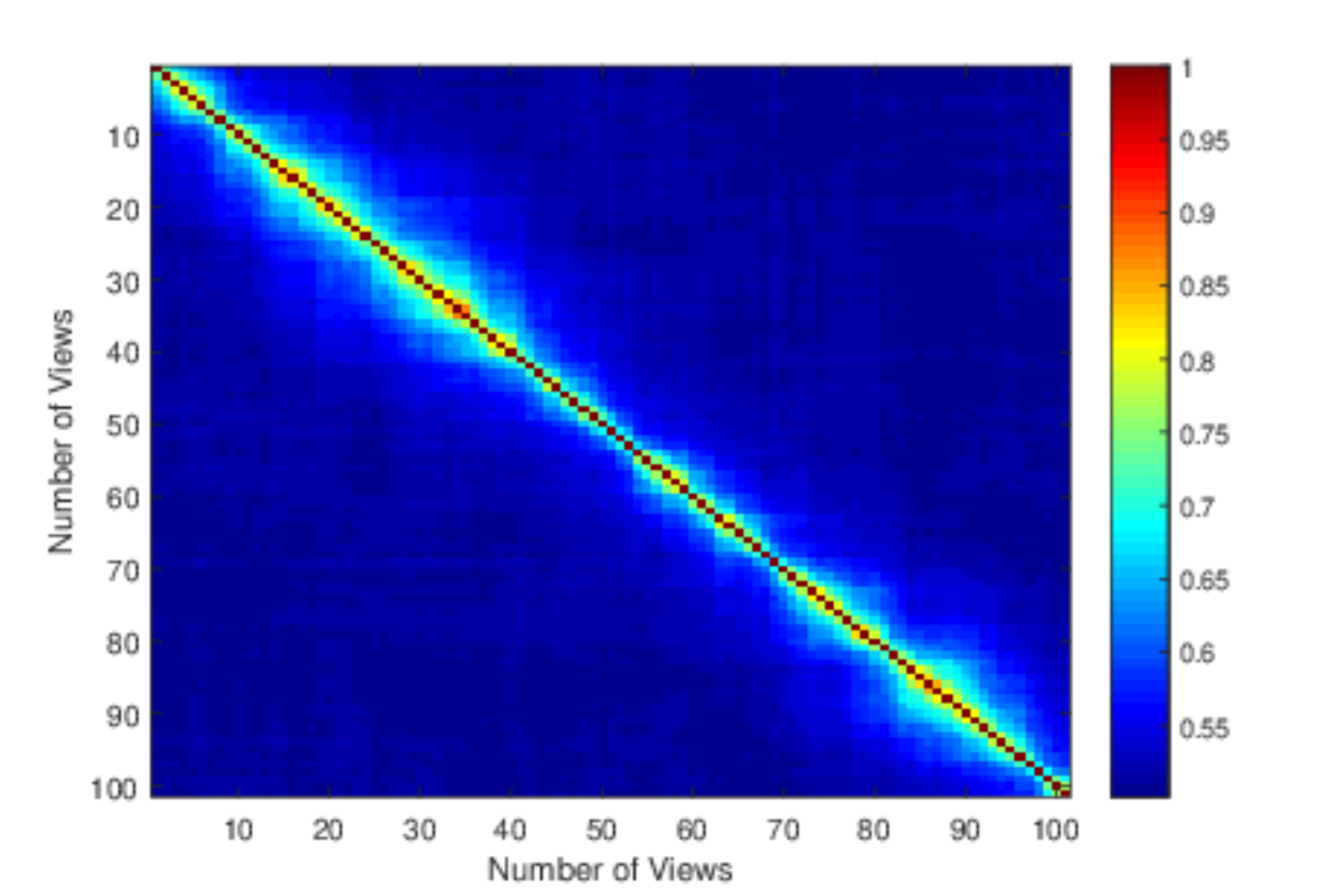}}
\vspace{-10pt}
\caption{Kernel Matrices Visualizations for EDTWK and WLSK Kernels.} \label{embeddingsKM}
\vspace{-20pt}
\end{figure}

To further reveal the effectiveness of the proposed EDTWK kernel, in this subsection we also draw the kernel matrices of both the proposed EDTWK kernel and the competitive alternative WLSK kernel. Note that, for another competitive QK kernel, we can observe the similar phenomenon with the WLSK kernel, thus we only exhibit the experiment with the WLSK kernel. The kernel matrices are computed between the networks belonging to the Newcentury crisis period and the Lehman crisis period, as well as that belonging to all the 6004 trading dats of the NYSE dataset. The kernel matrix visualization results are shown in Fig.\ref{embeddingsKM}, where both the x-axis and y-axis represent the time steps. Note that, to compare the two kernels in the same scaled Hilbert space, we consider the normalized version of both the kernels as $$k_n(G_p,G_q)=\frac{k(G_p,G_q)}{\sqrt{k(G_p,G_p)k(G_q,G_q)}},$$ where $k_n$ is the normalized kernel, and $k$ is either the EDTWK kernel or the WLSK kernel. As a result, the kernel values are all bounded between $0$ to $1$, and the colour bar beside each subfigure of Fig.\ref{embeddingsKM} represents the kernel value of the kernel matrix. Through Fig.\ref{embeddingsKM}, we observe that the kernel values tend to decrease when the elements of the kernel matrix are far away from the trace of the matrix. This is because such elements of the kernel matrix are computed between time-varying financial networks having long time spans, there are more structure changes when the network evolves with a long time variation. Thus, both the EDTWK and the WLSK kernels can reflect the structural evolution of the financial networks with time. However, on the other hand, we observe that the kernel values associated with the WLSK kernel tend to suddenly drop down when the element is a little far from the trace. By contrast, the kernel values associated with the proposed EDTWK kernel tend to gradually decrease when the element gets far away from the trace. This observation reveal the reason why only the kPCA embeddings through the proposed EDTWK kernel can form clear trajectory with time variation in Fig.\ref{embeddingsSPC}. Therefore, only the proposed EDTWK kernel can well distinguish and understand the structural changes of the network structures evolving with a long time period. Note that, we can observe similar results if we explore the proposed kernel and the remaining methods on the alternative financial crisis mentioned in Section~\ref{exp:s1}.

All the above experiments demonstrate the effectiveness of the proposed kernel. The reasons for the effectiveness are fourfold. First, unlike the WLSK kernel, the proposed kernel can directly accommodate the time-varying networks that are complete weighted graphs. Second, unlike the GC method that computes vectorial network characteristics and tends to approximate the time-varying networks in a low dimensional pattern space, the proposed kernel can represent the network structures in a high dimensional Hilbert space and thus better preserve the network characteristics. Third, unlike the GAK for original time series, the proposed kernel for time-varying networks is based on the dominant entropy time series and reflects richer correlated information than the original time series. Fourth, unlike the QJSK and FLK kernels, only the proposed kernel can capture reliable financial information through the dominant entropy time series.

\subsection{Classifications from C-SVMs}\label{exp:s5}
\begin{table*}
\centering {
\scriptsize
\caption{Classifying Time-varying Networks through Different Kernels Associated with C-SVMs.}\label{T:ClassificationGK}
\vspace{0pt}
\begin{tabular}{|c||c||c||c||c||c|}

  \hline
 ~Datasets~        & ~Black Monday  ~ & ~Newcentury~     & ~Lehman~           & ~Asia 1997~              & ~Dot-com~    \\ \hline \hline

 ~\textbf{EDTWK}~  & ~$\textbf{91.47}\pm0.54$~ & ~$\textbf{88.79}\pm0.63$~ & ~$\textbf{88.00}\pm0.65$~   &  ~$\textbf{90.82}\pm0.57$~   & ~$\textbf{91.03}\pm0.61$~  \\ \hline

  ~QJSK~           & ~$89.71\pm0.78$~ & ~$88.32\pm0.74$~ & ~$87.85\pm0.81$~   &  ~$89.88\pm1.11$~   & ~$89.17\pm0.72$~  \\ \hline

  ~WLSK~           & ~$88.70\pm0.51$~ & ~$88.30\pm0.80$~ & ~$86.36\pm0.88$~   &  ~$85.44\pm0.91$~   & ~$88.03\pm0.71$~  \\   \hline

  ~QK~           & ~$90.10\pm0.51$~ & ~$88.50\pm0.80$~ & ~$87.69\pm0.88$~   &  ~$90.21\pm0.91$~   & ~$89.79\pm0.71$~  \\   \hline

\end{tabular}
} \vspace{-10pt}
\end{table*}

In this subsection, we validate the effectiveness of the proposed EDTWK kernel on classification tasks. Specifically, we explore whether the proposed kernel can be used to correctly classify the time-varying financial networks into corresponding stages of each financial crisis period. The crisis periods for evaluations include the Black Monday period, the Dot-com Bubble period, the Newcentury Financial crisis period, the Lehman Crisis period and the 1997 Asia Financial crisis period. Since the Enron crisis is not an emergency crisis and continued for many trading days, we do not perform the classification evaluation on this crisis. For each of these selected financial crisis periods, we utilize 100 trading days around the particular day when the crisis happens, i.e., we respectively select 50 days before and after the crisis event. For each crisis, we sequentially divide the 100 continuous trading days into 10 stages and each stage contains 10 trading days, i.e., the time-varying financial networks of the 100 trading days are sequentially separated into 10 classes. For each financial crisis period, we calculate the kernel matrix between the financial networks of the trading days. Moreover, for the proposed kernel, we perform 10-fold cross-validation using the C-Support Vector Machine (C-SVM) Classification to compute the classification accuracies, using LIBSVM~\cite{ChangLinSVM2001}. We use nine samples for training and one for testing. All the C-SVMs were performed along with their parameters optimized on each dataset. We repeat the whole experiment 10 times and report the average classification accuracies and standard errors in Table.\ref{T:ClassificationGK}. We also compare the proposed EDTWK kernel with the competitive WLSK and QK kernels~\cite{DBLP:journals/jmlr/ShervashidzeVPMB09,DBLP:journals/ijon/CuiBZWH19}, as well as the QJSK kernel~\cite{DBLP:journals/pr/Bai0TH15}, and the evaluations associated with these kernels follow the same experimental setup of the proposed kernel. The experiment results are also reported in Table.\ref{T:ClassificationGK}. It is clear that the proposed EDTWK kernel outperforms the alternative state-of-the-art graph kernels on the evaluation of any crisis period, and the proposed kernel can well classify each financial network into correct time-varying stages. This evaluation demonstrates that the proposed kernel has better ability to understand how the structures of the financial networks evolve with time.

\section{Conclusion and Future Work}\label{s5}
In this paper, we have proposed a new dynamic time warping framework inspired kernel, namely the Entropic Dynamic Time Warping Kernels between time-varying financial networks for multiple co-evolving financial time series analysis. Specifically, for a family of time-varying financial networks with each vertex representing the individual time series of a stock and each edge between pairwise series representing the correlation, we have computed the commute time matrix on each of the network structures and shown how this matrix allows us to identify a dominant correlated stock set as well as the associated dominant probability distribution of these stocks belonging to this set. Based on the probability distribution, we have represented each original network as dominant Shannon entropy time series. With the dominant entropy time series for each pair of financial networks to hand, the proposed kernel has been defined through the dynamic time warping based global alignment kernel between the entropy time series. We have shown that the proposed kernel bridges the gap between graph kernels and the classical dynamic time warping framework for time series analysis. Experiments on time-varying networks extracted from New York Stock Exchange (NYSE) database demonstrate the effectiveness.

In this work, we have identified the most strongly correlated stock trading patterns based on the compute-time  between pairs of stocks in a network of trading relationships. Commute time implicity averages over all paths connecting a pair of stocks in the network, and not just the first-order nearest neighbour relations. This renders it robust to missed or erroneously inferred correlation relations in the time series of the stock closing prices. In the future, we will explore the use of hypergraph representations employing the relationships between multiple stocks. Here we will use the commute time to define the groups of stock and develop a hypergraph kernel to measure the relationships between groups.


\section*{Acknowledgments}
This work is supported by the National Natural Science Foundation of China (Grant no. 61976235, 61602535 and 61503422), the Open Project Program of the National Laboratory of Pattern Recognition (NLPR), and the program for innovation research in Central University of Finance and Economics. Primary Contract Authors: Lu Bai (bailucs@cufe.edu.cn) and Lixin Cui (cuilixin@cufe.edu.cn). Lu Bai and Lixin Cui have equal contributions.

\balance


\bibliographystyle{IEEEtran}
\bibliography{ijcai17}

\end{document}